\documentclass[twocolumn,superscriptaddress,showpacs,amssymb,amsmath,amsfonts,aps]{revtex4}
\input{epsf}
\usepackage{epsfig}
\usepackage{latexsym}
\usepackage{amssymb}
\usepackage{graphicx}
\usepackage[dvips]{color}

\begin{document}
\date{\today}
\title {\large {\bf Model Analysis of the $e p \rightarrow e'p\pi^+\pi^-$ 
Electroproduction Reaction on the Proton.}} 
\newcommand*{\JLAB}{Thomas Jefferson National Accelerator Facility, Newport
News, Virginia 23606, USA} 
\newcommand*{\ANL}{ Physics Division, Argonne National Laboratory, 
Argonne, Illinois 60439, USA}
\newcommand*{\EBAC}{Excited Baryon Analysis Center,Thomas Jefferson National
Accelerator Facility, Newport News, Virginia 23606,  USA}
\newcommand*{\MOSCOW}{Moscow State University, Skobeltsyn Institute of 
Nuclear Physics, 119899 Moscow, Russia} 
\newcommand*{\MOSCOWW}{Moscow State University,Physics Department, 119899 Moscow, Russia}

\author {V.I.~Mokeev} 
\affiliation{\JLAB} \affiliation{\MOSCOW}
\author {V.D.~Burkert} 
\affiliation{\JLAB}
\author {T.-S.H.~Lee}
\affiliation{\ANL} \affiliation{\EBAC}
\author {L.~Elouadrhiri} 
\affiliation{\JLAB}
\author {G.V.~Fedotov} 
\affiliation{\MOSCOW}
\author {B.S.~Ishkhanov}
\affiliation{\MOSCOW} \affiliation{\MOSCOWW}

\begin{abstract}
Recent CLAS data on the $p\pi^+\pi^-$
electroproduction off protons at 1.3$<$W$<$1.57 GeV and
0.25$<$$Q^{2}$$<$0.6 GeV$^{2}$ have been analyzed using a meson-baryon 
phenomenological 
model. By fitting nine 1-fold differential cross section data for each $W$ and
$Q^{2}$ bin,   
the charged double pion 
electroproduction mechanisms are identified from their manifestations in 
the observables. We have extracted the cross sections from amplitudes
of each of the considered
isobar channels as well as from their coherent sum.
We also obtained non-resonant 
partial wave amplitudes of all contributing isobar channels which could be 
useful for advancing a 
complete coupled-channel analysis of all meson electroproduction data.
\end{abstract}


\maketitle
\normalsize

\section{Introduction}

Experiments with the
CLAS detector at Thomas Jefferson National Accelerator Facility (JLab)
have accumulated extensive and accurate data of 
meson electroproduction reactions on protons.
Detailed experimental data  have now become 
available for the $p\pi^0$, $n\pi^+$ and $p\pi^{+}\pi^{-}$ exclusive channels
\cite{Joo1,Joo2,Joo3,Egiyan,Ungaro,Smith,Park,Biselli,Ri03,Fe07a,Fe07}.
In addition first electroproduction data for the channels with smaller 
cross sections, such as the $p\eta$, 
and $KY$, have also been
obtained with CLAS 
\cite{De07,Thom01,Carm09,Carm07,Ambr07}. Review of these CLAS experimental results may be found
in the papers \cite{Bu07,Bu07a,Bu05,Bu04}. 
These data combined will 
allow us to determine the
electromagnetic transition amplitudes for the majority of nucleon 
resonances ($N^*$)
in a wide range of photon virtualities $Q^{2}$ from 0.2 to 5.0 GeV$^{2}$. 
The $Q^{2}$ evolution of these amplitudes contains information on the 
relevant degrees of freedom
in the nucleon resonance structure at varying distance scales. It also allows 
to explore the
strong interaction mechanisms responsible for baryon formation and their
relationship to QCD \cite{Ro07,Ch09,Rich07,Lin:2008pr,Bra09,Brod04,Me02,Wp08}.

The $N\pi$ and $N\pi\pi$ exclusive  channels are 
two major contributors 
to the $\gamma N$ and $N(e,e')$  reactions
in the resonance excitation region. They are strongly coupled through hadronic
interactions of the final $N\pi$ and $N\pi\pi$ states \cite{Bu07,Pe02}
Thus an understanding of 
the reaction mechanisms for these two channels
is vital for extracting the $N$-$N^*$ transition amplitudes at various photon
virtualities. It is also a necessary first step towards 
 the exploration of $N^{*}$'s in the mass region above 1.5
GeV using other exclusive channels with
smaller
cross sections, such as $p\eta$, $p\omega$, and $KY$. 
The cross sections of these weaker channels 
can be affected considerably by the $N\pi$
and $N\pi\pi$ channels because of unitarity condition.
The mechanisms of the $N\pi$ electroproduction have been extensively studied. The most
comprehensive recent studies are by the efforts at MAMI \cite{MAID,Dr08}, GWU
\cite{Ar08} and
JLab \cite{Az08a,Bu08c,Az05-1,Az03}. In contrast, the 
study of  $N\pi\pi$ electroproduction mechanisms is still limited.
In this work we try to improve the situation by
analyzing the recently published CLAS data of $p\pi^+\pi^-$ channel 
\cite{Fe07a}.

The world data  on $p\pi^{+}\pi^{-}$ electroproduction in the 
resonance excitation region were rather
scarce before the CLAS experiments. 
They were limited to integrated cross sections, 
various invariant mass distributions, 
and $\pi^{-}$ angular distributions \cite{Wa78}. 
Furthermore, the large binning in $W$ and $Q^{2}$ does not allow the determination 
of $N^{*}$ parameters from these data. 

First detailed  
measurements of $p\pi^+\pi^-$ electroproduction cross sections  
 with CLAS were reported in \cite{Ri03}.
The new  CLAS $p\pi^+\pi^-$ data \cite{Fe07a,Fe07} use binnings of 
$\Delta W =25$ MeV and
$\Delta Q^2 =$ 0.05 GeV$^2$, and are much more precise than the
previous data. For the first time nine sets of 1-fold differential 
cross sections were determined in the region of
1.3 $<$ W $<$ 1.6 GeV and 0.25  $<$ Q$^{2}$ $<$ 0.60 GeV$^{2}$.
They consist of invariant mass distributions of the final
$\pi^{-} \pi^{+}$, $\pi^{+} p$, and $\pi^{-} p$ and the angular
distributions for all final state particles, as will be described in the
Section~\ref{xsect}.
All cross sections measured 
in this reaction and in others CLAS
experiments may be found in
\cite{db07}.

Several models have been developed 
for analyzing the 
double pion photoproduction \cite{Os1,La96,Hi03,Fi05,Ki06} and electroproduction
\cite{Os2}  reactions
in the nucleon resonance excitation region,
beginning with the pioneering effort of \cite{So71}. They are 
based on the tree-diagrams of effective meson-baryon Lagrangians and have been 
reasonably successful in describing the very limited  data of the
fully integrated cross sections and the
invariant mass distributions. It remains to be seen, to what  extent 
these models can describe the recent CLAS data. A comparison between these model
predictions and $N\pi\pi$ cross sections (e.g.
~\cite{As03}) clearly showed the necessity for further improvements of 
reaction models in order to isolate the resonant contribution and
to determine nucleon resonance parameters from the data. 
Comprehensive CLAS data on various  differential $p\pi^+\pi^-$ cross
open up new opportunities for phenomenological data analysis. 
By studying 
the kinematic dependencies
of the differential cross section and their correlations we are able to establish
the presence and strength of the relevant reaction mechanisms and to achieve
reliable
separation of resonant and non-resonant contributions.
\begin{figure*}[ht]
\begin{center}
\hspace{-4cm}
\epsfig{file=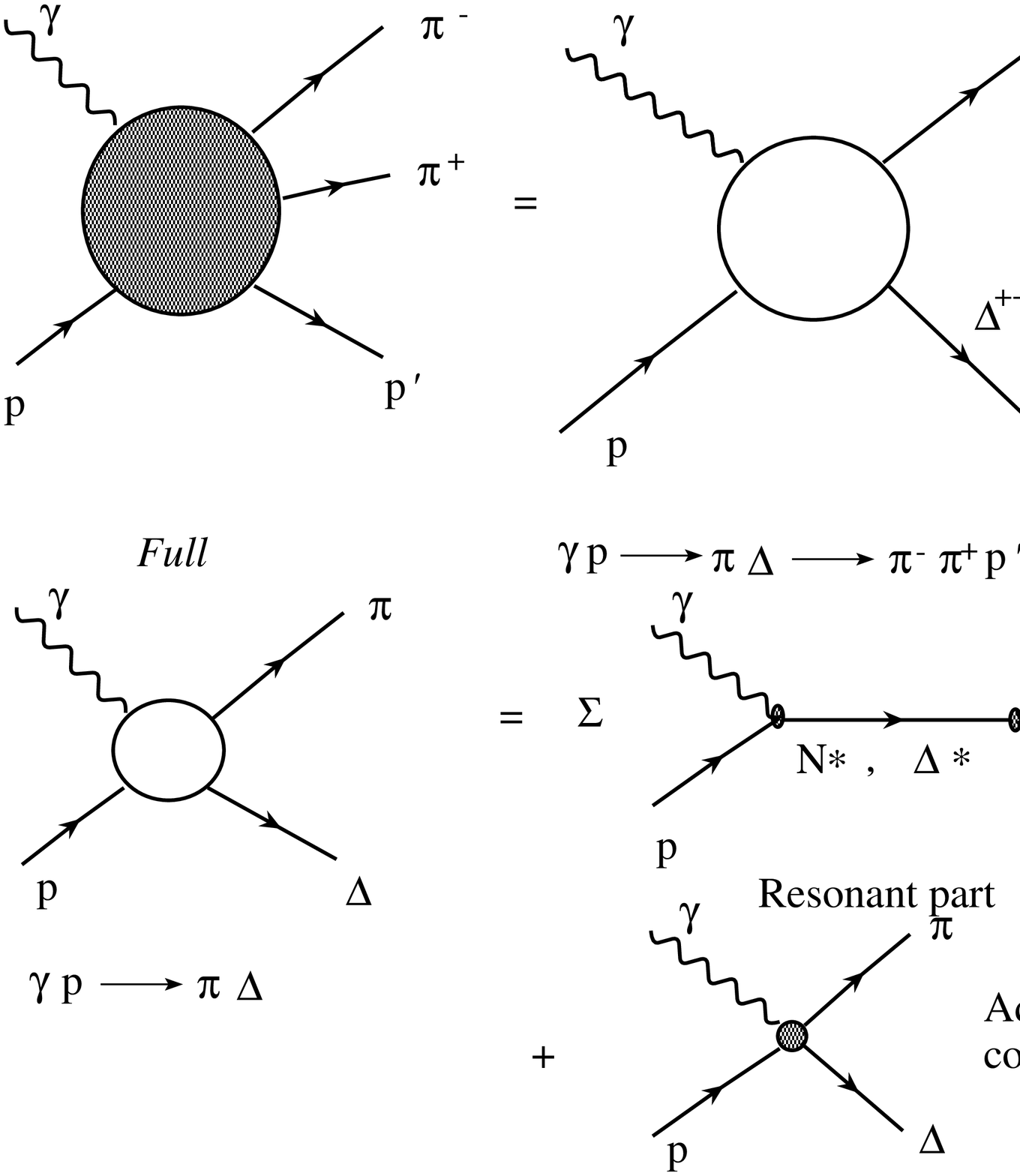,width=8cm}
\end{center}
\begin{center}
\hspace{-4cm}
\epsfig{file=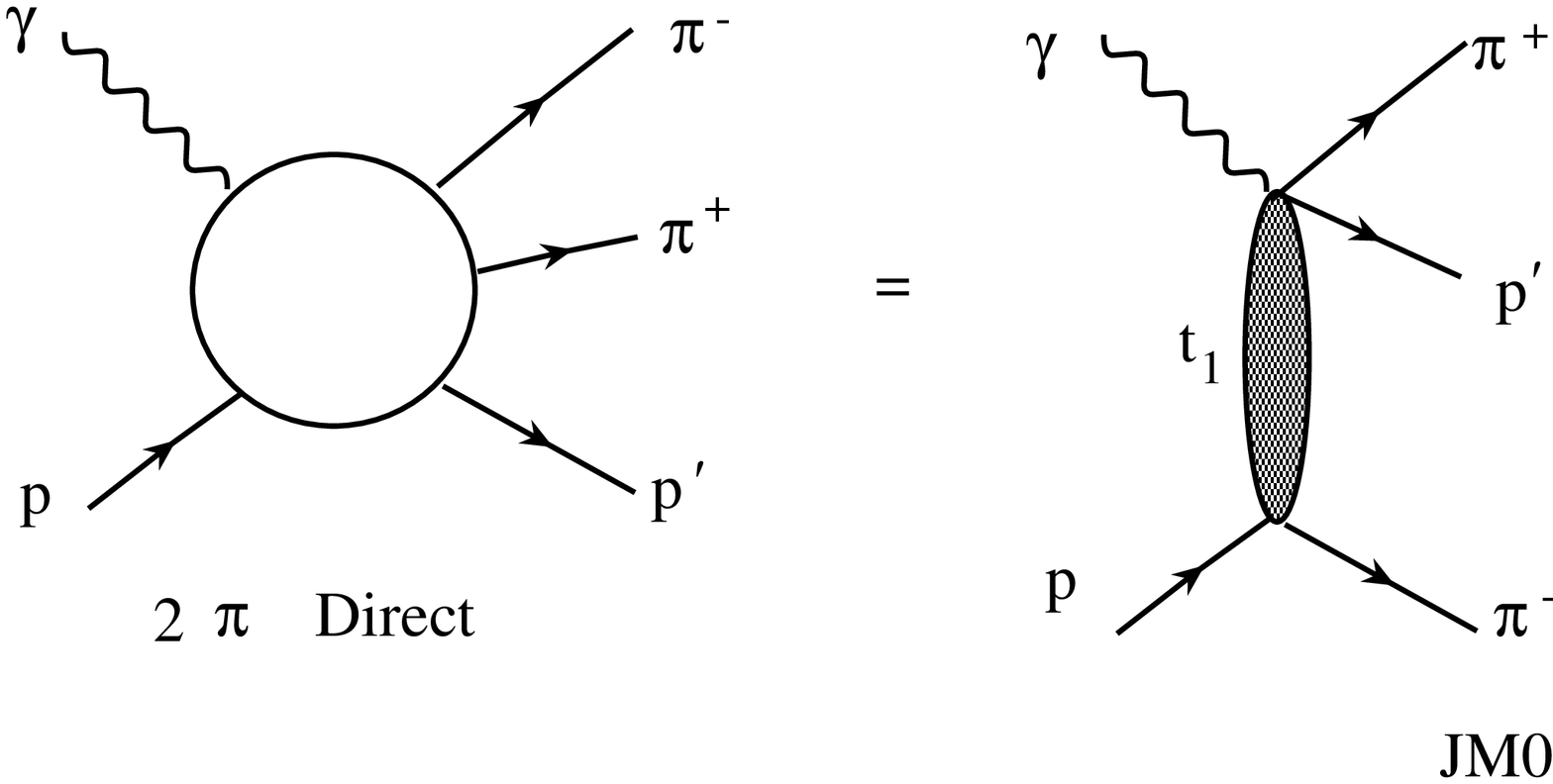,width=8cm}
\end{center}
\vspace{0.6cm}
\caption{\small The mechanisms of JM05 model \cite{Mo06,Az05} contributing to
$p\pi^{+}\pi^{-}$ electroproduction in kinematic area covered by recent CLAS
measurements \cite{Fe07a}: W $<$ 1.6 GeV and 0.25 $<$ $Q^{2}$ $<$
0.6 GeV$^{2}$}
\label{mech_low_06}
\end{figure*}

In this paper we apply a 
 phenomenological approach 
developed in the past several years by 
the  Jefferson Laboratory-Moscow
State University (JM) collaboration 
\cite{Ri00,Mo01,Mo03,Sh07,Az05,Mo06,Mo06-1,Mo07,Mo09,Fe07b}, to analyze
the CLAS 
data \cite{Fe07a,Fe07} on $p\pi^+\pi^-$ electroproduction 
at $W$ $<$ 1.6 GeV.     
Within the JM model developed up to 2005, called JM05,
the major part of $p\pi^+\pi^-$ production  at $W$ $<$ 1.6 GeV is due to 
contributions from the $\pi \Delta$ isobar channels.
The $\Delta$ (1232) resonance
is clearly seen in all $\pi^{+} p$ mass distributions at W $>$ 1.4
GeV, while other mass distributions do not show any structures. 
The contributions from all other isobar channels $p\rho$,
$\pi^{+}D_{13}^{0}(1520)$, $\pi^{+}F_{15}^{0}(1685)$ and
$\pi^{-}P_{33}^{++}(1640)$ included in JM05 
\cite{Sh07,Az05,Mo06,Mo06-1} are negligible in the near-threshold and 
sub-threshold 
regions. Thus these channels are not included to this work.
The $\gamma_{v} p \rightarrow p\pi^+\pi^-$ 
production amplitude within JM05 can then be written as
\begin{eqnarray}
T_{\gamma_{v} N,\pi\pi N} =  T^{\pi\Delta}_{\gamma_{v}N, \pi\pi N} 
+T^{dir}_{\gamma_{v}N, \pi\pi N}
\label{eq:full-t}
\end{eqnarray}
with
\begin{eqnarray}
T^{\pi\Delta}_{\gamma_{v} N,\pi\pi N} = [t^R_{\gamma_{v} N,\pi\Delta}+
 t^{Born}_{\gamma_{v} N,\pi\Delta} 
+ t^{c}_{\gamma_{v}N,\pi\Delta}]G_{\Delta}\Gamma_{\Delta,\pi N}\,
\label{eq:pid-t}
\end{eqnarray}
where $G_{\Delta}$ is the propagator of the $\Delta$ intermediate
state , and the vertex function
$\Gamma_{\Delta,\pi N}$ describes the
$\Delta (1232) \rightarrow \pi N$ decay. Explicit expressions for  these
amplitudes may be found in Appen. I-III.
Contributing mechanisms in equations (\ref{eq:full-t},\ref{eq:pid-t})are illustrated
in Fig.~\ref{mech_low_06}. The diagram 
$\gamma N \rightarrow N^*,\Delta^* \rightarrow \pi\Delta$ in the second row is 
the resonant term $t^{R}_{\gamma_{v} N,\pi\Delta}$ 
in Eq.(\ref{eq:pid-t}).
It is parameterized as a  Breit-Wigner
form \cite{Ri00}  and calculated from
all well established $N^*$, $\Delta^{*}$ states with
masses less than 2.0 GeV which have hadronic decays to the $N\pi$$\pi$ final 
states. 
The non-resonant term $t^{Born}_{\gamma_{v} N,\pi\Delta}$
is calculated from the
 well established Born terms 
 of $\gamma_{v} N \rightarrow \pi \Delta$~\cite{So71,Ri00}. 
 Their amplitudes are
 presented in Appendix I. 
The additional contact term $t^{c}_{\gamma_{v} N,\pi\Delta}$ was
introduced in ~\cite{Mo06,Mo06-1,Mo07}  to
account  phenomenologically for 
the other possible production mechanisms through the $\pi \Delta$
intermediate states, as well as for FSI effects. Parametrization of these
amplitudes may be found in Appendix II.
The diagrams in the bottom of Fig.~\ref{mech_low_06} represents
the direct term $T^{dir}_{\gamma_{v} N,\pi\pi N}$ in
Eq.(\ref{eq:full-t}) which was introduced in \cite{Az05,Mo06-1} to
describe the direct $\gamma N \rightarrow
N\pi\pi$ mechanisms. It was  parameterized in JM05 \cite{Az05,Mo06-1}
in terms of a contact vertex and a particle-exchange amplitude .
The parameterization of this term in JM05 will be given explicitly 
in Section \ref{jlabmsugen}.

\begin{figure}[b]
\begin{center}
\includegraphics[width=8cm]{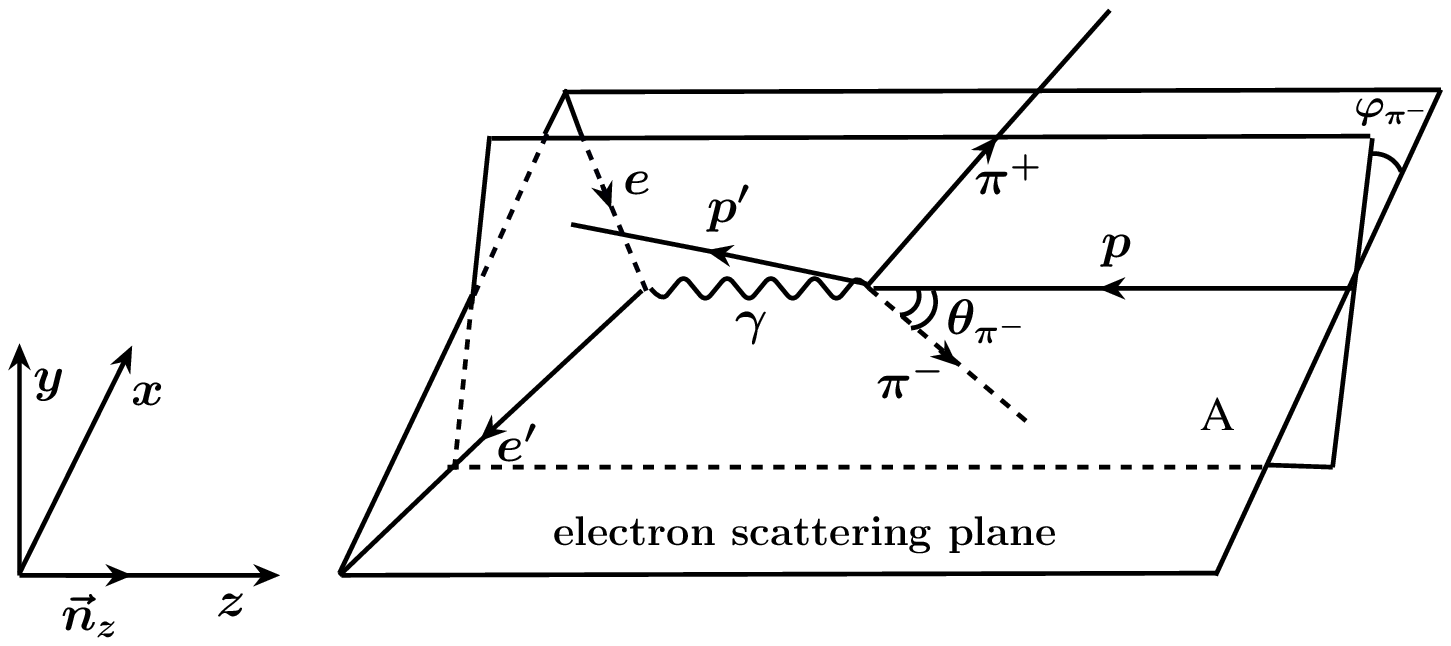}
\includegraphics[width=8cm]{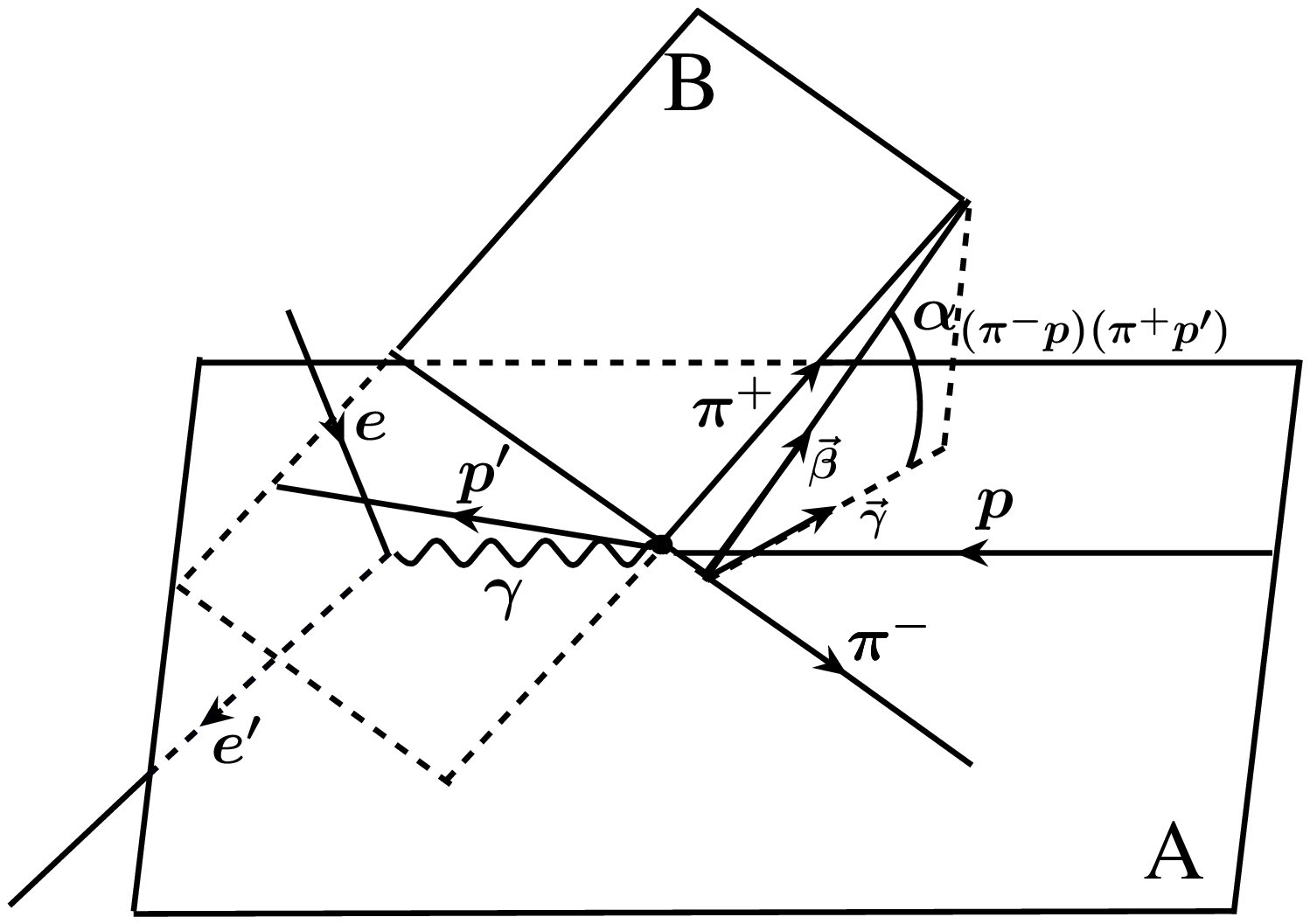}
\caption{\small The kinematics variables for
description  of
$e p \rightarrow e' p' \pi^{+} \pi^{-}$ reaction in the CM frame of the final
hadrons (first assignment presented
in the Section~\ref{xsect}). The top plot shows
$\pi^{-}$ spherical angles $\theta_{\pi^{-}}$ and $\varphi_{\pi^{-}}$
while the bottom plot
shows angle $\alpha_{[p\pi^{-}][p'\pi^{+}]}$  between two planes: one of them (plane A)
is defined by the 3-momenta of the initial proton and the final $\pi^{-}$,
a second (plane B) is defined by the 3-momenta of the two others final
hadrons $\pi^{+}$ and p.} \label{kinematic}
\end{center}
\end{figure}

\begin{figure*}[ht]
\vspace{2cm}
\begin{center}
\epsfig{file=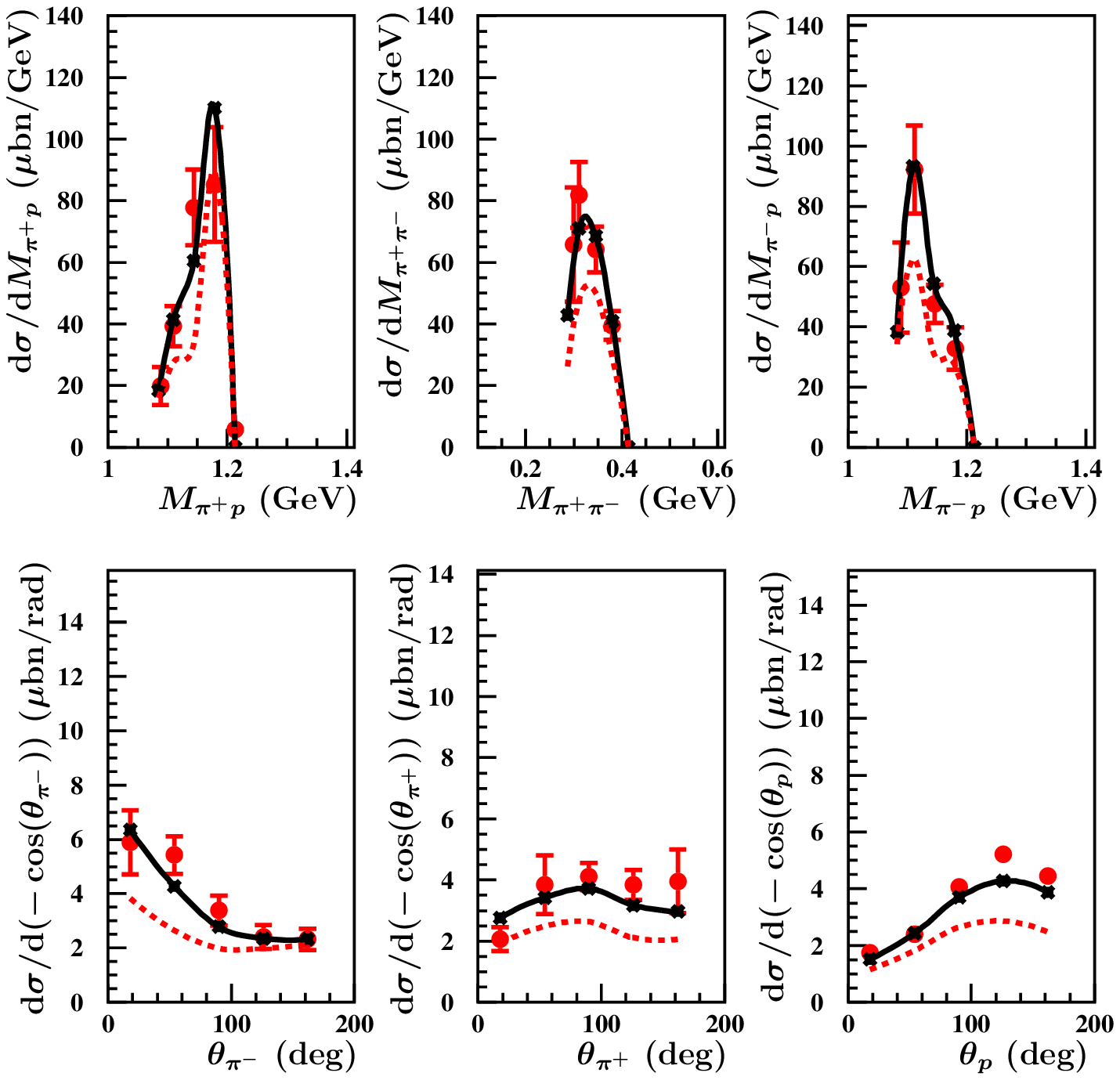,width=8cm}
\hspace{1cm}
\epsfig{file=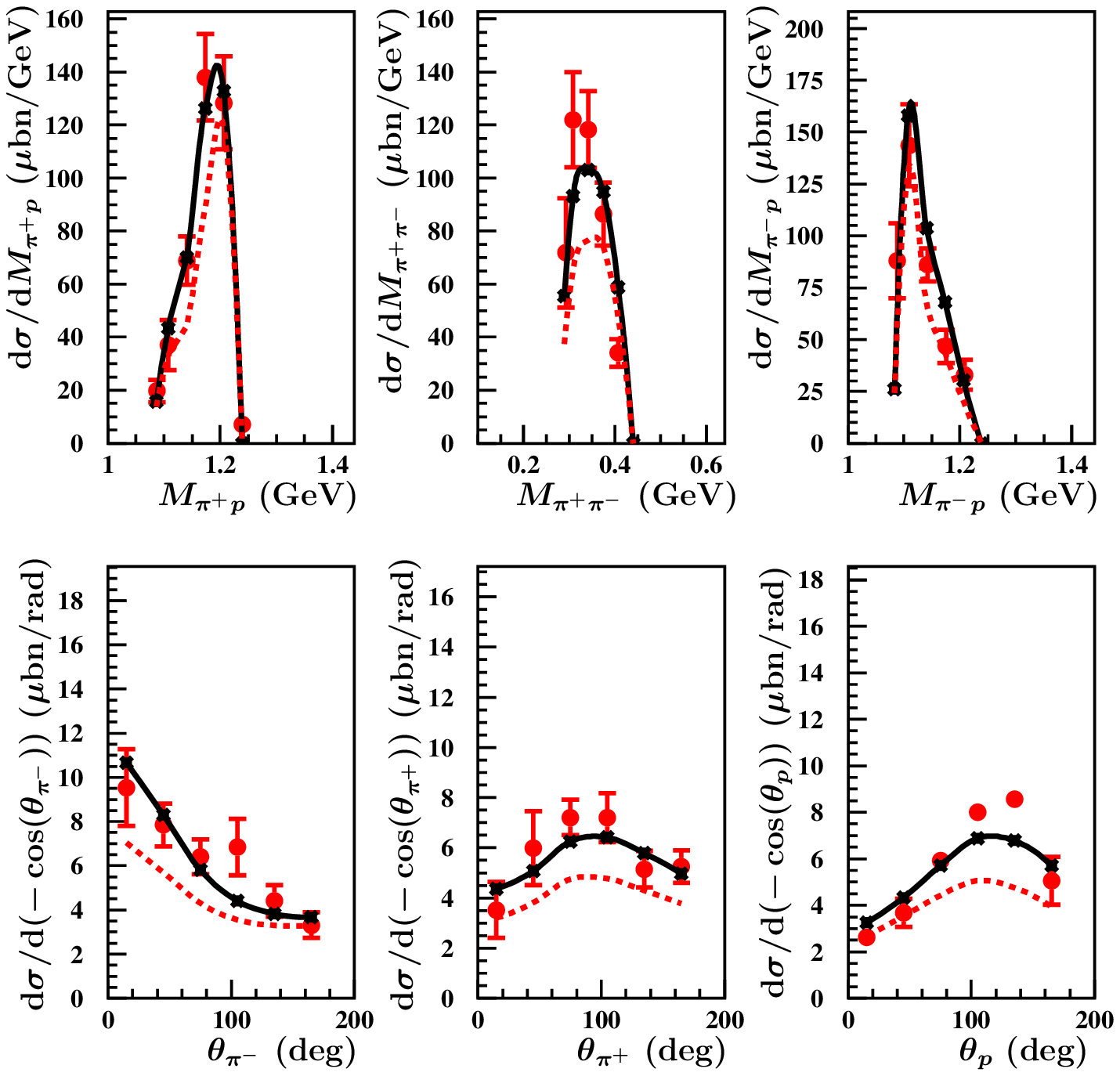,width=8cm}
\end{center}
\vspace{0.2cm}
\caption{\small (color online) Manifestation of additional contributions to the 
$\pi \Delta$ isobar channels with respect to the resonant and Born terms, 
parametrized by extra contact terms, described in the Section~\ref{jlabmsugen}. 
The data \cite{Fe07a}
at W=1.33 GeV (six panels at the left side), W=1.36 GeV (six panels at the
right side) and $Q^{2}$=0.425 $GeV^{2}$ are shown by full symbols.
Full JM06 calculations are 
shown by solid lines, while the same cross sections, calculated taking
off the
contributions from 
additional contact terms in $\pi \Delta$ channels, are shown by dotted lines.
 }
\label{contact}
\end{figure*}

\begin{figure}[t]
\begin{center}
\includegraphics[width=9cm]{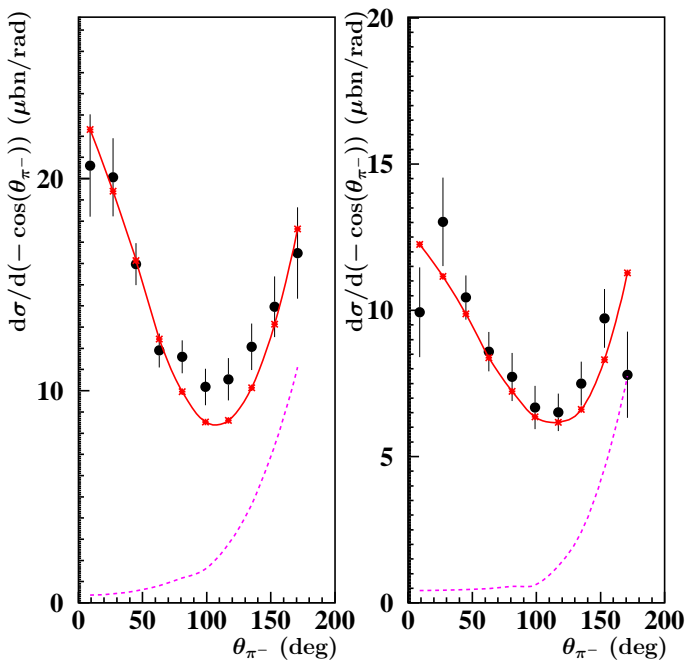}
\caption{\small(color online) A signature of direct 2$\pi$ production mechanisms, 
seen in the previous 
CLAS $p\pi^+\pi^-$ data \cite{Ri03}. 
The  data on $\pi^{-}$ angular distributions at 
W=1.49 GeV, $Q^{2}$=0.65 $GeV^{2}$ (left) and at W=1.49 GeV, 
$Q^{2}$=0.95 $GeV^{2}$ (right) are shown by points with error bars. 
The full calculation are
shown by solid
lines.The dashed lines
correspond to the
contributions from JM05 direct 2$\pi$ production \cite{Az05}.}
\label{dhq2}
\end{center}
\end{figure}

\begin{figure*}[ht]
\begin{center}
\epsfig{file=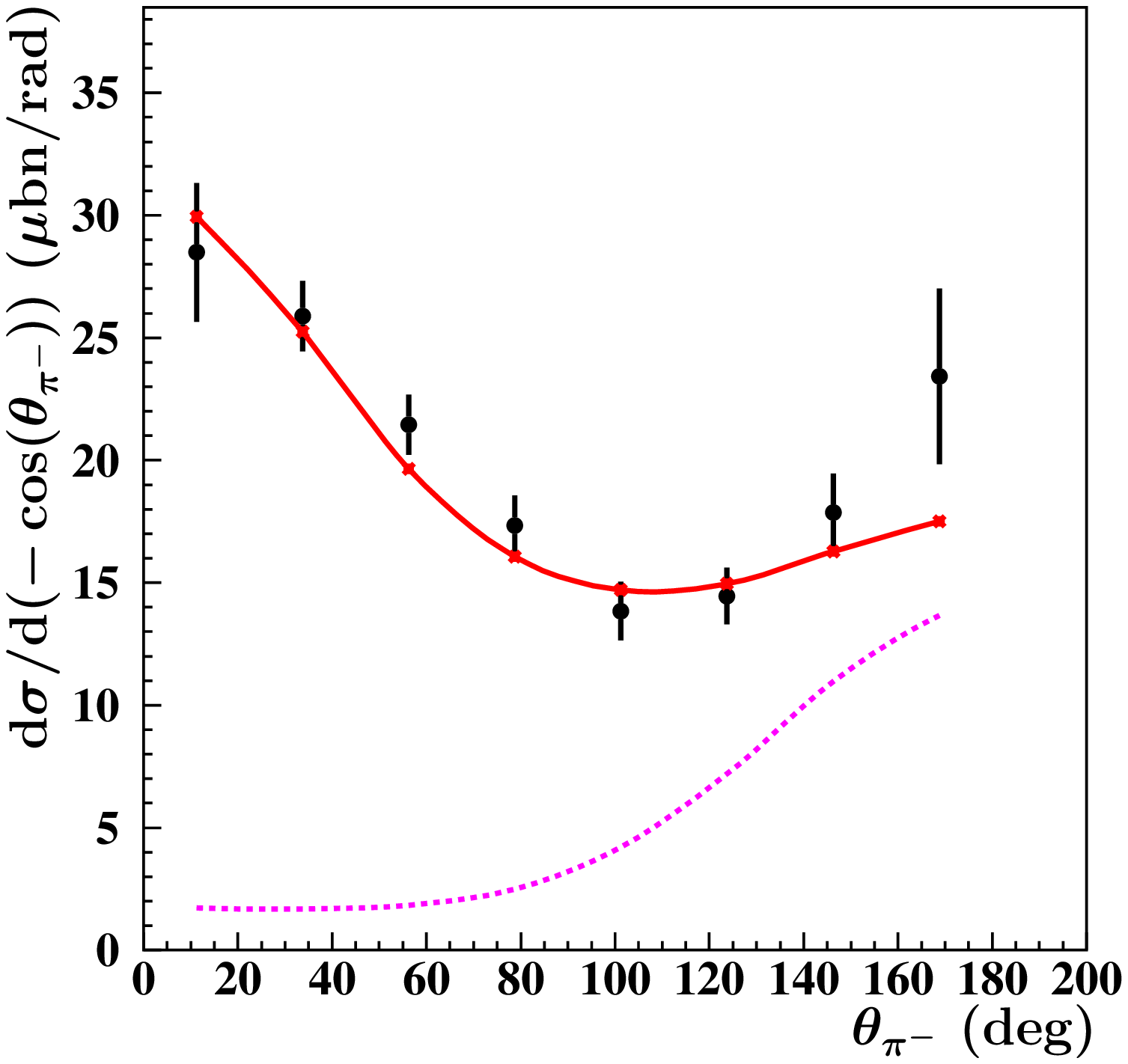,width=7.cm}
\epsfig{file=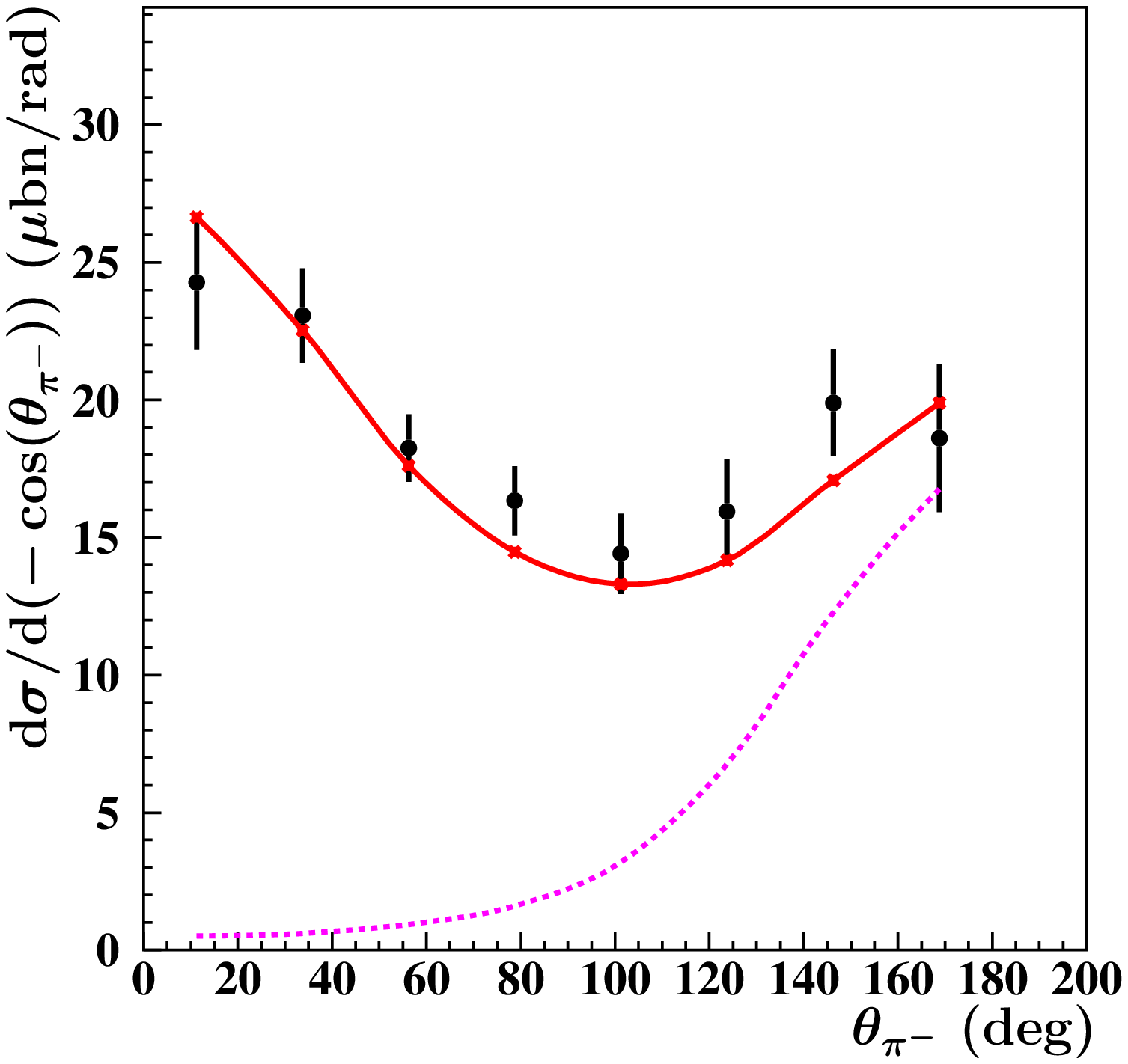,width=7.cm}
\epsfig{file=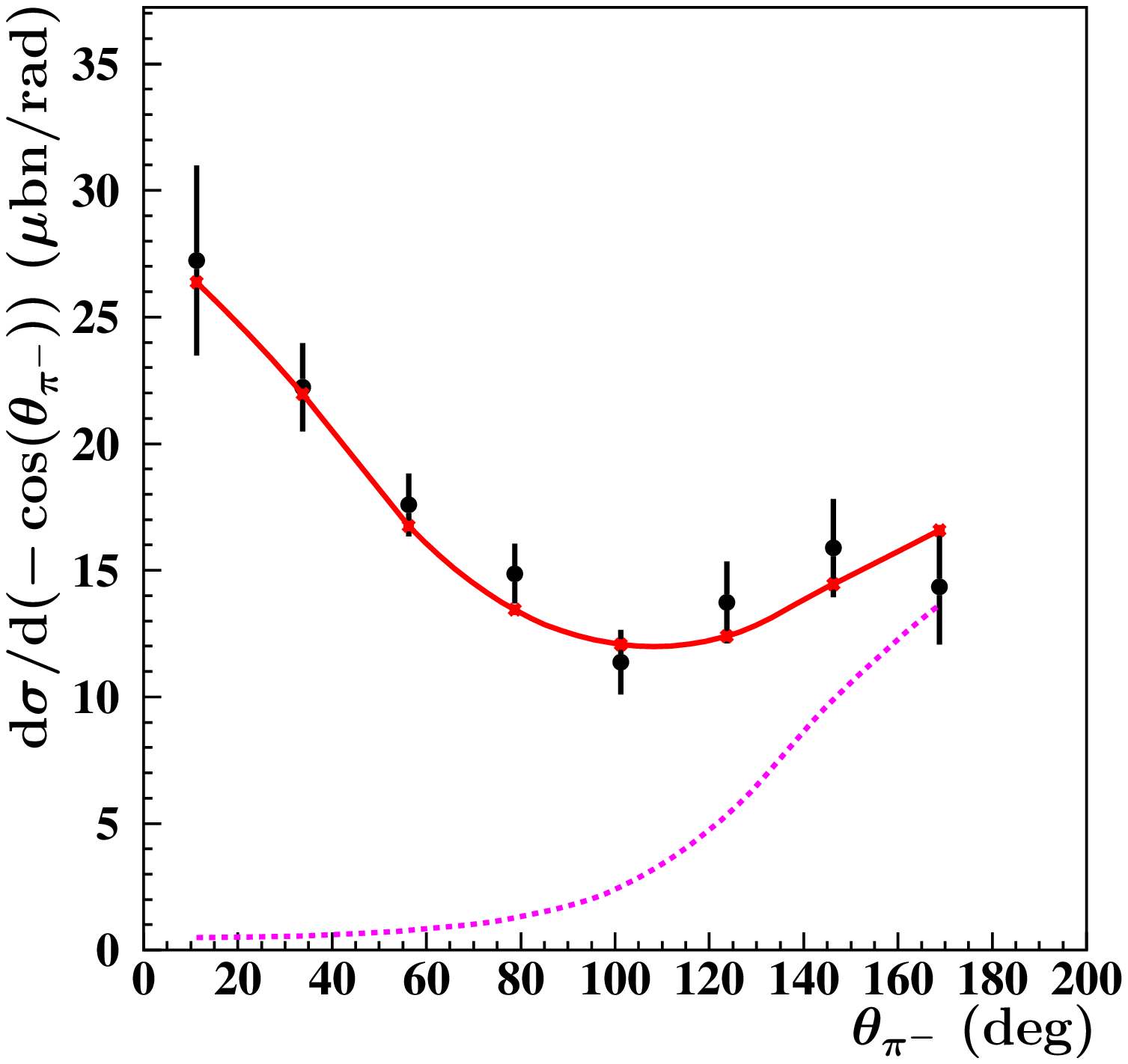,width=7.cm}
\epsfig{file=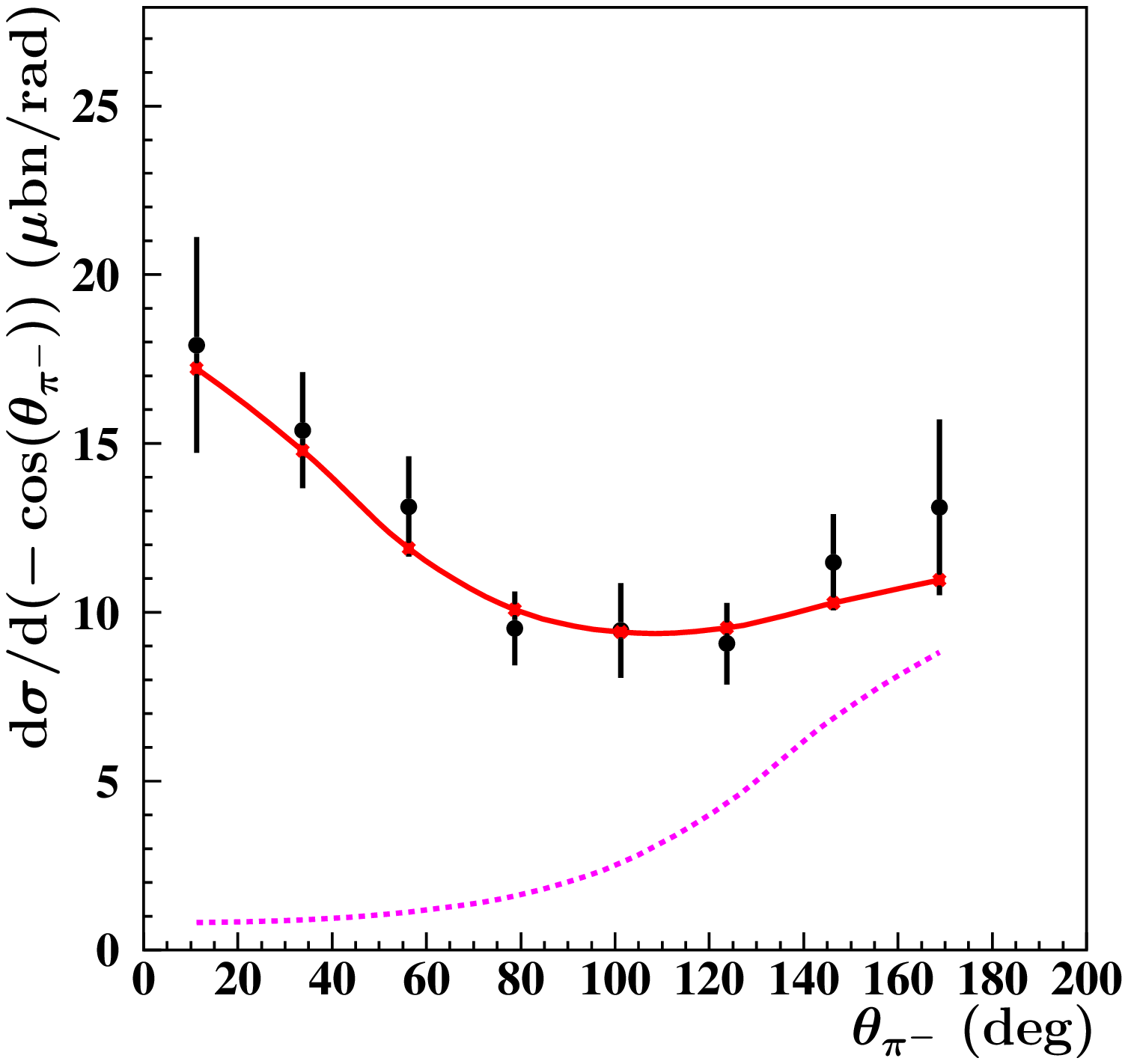,width=7.cm}
\end{center}
\caption{\small(color online) A signature of direct 2$\pi$ production in 
the recent
CLAS $p\pi^+\pi^-$ data at low W and $Q^{2}$ \cite{Fe07,Fe07a}. 
$\pi^{-}$ C.M. angular distributions 
at W=1.49
GeV, $Q^{2}$=0.325 $GeV^{2}$ (top left), W=1.49 GeV, $Q^{2}$=0.375 $GeV^{2}$
(top right), W=1.49 GeV, $Q^{2}$=0.425 $GeV^{2}$ (bottom left), 
W=1.46 GeV, $Q^{2}$=0.525 $GeV^{2}$ (bottom right) and their fit within 
the framework of JM05 model \cite{Mo06}. Full calculations are shown by solid
lines, 
while
the contributions from direct 2$\pi$ processes of JM05 model are shown by 
dash lines.}
\label{2pidirlowq2}
\end{figure*}

In this paper, we have further improved the JM05 model in order
to fit the
 new CLAS data on $p\pi^+\pi^-$ electroproduction cross sections.
The CLAS $p\pi^+\pi^-$ data  allowed us to extend considerably
our knowledge of the direct production
term $T^{dir}_{\gamma_{v} N,\pi\Delta}$ of Eq.(\ref{eq:full-t})
in the low W and low Q$^{2}$ region. 
The resulting model, called JM06, describes successfully all available CLAS 
and world $p\pi^+\pi^-$ electroproduction data 
 at W $<$ 1.6 GeV and  $Q^{2}$ from 0.25 
 to 0.6 GeV$^{2}$. A good description of the data of
nine  differential cross sections
 with rather different manifestation of contributing mechanisms in these
 observables enables us to establish all
 essential  $p\pi^+\pi^-$ electroproduction mechanisms in this kinematic domain. 
 In particular, it allows us to isolate the resonant contributions 
for determining the
electrocouplings of the P$_{11}(1440)$ and D$_{13}(1520)$
states. Within JM06, we find that
these two N$^{*}$ states are the main contributors to the resonant part of the 
$p\pi^+\pi^-$ cross sections at W $<$ 1.6 GeV.

In Section \ref{xsect}, we define the kinematics of $p\pi^+\pi^-$ production.
The JM06 model will be specified in Section \ref{jlabmsugen}.
The fitting procedures are explained in Section \ref{fit}.
Section \ref{results} is devoted to discuss the results. A summary and outlook are given in
Section \ref{concl}.
 
\section{Kinematics and cross sections of $p\pi^+\pi^-$ Electroproduction}
 \label{xsect}
At a given invariant mass $W$ and photon virtuality $Q^2$,
the cross section of the $\gamma_{v} p \rightarrow p\pi^+\pi^-$ reaction
can be written as
\begin{eqnarray}
\frac{d\sigma}{d\nu d\Omega_{e'}} = \Gamma_{v} \int \frac{d^5\sigma}{d^5\tau} d^5\tau
\label{eq:sigma-e}
\end{eqnarray}
where  $\nu$ is virtual
photon energy in the lab. frame, $\Gamma_{v}$ is the virtual photon flux defined 
by the momenta of incoming
and outgoing electrons 
$^{[1]}$  
\footnotetext[1]{explicit expressions for invariant $\nu$ and virtual photon
flux $\Gamma_{v}$ may be found in 
\cite{Am89}} 
and $d^5\tau$ is the phase space volume of the
five independent variables in the C.M. system of the final
$p \pi^{+} \pi^{-}$ state. There are
many possible choices~\cite{Byc} of the
five independent variables.
Defining $M_{\pi^+p}$, $M_{\pi^-p}$, and
$M_{\pi^+\pi^-}$ as invariant mass variables of
 the three possible two-particle pairs in the $p\pi^{+}\pi^{-}$ system,
we adopt the following three assignments:

\begin{enumerate}
\item $d^5\tau_1 =dM_{p\pi^+}dM_{\pi^+\pi^-}d\Omega_{\pi^-}
d\alpha_{[p'\pi^{+}][p\pi^{-}]}$, where
$\Omega_{\pi^-}$ ($\theta_{\pi^{-}}$, $\varphi_{\pi^{-}}$) are
the final $\pi^-$
spherical angles with respect to the direction of virtual photon, and
$\alpha_{[p' \pi^{+}]p\pi^{-}]}$ is the angle
between the plane B defined by the momenta of
the final $p'\pi^{+}$ pair and the plane A defined by
the initial proton and the final $\pi^{-}$ (see Fig.~\ref{kinematic});
\item $d^5\tau_2
=dM_{p\pi^+}dM_{\pi^+\pi^-}d\Omega_{p'}d\alpha_{[p'p][\pi^{+}\pi^{-}]}$,
where 
 $\Omega_{p'}$ ($\theta_{p'}$, $\varphi_{p'}$) are the final proton
spherical angles with respect to the direction of virtual photon, and
 $\alpha_{[p'p][\pi^{+}\pi^{-}]}$ is the angle
between the plane B' defined by the momenta of
$\pi^{+}\pi^{-}$ pair and the plane A'  defined by the momenta of the
initial and final protons ;
\item $d^5\tau_3 =dM_{p\pi^+}dM_{p\pi^-}d\Omega_{\pi^+}
d\alpha_{[p'\pi^{-}][p\pi^{+}]}$, where 
$\Omega_{\pi^+}$ ($\theta_{\pi^{+}}$, $\varphi_{\pi^{+}}$)
are 
the final $\pi^+$
spherical angles with respect to the direction of virtual photon,
and
$\alpha_{[p'\pi^{-}][p\pi^{+}]}$ is the angle 
between the plane B''  defined by the momenta of
the final $p'\pi^{-}$ pair and the plane A'' defined by
the initial proton and the final $\pi^{+}$.
\end{enumerate}

The emission angles for the final state particles in the case of first assignment 
are shown in  Fig.~\ref{kinematic}.
This choice of the kinematical variables is 
most suitable for describing  $p\pi^+\pi^-$ electroproduction through the 
$\pi^{-}\Delta^{++}$ intermediate state, which is the dominant contributor
of all isobar channels in  the kinematical region covered by the considered 
data.
 For the others assignments the emission
angles of the final hadrons are similar to the ones given in Fig.~\ref{kinematic}.
The relations between the  momenta of the final state hadrons and the five
variables of the first assignment can be found in \cite{Fe07a}.

The limited statistics of the available data does not allow the use of
correlated five-fold differential cross sections in the analysis. 
Instead, we  use the differential cross sections depending
on only one final kinematic variable,
obtained from integrating the 5-fold differential cross sections over four
other independent kinematic variables.  These differential 
cross sections were obtained with reasonable statistical 
accuracy in ~\cite{Ri03,Fe07,Fe07a,Fe07b}. They are defined
as:.   
$$
\frac{d\sigma}{dM_{\pi^{+}\pi^{-}}} =
\int\frac{d^{5}\sigma}{d^{5}\tau_{2}}
dM_{\pi^{+}p}d\Omega_{\pi^{-}}d\alpha_{[p\pi^{-}][p'\pi^{+}]} 
$$
$$
\frac{d\sigma}{dM_{\pi^{+}p}} =
\int\frac{d^{5}\sigma}{d^{5}\tau_{2}}
dM_{\pi^{+}\pi^{-}}d\Omega_{\pi^{-}}d\alpha_{[p\pi^{-}][p'\pi^{+}]} 
$$
$$
\frac{d\sigma}{dM_{\pi^{-}p}} =
\int\frac{d^{5}\sigma}{d^{5}\tau_{3}}
dM_{\pi^{+}p}d\Omega_{\pi^{+}}d\alpha_{[p\pi^{+}][p'\pi^{-}]} 
$$
for invariant mass distributions, and 
$$
\frac{d\sigma}{d(-cos\theta_{\pi^-})} =
\int\frac{d^{5}\sigma}{d^{5}\tau_{2}}
dM_{p\pi^+}dM_{\pi^+\pi^-}d\varphi_{\pi^-}
d\alpha_{[p\pi^{-}][p'\pi^{+}]}
$$
\begin{eqnarray}
\label{inegr5diff}
\frac{d\sigma}{d(\alpha_{[p\pi^{-}][p'\pi^+]})} =
\int\frac{d^{5}\sigma}{d^{5}\tau_{2}}
dM_{p\pi^+}dM_{\pi^+\pi^-}d\Omega_{\pi^-}
\end{eqnarray}
for angular distributions. Other distributions
 for the angles $\theta_{\pi^+}$,
$\theta_p$, and $\alpha_{[p\pi^{+}][p'\pi^-]}$, 
$\alpha_{[p p'][\pi^+\pi^-]}$ 
 are similarly defined. Overall nine independent 1-fold differential cross
 section in each bin of $W$ and $Q^{2}$ were included in the analysis.


\section{\label{anal} JM06 Model}
\label{jlabmsugen}

As a starting point, we described the recent CLAS data \cite{Fe07a} at 
1.3 $<$ W $<$ 1.6 GeV and 0.25  $<$ Q$^{2}$ $<$ 0.60 GeV$^{2}$ within the  
JM05 model~\cite{Az05,Mo06}, which is
briefly outlined  in Section I. The relations between JM model amplitudes
and charged double  pion electroproduction cross sections are presented 
in Appendix IV.
Here we only specify the input to our calculations
and indicate improvements incorporated in JM06 for the analysis presented in this paper.

The nucleon resonances included in the
resonant term $t^R$ of Eq.(\ref{eq:pid-t})
are listed in 
Table~\ref{nstlist}. The $3/2^+(1720)$ state
observed in the analysis of
CLAS $p\pi^+\pi^-$ electroproduction data~\cite{Ri03} is also
included there.
The $N^{*}$ hadronic decay widths, branching fraction to 
$\pi \Delta$ and $\rho p$ final states were taken in part from analysis 
of hadroproduction data 
\cite{Vrana} (numbers in Table~\ref{nstlist}).
Hadronic parameters for other states needed in the calculations of
resonant amplitude $t^R$ (symbols "var" in Table~\ref{nstlist})
were taken from analysis
\cite{Mo06} of the CLAS data \cite{Ri03}, which covered 1.4 $<$ $W$ $<$ 2.0 GeV
and 0.5 $<$ $Q^{2}$ $<$ 1.5 GeV$^{2}$.
The initial values of the
electromagnetic form factors,
$A_{1/2}(Q^2)$, $S_{1/2}(Q^2)$, $A_{3/2}(Q^2)$ were
estimated from  interpolations \cite{Az05-1,Burk2003,Mo06} of CLAS and world 
data to the considered $Q^2$ region covered by the data \cite{Fe07a} and further adjusted 
in fitting the  data. With these specifications, 
it was found that the fits  at low $W$ and $Q^2$~\cite{Fe07,Fe07a}
are only sensitive to
the first two $N^{*}$'s listed in the Table~\ref{nstlist}.
The contributions from the higher masses 
$N^{*}$, $\Delta^{*}$ states are not varied in the fits of recent CLAS data
\cite{Fe07a} collected at 1.3 $<$ W $<$ 1.6 GeV and 
0.25  $<$ Q$^{2}$ $<$ 0.60 GeV$^{2}$.

\begin{table}
\begin{center}
\begin{tabular}{|c|c|c|c|c|}
\hline
$N^{*},\Delta^{*}$ & $M,$ & $\Gamma_{tot}$ &
$BF_{\pi\Delta}$ & $BF_{\rho p}$ \\
 &  GeV  &  GeV  &
 \%  & \% \\
\hline
$P_{11}(1440)$ & $1.440$ & $var$ & $var$ & $var$ \\
$D_{13}(1520)$ & $1.520$ & $var.$ & $var.$ & $var.$ \\
$S_{31}(1620)$ & $1.620$ & $0.150$ & $62.$ & $29.$ \\
$P_{33}(1600)$ & $var$ & $var.$ & $var.$ & $var.$ \\
$S_{11}(1650)$ & $1.650$ & $0.167$ & $2.$ & $3.$ \\
$D_{15}(1675)$ & $1.675$ & $0.160$ & $53.$ & $0.$ \\
$F_{15}(1680)$ & $1.680$ & $0.130$ & $22.$ & $7.$ \\
$D_{13}(1700)$ & $var.$ & $var.$ & $var.$ & $var.$ \\
$D_{33}(1700)$ & $1.700$ & $0.300$ & $78.$ & $8.$ \\
$P_{13}(1720)$ & $var.$ & $var.$ & $var.$ & $var.$ \\
$3/2^{+}(1720) cand.$ & $var.$ & $var.$ & $var.$ & $var.$ \\
\hline
\end{tabular}
\caption{\label{nstlist} List of resonances included  and theirs hadronic
properties: total decay widths $\Gamma_{tot}$, branching fractions (BF) to $\pi
\Delta$ and $\rho p$ final states.
The quoted values are taken from Review of Particle Properties.
The quantities labeled as "var." correspond to the variable parameters fit
to the CLAS $p\pi^+\pi^-$ data \cite{Ri03}.}
\end{center}
\end{table}

\begin{figure}[htp]
\begin{center}
\includegraphics[width=9cm]{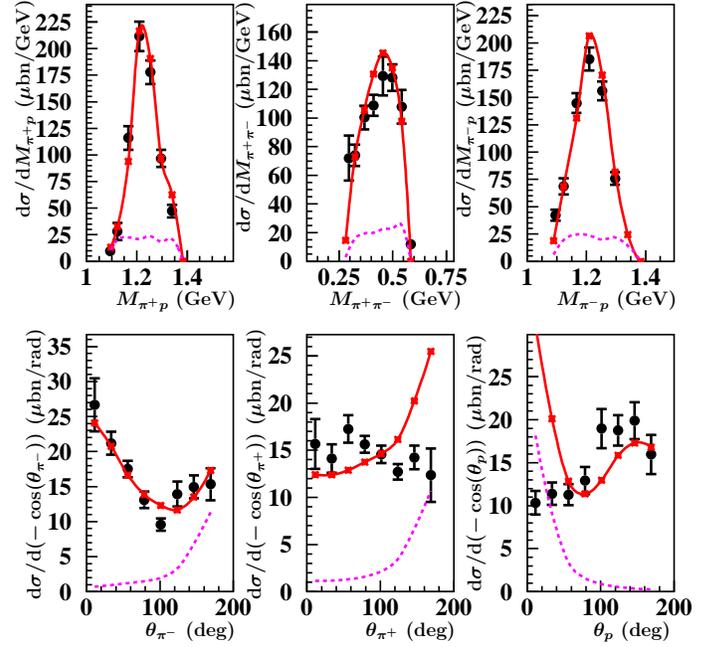}
\caption{\small(color online) The recent CLAS $p\pi^+\pi^-$ data at W=1.51 GeV,
$Q^2$=0.43 GeV$^2$ and their fit
within the framework of JM05 model shown by solid lines.
The
contributions from direct 2$\pi$ production, resulting in
discrepancies in the description of $\pi^{+}$ and p angular distributions,
 are shown
by dashed lines.}
 \label{fitjm05}
\end{center}
\end{figure}

\begin{figure*}[ht]
\epsfig{file=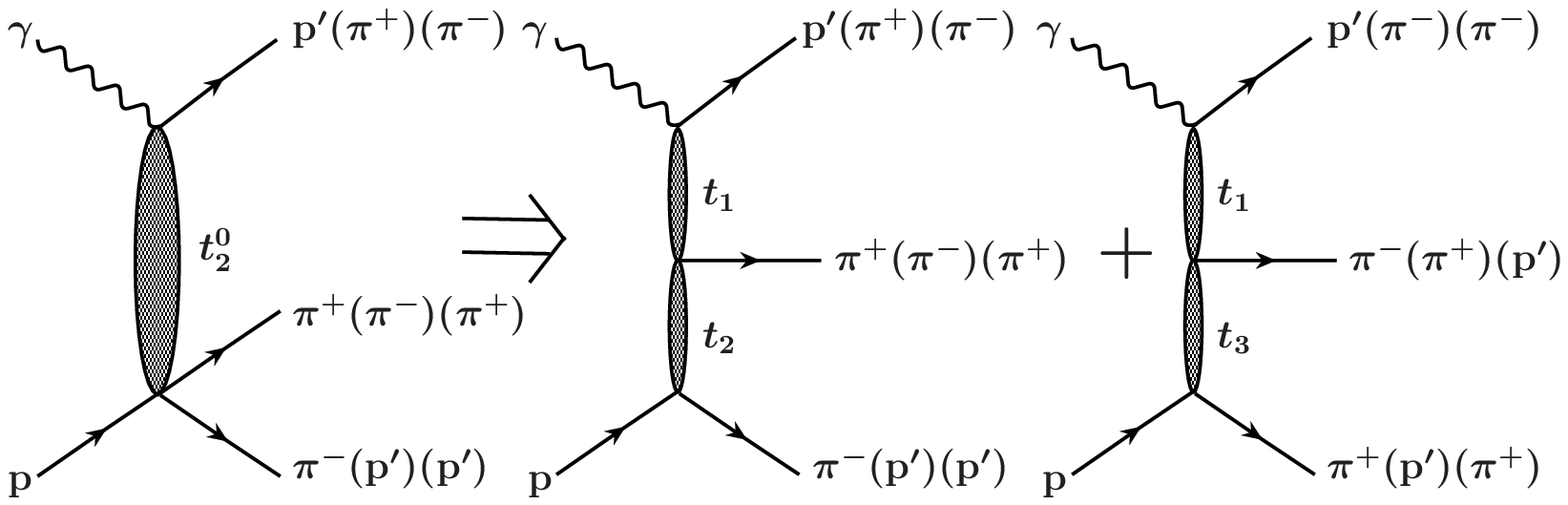,width=14.cm}
\epsfig{file=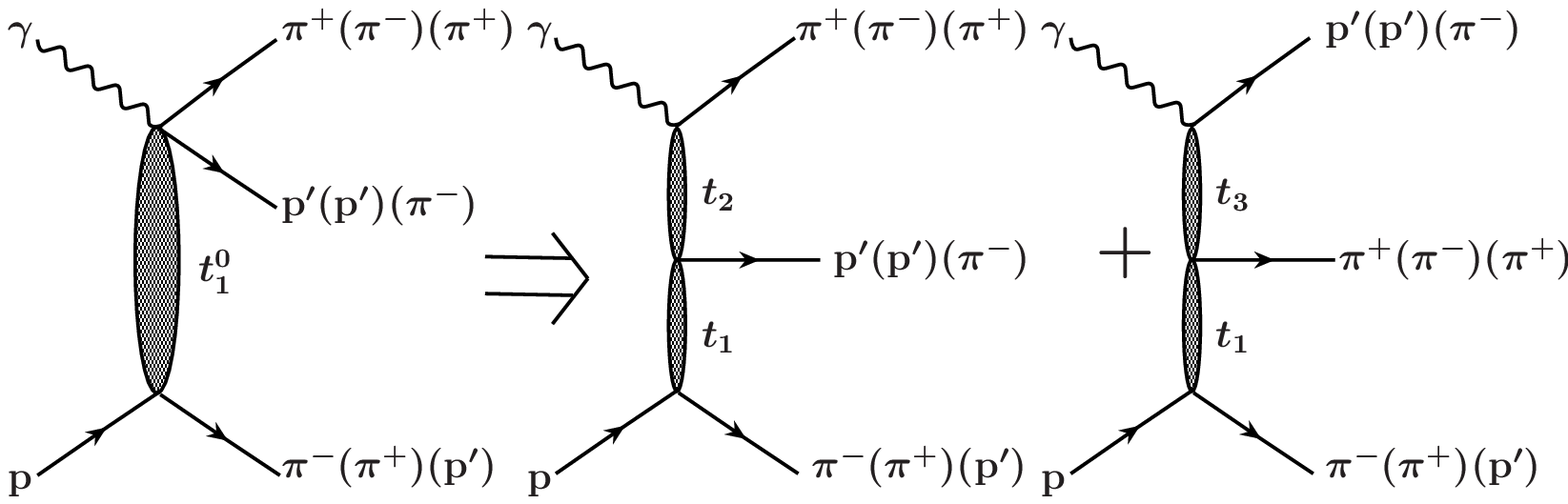,width=14.cm}
\caption{\small Direct 2$\pi$ production mechanisms
of JM06 model. Top (bottom): the diagrams on the left side, describing 
direct 2$\pi$ production mechanisms 
of JM05 model, were replaced by diagrams, shown in the middle and right
parts of the plot. Contact 
interaction in down (upper) vertex was replaced by additional 
particle-exchange amplitudes. Sets of diagrams in the middle and right parts of
the plot correspond to 
various assignments
for the four momenta squared running over propagators in  
exchange amplitudes (see definitions Eqs. (\ref{topimprborn},\ref{bottombornimp}) 
of $t_{i}$ (i=1,2,3) 
for diagrams in top$/$bottom rows
respectively).}
\label{diag1m}
\end{figure*}

Detailed expressions of the 
term $t^{Born}$ in Eq.(\ref{eq:pid-t})
may be found in \cite{Ri00} and are summarized in Appendix I,
It contains the  minimal set of Reggeiezed 
Born terms.  The initial and final state interactions in $t^{Born}$
are taken into account in absorptive approximation \cite{Go}
by using a
procedure developed in \cite{Ri00} to
evaluate the absorptive coefficients 
from the data on $\pi N$-scattering.

An analysis of the CLAS $p\pi^+\pi^-$ data at $Q^{2}$ $>$ 0.5
GeV$^{2}$ \cite{Ri03}
revealed the need of an additional contributions to $\pi \Delta$ isobar
channels. These contributions were parametrized by the contact terms
$t^c$ of Eq.(\ref{eq:pid-t}) \cite{Mo06,Mo06-1} that have different Lorentz
structure with respect to the contact interactions in the described above Born
terms. The contributions from these
additional contact terms were confirmed in the analysis of the recent CLAS 
data \cite{Fe07,Fe07a} at $Q^{2}$ $<$ 0.5 GeV$^{2}$. This is demonstrated in
 Fig.~\ref{contact} where the differences between the solid and
dashed curves are from the  $t^c$ term. The additional contact 
terms $t^c$ are given in Appendix II.

Within JM05 it was found that the $p\pi^+\pi^-$ production through
the isobar channels account for 70 to 90 \%
of the fully integrated cross sections in
the nucleon resonance region. The remaining parts 
were assumed to be due to direct 2$\pi$ production mechanism that
 the final $p\pi^{+}\pi^{-}$ 
state is produced without forming an unstable hadron in the intermediate state.
Initially these direct $2\pi$ production contributions
were parametrized by the 3-body phase space \cite{Mo01,Mo03}. 
Their strengths in each interval of
$W$ and $Q^{2}$ were adjusted to fit the data.
However, the CLAS data on C.M. $\pi^{-}$ angular distributions shown in
Figs.~\ref{dhq2} and \ref{2pidirlowq2} revealed  steep slopes at the backward 
C.M. $\pi^{-}$ emission angles. Such behavior
 is clearly incompatible with the
3-body phase space parametrization of the direct 2$\pi$ pion production
mechanisms. We improve this part by considering
the particle-exchange
processes illustrated on  the
left side of Fig.~\ref{diag1m}.
The direct $\gamma_{v} N \rightarrow \pi\pi N$ term $T^{dir}$ of
Eq.(\ref{eq:full-t}) 
is then parameterized as \cite{Az05}:
\begin{eqnarray}
T^{dir} &=& 
 A(W)\varepsilon_{\mu}(q_{\gamma})\overline{U}_{p'}(P_{p'})
\gamma^{\mu}U_{p}(P_{p})
((P_{1}P_{2})e^{b(t^{0}_{1}-t^{0}_{1\;max})} \nonumber \\
&+& (P_{2}P_{3})e^{b(t^{0}_{2}-t^{0}_{2\;max})})
\label{direct2pi}
\end{eqnarray}
with
\begin{eqnarray}
t^{0}_{1}&=& (P_p - P_3)^2  \nonumber \\
t^{0}_{2}&=& (q_\gamma - P_1)^2 
\label{num} 
\end{eqnarray}
where  $U_{p}$, $\overline{U_{p'}}$ are  the  spinors of
the initial and final protons. The first and the second terms in Eq.
(\ref{direct2pi}) describe the contributions from the bottom and top diagrams
on the left side of Fig.~\ref{diag1m}. The $P_{i}$ ($i$=1,2,3) in
Eqs.(\ref{direct2pi})-(\ref{num})
stand for
the final hadron 4-momenta, shown  
in the bottom/top left 
sides of Fig.~\ref{diag1m} by upper middle and down legs, respectively.
The variables $t^{0}_{j}$ (j=1,2) in Eq.(\ref{num}) are four 
momenta-transfer square in exchange processes shown on the left sides of
 bottom and top rows of Fig.~\ref{diag1m}. The maximum values of 
 $t^{0}_{1}$ and $t^{0}_{2}$ are defined as $t^{0}_{1\;max}$, $t^{0}_{2\;max}$, respectively. 
The $\epsilon_\mu\bar{U}\gamma^\mu U$ 
represents a Lorentz invariant coupling between the incident virtual photons and
protons; the propagators of the unspecified exchange particles as well
as other couplings with external momenta are
parameterized as $((P_{1}P_{2})e^{b(t^{0}_{1}-t^{0}_{1\;max})}$ 
(for the bottom diagram on the left side of Fig.~\ref{diag1m}) and
$(P_{2}P_{3})e^{b(t^{0}_{2}-t^{0}_{2\;max})})$ 
(for the top diagram on the left side of Fig.~\ref{diag1m}), the strength $A(W)$ is 
adjusted in the fit.

\begin{figure*}
\begin{center}
\includegraphics[width=15cm]{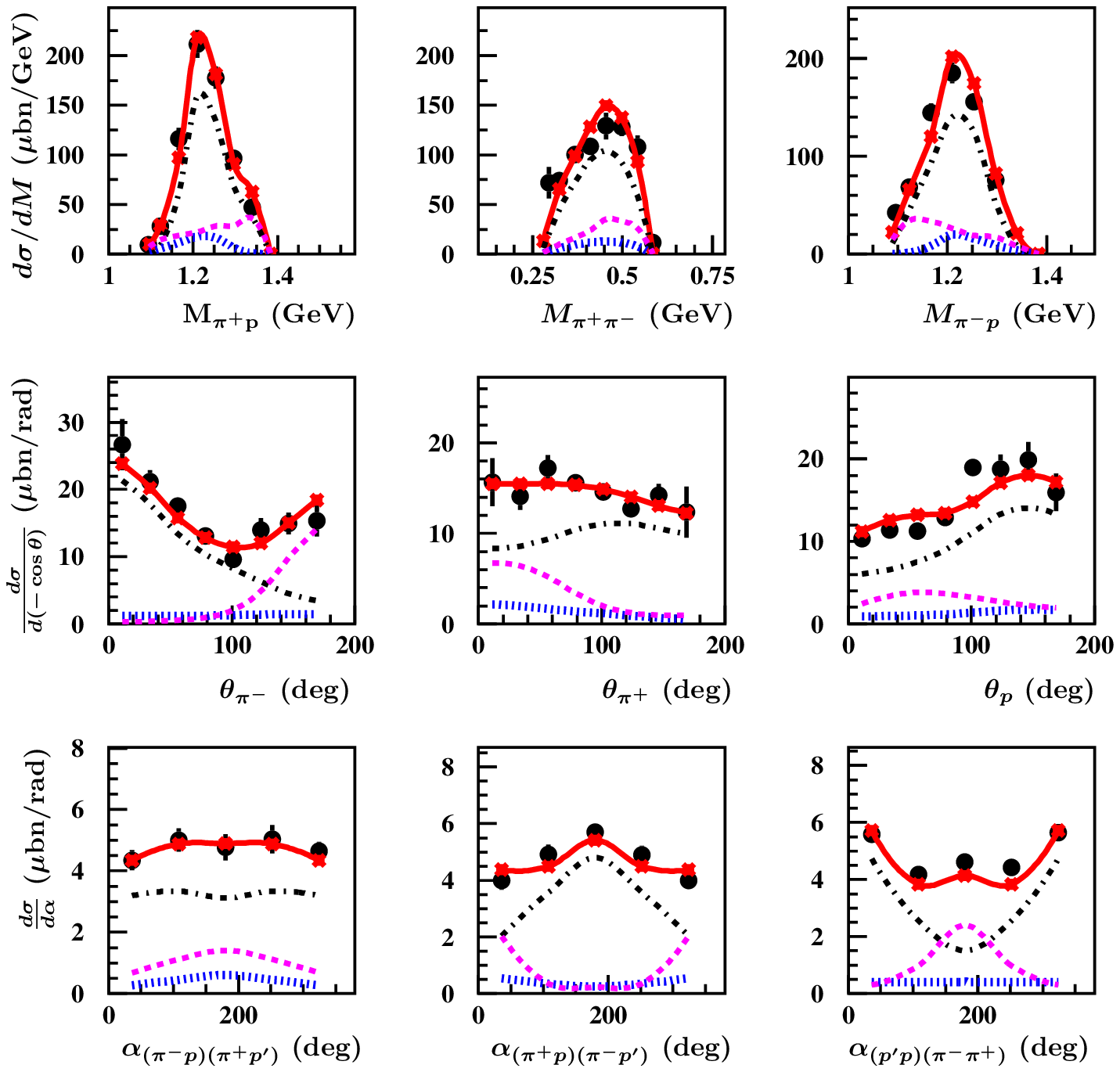}
\caption{\small(color online) Fit of recent CLAS charged double pion data
\cite{Fe07a} at $W$= 1.51 GeV and $Q^{2}$=0.43 GeV$^{2}$ after discussed in the Section \ref{jlabmsugen}
improvements in parametrization of direct 2$\pi$
production mechanisms. Full JM06 calculations are shown by solid lines, while
the contributions from $\pi^{-} \Delta^{++}$, $\pi^{+} \Delta^{0}$ isobar
channels and direct 2$\pi$ production are shown by dot-dash, dot and dash
lines respectively.} 
\label{9sectok}
\end{center}
\end{figure*}

\begin{figure*}[htp]
\begin{center}
\includegraphics[width=15cm]{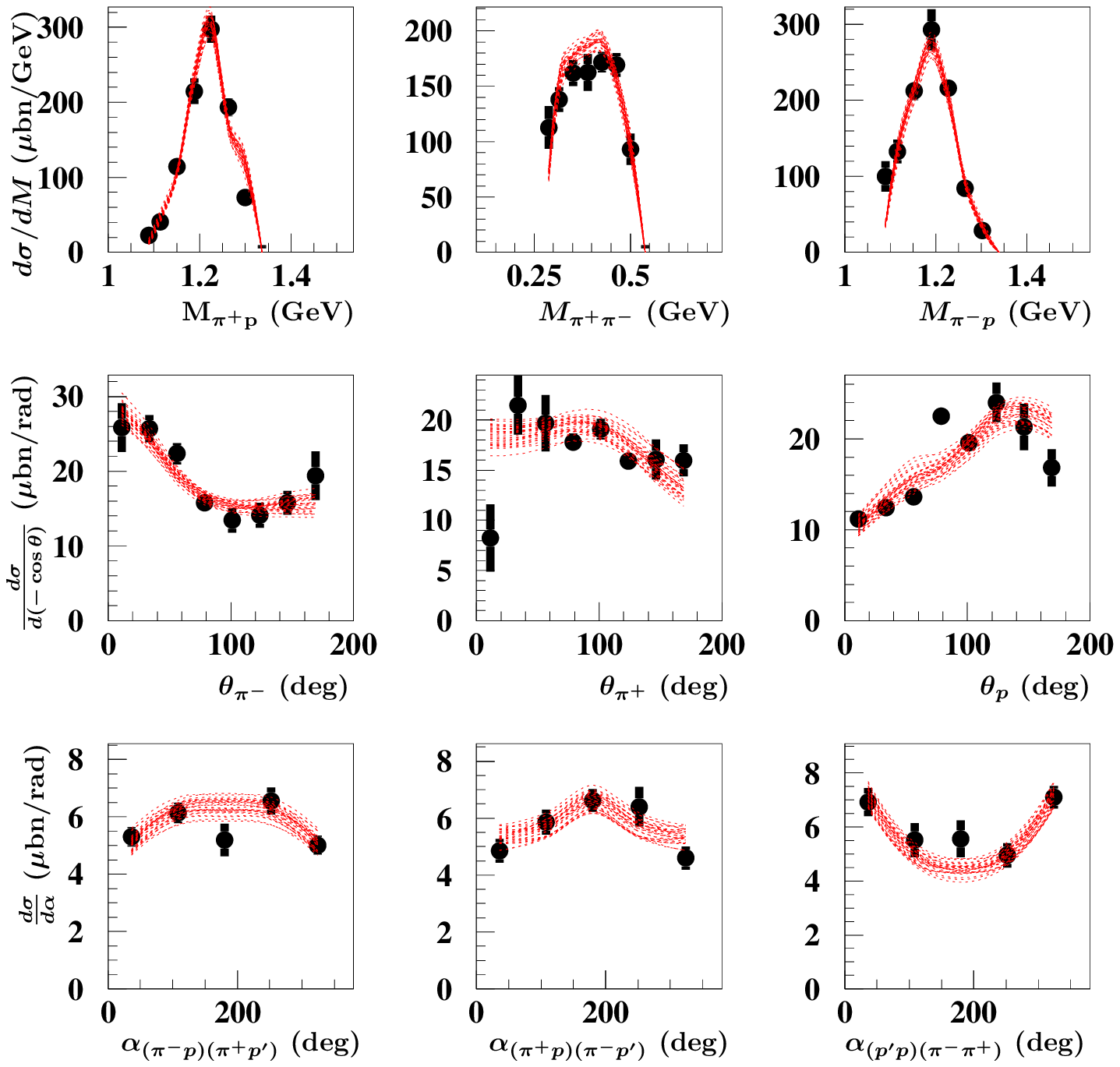}
\caption{\small (color online)  Fit of CLAS data on 
$p \pi^+ \pi^-$ electroproduction 
\cite{Fe07a} at W=1.46 GeV and Q$^{2}$=0.33 GeV$^{2}$ within 
the framework of JM06 model. Selected in the fitting procedure calculated
differential cross sections are shown by bunches of dashed lines}
\label{fit_9sec}
\end{center}
\end{figure*}

The direct 2$\pi$ production mechanisms, $T^{dir}$ 
as parameterized in Eqs.(5)-(6), are  needed to         
describe the three invariant masses and the
$\pi^{-}$ angular distribution obtained in \cite{Ri03}. In particular they allowed us to describe 
$\pi^{-}$ angular distributions at
the backward C.M. angles and $Q^{2}$ $>$ 0.5 GeV$^{2}$.
This is shown in Fig.~\ref{dhq2}.
Here we note that the steep enhancement at backward angles
for the $\pi^-$  differential cross sections is generated from
the exponential form of the parameterization.
In Fig.~\ref{2pidirlowq2} we show that 
the direct 2$\pi$ production term $T^{dir}$     
is further established in fitting  the 
data on $\pi^{-}$ angular
distribution at $Q^{2}$ $<$ 0.5 GeV$^{2}$. 

%

The JM05 model, described above, was unable  to reproduce the data on 
$\pi^{+}$ and  $p$ angular distributions.
This failure has become evident at W $>$ 1.40 GeV and 
for all photon virtualities of \cite{Fe07,Fe07a}.   
One example is shown 
in Fig.~\ref{fitjm05}.
We see that the invariant mass distributions and 
$\pi^{-}$ angular 
distribution can be reproduced reasonably, while
considerable differences remain for the $\pi^{+}$ and  p angular distributions 
remain.
It is therefore necessary to
improve the parameterization of $T^{dir}$.


In the JM06 model employed in this work, the contact interactions of the final
hadron and the initial particles in two diagrams of the left side of
Fig.~\ref{diag1m} were replaced by additional exchange processes shown in
four diagrams on the right side of Fig.~\ref{diag1m}. The 
propagators in both unspecified exchange mechanisms are parametrized by the
same exponential functions of the momentum-transfer variables $t_i$ and
their
slope
parameter $b= 4.0$ GeV$^{-2}$ is determined in the fit to the data.
In this way we get the following parameterization for the two diagrams shown 
on the
top right part of Fig.~\ref{diag1m}: 
\begin{eqnarray}
&T^{dir}= A_{k}(W)\varepsilon_{\mu}(q_{\gamma})\overline{U}_{p'}(P_{p'})
\gamma^{\mu}U_{p}(P_{p})\times\nonumber \\ 
&(P_{2}P_{3})e^{b(t_{1}-t_{1\;max})} \times \nonumber \\  
& \lbrace \alpha_{1}(W)e^{b(t_{2}-t_{2\;max})}  +  \alpha_{2}(W)e^{b(t_{3}-t_{3\;max})}
\rbrace \nonumber,\\  
\label{topimprborn}
\end{eqnarray}
with
\begin{eqnarray}
\label{titop}
t_{1} & = & (q_{\gamma} - P_{1})^{2};  \nonumber \\  
t_{2} & = & (P_{p} - P_{3})^{2};   \\  
t_{3} & = & (P_{p} - P_{2})^{2}   \nonumber,
\end{eqnarray}
where $P_{i}$  (i=1,2,3) stand for four-momenta of the three hadrons, 
shown in the top middle diagram 
of Fig.~\ref{diag1m} by upper middle and down legs, respectively, and
$P_{p}$,
$P_{p'}$ are four-momenta of the initial and final state proton. 
 The $q_\gamma$ represents the four-momentum of the initial photon,
and  $\epsilon_{\mu}$
is the initial photon vector.
The scalar products ($P_{2}$$P_{3}$) in Eq.(\ref{topimprborn}) and 
($P_{1}$$P_{2}$) in (\ref{bottombornimp}) were implemented in order 
to better describe invariant mass distributions for the final state hadrons.

Two terms in the parametrization (\ref{topimprborn}) correspond to the  diagrams
in the middle
and right sides of the top row in the Fig.~\ref{diag1m}. The relative 
contributions from these
diagrams  $\alpha_{1}(W)$
and $\alpha_{2}(W)$ are determined in fits to the data under restriction 
$\alpha_{1}(W)+\alpha_{2}(W)=1$. Each diagram accounts for three
processes, corresponding to
various assignments of the final state hadrons shown on 
the right side of Fig.~\ref{diag1m} diagrams.
Full amplitude is evaluated from the coherent sum 
of all contributions.

Similarly, the parameterization of the amplitudes for diagrams on 
the right-hand-side of the bottom row in Fig.~\ref{diag1m} 
are from modifying Eq.(\ref{direct2pi})
of JM05 model :
\begin{eqnarray}
T^{dir}=A(W)\varepsilon_{\mu}(q_{\gamma})\overline{U}_{p'}(P_{p'})\gamma^{\mu}U_{p}(P_{p})(P_{1}P_{2}) \times  \nonumber \\ 
e^{b(t_{1}-t_{1\;max})}   
\lbrace \alpha_{1}(W)e^{b(t_{2}-t_{2\;max})}(t_{2}-t_{2\;max})  
+    \nonumber \\   
 \alpha_{2}(W)e^{b(t_{3}-t_{3\;max})}(t_{3}-t_{3\;max})\rbrace,  
\label{bottombornimp}
\end{eqnarray}
where
\begin{eqnarray}
\label{tibottom}
t_{1} & = & (P_{p} - P_{3})^{2};  \nonumber \\  
t_{2} & = & (q_{\gamma} - P_{1})^{2}; \\  
t_{3} & = & (q_{\gamma} - P_{2})^{2}.  \nonumber,
\end{eqnarray}
and $P_{i}$  (i=1,2,3) stand for four-momenta of the three final hadrons, 
shown in the bottom middle diagram 
of the Fig.~\ref{diag1m} by upper middle and down legs, respectively.
It is found that the additional exponential function and the factors
$(t_2-t_{2 max})$ and $(t_3-t_{3 max})$ in Eqs.(\ref{topimprborn},\ref{bottombornimp})
are essential to remove the
discrepancies of $\pi^+$ and $p$ angular distributions in backward and forward
angles, respectively, seen in Fig.~\ref{fitjm05}. After implementation of
the improvements in parametrization of direct 2$\pi$ production
mechanisms, we succeeded in describing all differential cross section 
of the CLAS $p\pi^+\pi^-$ data. One example is shown in Fig.~\ref{9sectok}

\begin{table}
\centering
\caption{The intervals of $\chi^{2}$/(data points) values corresponding to 
closest to the data calculated
differential cross sections, selected in the fit. \label{chi}}

\begin{tabular}{|c|c|}
\hline
$Q^{2}$ interval & \\ 
GeV$^{2}$  &$\chi^{2}$/(data points)   \\
  &   \\
\hline
0.25-0.4 & 2.71-2.80 \\
\hline
0.4-0.5 & 1.77-1.87 \\
\hline
0.5-0.6 & 1.39-1.65 \\
\hline

\end{tabular}
\end{table}

\begin{table}
\centering
\caption{$\chi^{2}/d.p.$ from the fit of all invariant mass and 
$\theta_{i}$ (i=$\pi^{-}$,$\pi^{+}$,p) angular
distributions within the framework of JM06 model.}
\label{pizdato}
\begin{tabular}{|c|c|c|c|c|c|c|c|}
\hline
$Q^{2}$, GeV$^{2}$ &  & & & & & & \\
bin center  & 0.275 & 0.325  & 0.375  & 0.425 & 0.475 & 0.525 & 0.575 \\
\hline
 &  &  &  & & &  & \\
$\chi^{2}/$d.p. & 2.74 & 2.01 & 1.90 & 1.60 & 1.37 & 1.63 & 1.25 \\
 &  &  &  & & &  & \\
\hline

\end{tabular}
\end{table}

\begin{table}
\centering
\caption{The partial waves for coherent sum of Born  and additional 
contact terms in $\pi^{-} \Delta^{++}$ 
channel of total angular
momentum J  and for helicities of the initial and final particles $\lambda_{\gamma}$=1,
$\lambda_{p}$=1/2, $\lambda_{\Delta}$=3/2. The amplitudes were evaluated at Q$^{2}$=0.275 
GeV$^{2}$ and for running $\Delta$ mass 1.14 GeV  \label{lg1lp12ld32}}
\vspace{0.2cm}
\begin{tabular}{|c|c|c|c|}
\hline
W, GeV &  J & Real part for & Imaginary part   \\
 & & T$^{J}$ Born & for T$^{J}$ Born \\
 &  &and contact   terms     &and contact  terms \\
\hline
 1.31& 0.5&    0.&    0.\\
 1.34& 0.5&    0.&  0.\\
 1.36& 0.5&     0.&  0.\\
 1.39& 0.5&     0.&  0.\\
 1.41& 0.5&    0.&  0.\\
 1.44& 0.5&    0.&  0.\\
 1.46& 0.5&     0.&  0.\\
 1.49& 0.5&    0.&  0.\\
 1.51& 0.5&     0.&  0.\\
 1.54& 0.5&     0.&  0.\\
 1.56& 0.5&    0.&  0.\\
 1.31& 1.5&   -14.63& -5.31\\
 1.34& 1.5&     -14.89& -5.42\\
 1.36& 1.5&     -15.04& -5.48\\
 1.39& 1.5&   -14.70& -5.48\\
 1.41& 1.5&    -13.82& -5.45\\
 1.44& 1.5&   -11.38& -5.38\\
 1.46& 1.5&     -10.15& -5.28\\
 1.49& 1.5&  -10.56& -5.16\\
 1.51& 1.5&   -9.23& -5.00\\
 1.54& 1.5&   -8.69& -4.83\\
 1.56& 1.5&   -8.09& -4.64\\
 1.31& 2.5&     -3.59& -0.58\\
 1.34& 2.5&    -5.05& -0.87\\
 1.36& 2.5&    -6.22& -1.14\\
 1.39& 2.5&    -7.13& -1.39\\
 1.41& 2.5&   -7.78& -1.63\\
 1.44& 2.5&   -7.95& -1.87\\
 1.46& 2.5&    -8.21& -2.09\\
 1.49& 2.5&    -8.76& -2.31\\
 1.51& 2.5&    -8.77& -2.51\\
 1.54& 2.5&     -8.90& -2.70\\
 1.56& 2.5&   -8.93& -2.88\\

\hline

\end{tabular}
\end{table}

\begin{table}
\centering
\caption{The partial waves for coherent sum of Born  and additional 
contact terms in $\pi^{-} \Delta^{++}$ 
channel of total angular
momentum J  and for helicities of the initial and final particles $\lambda_{\gamma}$=1,
$\lambda_{p}$=1/2, $\lambda_{\Delta}$=1/2. The amplitudes were evaluated at Q$^{2}$=0.275 
GeV$^{2}$ and for running $\Delta$ mass 1.14 GeV  \label{lg1lp12ld12}}
\vspace{0.2cm}
\begin{tabular}{|c|c|c|c|}
\hline
W, GeV &  J & Real part for & Imaginary part   \\
 & & T$^{J}$ Born & for T$^{J}$ Born \\
 &  & and contact  terms     &and contact  terms \\
\hline
 1.31& 0.5&    -0.09&    0.73\\
 1.34& 0.5&    -0.12&  1.04\\
 1.36& 0.5&     1.25&  1.28\\
 1.39& 0.5&     1.36&  1.47\\
 1.41& 0.5&    2.18&  1.63\\
 1.44& 0.5&    3.19&  1.76\\
 1.46& 0.5&     3.63&  1.85\\
 1.49& 0.5&    3.72&  1.92\\
 1.51& 0.5&     4.65&  1.97\\
 1.54& 0.5&     4.73&  1.98\\
 1.56& 0.5&    4.52&  1.98\\
 1.31& 1.5&   -12.56& -4.37\\
 1.34& 1.5&    -11.99& -4.05\\
 1.36& 1.5&    -11.25& -3.74\\
 1.39& 1.5&   -10.41& -3.42\\
 1.41& 1.5&   -8.99& -3.09\\
 1.44& 1.5&   -6.18& -2.76\\
 1.46& 1.5&    -4.60& -2.41\\
 1.49& 1.5&  -4.60& -2.07\\
 1.51& 1.5&   -2.80& -1.72\\
 1.54& 1.5&   -2.04& -1.37\\
 1.56& 1.5&   -1.40& -1.03\\
 1.31& 2.5&    -3.86& -0.62\\
 1.34& 2.5&    -5.16& -0.89\\
 1.36& 2.5&    -6.05& -1.12\\
 1.39& 2.5&   -6.67& -1.33\\
 1.41& 2.5&  -6.96& -1.51\\
 1.44& 2.5&   -6.69& -1.68\\
 1.46& 2.5&    -6.53& -1.83\\
 1.49& 2.5&    -6.71& -1.96\\
 1.51& 2.5&   -6.23& -2.07\\
 1.54& 2.5&    -5.96& -2.15\\
 1.56& 2.5&  -5.64& -2.20\\

\hline

\end{tabular}
\end{table}

\begin{table}
\centering
\caption{The partial waves for coherent sum of Born  and additional 
contact terms in $\pi^{-} \Delta^{++}$ 
channel of total angular
momentum J  and for helicities of the initial and final particles $\lambda_{\gamma}$=1,
$\lambda_{p}$=1/2, $\lambda_{\Delta}$=-1/2. The amplitudes were evaluated at Q$^{2}$=0.275 
GeV$^{2}$ and for running $\Delta$ mass 1.14 GeV  \label{lg1lp12ldm12}}
\vspace{0.2cm}
\begin{tabular}{|c|c|c|c|}
\hline
W, GeV &  J & Real part for & Imaginary part   \\
 & & T$^{J}$ Born & for T$^{J}$ Born \\
 &  & and contact  terms     & and contact terms \\
\hline
 1.31& 0.5&     -0.77&  0.40\\
 1.34& 0.5&    -1.35&  0.42\\
 1.36& 0.5&     -1.24&  0.38\\
 1.39& 0.5&    -1.23&  0.31\\
 1.41& 0.5&     -0.96&  0.22\\
 1.44& 0.5&    -0.10&  0.12\\
 1.46& 0.5&     0.10&  0.009\\
 1.49& 0.5&     0.012& -0.109\\
 1.51& 0.5&      0.24& -0.22\\
 1.54& 0.5&      0.12& -0.34\\
 1.56& 0.5&    -0.10& -0.45\\
 1.31& 1.5&    -11.00& -3.67\\
 1.34& 1.5&     -10.06& -3.12\\
 1.36& 1.5&    -9.21& -2.63\\
 1.39& 1.5&   -8.42& -2.18\\
 1.41& 1.5&    -7.27& -1.76\\
 1.44& 1.5&    -4.98& -1.37\\
 1.46& 1.5&     -3.77& -1.01\\
 1.49& 1.5&   -3.34& -0.67\\
 1.51& 1.5&    -2.77& -0.37\\
 1.54& 1.5&   -2.42& -0.12\\
 1.56& 1.5&   -2.14&  0.11\\
 1.31& 2.5&     -3.41& -0.53\\
 1.34& 2.5&    -4.34& -0.72 \\
 1.36& 2.5&    -4.90& -0.86\\
 1.39& 2.5&    -5.24& -0.98\\
 1.41& 2.5&  -5.31& -1.07\\
 1.44& 2.5&   -4.92& -1.13\\
 1.46& 2.5&     -4.64& -1.17\\
 1.49& 2.5&     -4.71& -1.20\\
 1.51& 2.5&   -4.23& -1.19\\
 1.54& 2.5&     -3.96& -1.15\\
 1.56& 2.5&   -3.67& -1.09\\

\hline

\end{tabular}
\end{table}

\begin{table}
\centering
\caption{The partial waves for coherent sum of Born  and additional 
contact terms in $\pi^{-} \Delta^{++}$ 
channel of total angular
momentum J  and for helicities of the initial and final particles 
$\lambda_{\gamma}$=1,
$\lambda_{p}$=1/2, $\lambda_{\Delta}$=-3/2. The amplitudes were evaluated at Q$^{2}$=0.275 
GeV$^{2}$ and for running $\Delta$ mass 1.14 GeV  \label{lg1lp12ldm32}}
\vspace{0.2cm}
\begin{tabular}{|c|c|c|c|}
\hline
W, GeV &  J & Real part for & Imaginary part   \\
 & & T$^{J}$ Born & for T$^{J}$ Born \\
 &  &  and contact terms     & and contact terms \\
\hline
 1.31& 0.5&    0.&    0.\\
 1.34& 0.5&    0.&  0.\\
 1.36& 0.5&     0.&  0.\\
 1.39& 0.5&     0.&  0.\\
 1.41& 0.5&    0.&  0.\\
 1.44& 0.5&    0.&  0.\\
 1.46& 0.5&     0.&  0.\\
 1.49& 0.5&    0.&  0.\\
 1.51& 0.5&     0.&  0.\\
 1.54& 0.5&     0.&  0.\\
 1.56& 0.5&    0.&  0.\\
 1.31& 1.5&     -9.82& -3.29\\
 1.34& 1.5&    -8.70& -2.78\\
 1.36& 1.5&    -7.97& -2.42\\
 1.39& 1.5&  -7.25& -2.12\\
 1.41& 1.5&     -6.44& -1.88\\
 1.44& 1.5&   -4.98& -1.67\\
 1.46& 1.5&      -4.16& -1.49\\
 1.49& 1.5&  -4.23& -1.34\\
 1.51& 1.5&    -3.47& -1.20\\
 1.54& 1.5&   -3.16& -1.08\\
 1.56& 1.5&    -2.85& -0.97\\
 1.31& 2.5&      -2.44& -0.35\\
 1.34& 2.5&     -2.95& -0.43 \\
 1.36& 2.5&     -3.24& -0.47\\
 1.39& 2.5&    -3.37& -0.50\\
 1.41& 2.5&   -3.38& -0.52\\
 1.44& 2.5&   -3.16& -0.53\\
 1.46& 2.5&     -3.01& -0.53\\
 1.49& 2.5&    -3.04& -0.53\\
 1.51& 2.5&    -2.82& -0.52\\
 1.54& 2.5&     -2.70& -0.52\\
 1.56& 2.5&   -2.56& -0.51\\

\hline

\end{tabular}
\end{table}

\section{Fitting Procedures}
\label{fit}

The JM06 parameters for the non-resonant mechanisms, as well as 
the electrocouplings and hadronic decay
widths to the $\pi \Delta$ and $\rho p$ final states of nucleon resonances 
listed in Table~\ref{nstlist}  were varied simultaneously in the $\chi^2$-fits.
In this way we accounted for correlations of resonant and non-resonant contributions in the fits. 

For the non-resonant amplitudes, the $\pi \Delta$
Born terms are fixed in the fit. The magnitudes of 
the additional contact terms in $\pi^{-} \Delta^{++}$ and
$\pi^{+} \Delta^{0}$ isobar channels and  all direct
2$\pi$ production amplitudes defined in the previous section
were varied, applying different  
multiplicative factors to the different amplitudes.
These factors were the same for all 
$W$-bins inside a single $Q^2$-interval, while they were different in different
$Q^2$-bins.  Therefore, the $W$-dependence
of non-resonant mechanisms, established in preliminary adjustment to the CLAS data,
 remain unchanged in the fit, allowing
us to avoid additional $W$-dependent fluctuations of non-resonant mechanisms,
that may mask the $N^{*}$ contributions.
 The multiplicative factors were determined as random numbers 
normally distributed around
unity
 with 
$\sigma$ values in the range 
of 10 to 20 \%.

Electrocouplings of  $P_{11}(1440)$ and $D_{13}(1520)$ resonances  were varied
within 30~\% of their initial values, determined in a preliminary 
adjustment to the  CLAS data. Recent analyses of the CLAS $\pi^{+}n$ and $\pi^0p$
electroproduction data \cite{Az08a,Bu08c} revealed zero crossing in $Q^2$-evolution of
$P_{11}(1440)$ $A_{1/2}$ electrocoupling at photon virtualities $~$0.4-0.5
GeV$^2$. At these photon virtualities only non-zero $S_{1/2}$ electrocoupling of 
$P_{11}(1440)$ state affects
the $p\pi^+\pi^-$ CLAS data description, while negligible $P_{11}(1440)$ 
$A_{1/2}$ electrocoupling values are required in order to reproduce the
 final hadron angular distributions. 
Electrocoupling of other states 
listed in Table~\ref{nstlist} were fixed at the values 
obtained in a preliminary 
adjustment to the  CLAS data, as described in the previous Section. 
The $\pi \Delta$ and $\rho p$ hadronic decay widths for $P_{11}(1440)$ and
$D_{13}(1520)$ were also varied in a range, which corresponds to the total hadronic
decay width floating from 40
to 600 MeV.  
The total $N^{*}$ decay widths were calculated by summing-up 
the partial widths over all decay channels. 
Partial hadronic decay widths for $N^{*}$'s with
masses greater than 1.55 GeV were taken from the previous analyses of charged
double pion electroproduction data \cite{Az05,Mo06,Mo06-1}. They are 
independent of $Q^2$ and are in  good agreement
with the values reported in PDG~\cite{PDG08}.

With the model described above, we
 fit the CLAS data \cite{Fe07a} of nine differential cross sections of
 $ep \rightarrow e'p'\pi^+\pi^-$ electroproduction reaction, minimizing 
 $\chi^2$/d.p.. A special 
 procedure was developed in order to obtain not only the best data fit,
 corresponding to minimal $\chi^2$/d.p., but also to establish the bands 
 of calculated cross sections, that are compatible to the data within their 
 uncertainties. For each trial set of calculated cross sections 
 the $\chi^2$/d.p. value
 was estimated in point by point comparison between measured and calculated 
 nine 1-fold
 differential cross sections in all bins of $W$ and $Q^2$ covered by
 measurements. We selected in the fit all calculated 1-fold differential cross
 sections with  $\chi^2$/d.p less than maximal $\chi^2/d.p_{max}$, determined 
 so that the values of selected  1-fold differential cross
 section should be inside the data
 uncertainties for dominant part of the data points (see example 
 in Fig~\ref{fit_9sec}). 
 The values of resonant and non-resonant
 parameters of JM06 model assigned to one selected in the fit cross sections 
 were utilized
 in evaluation of various isobar channel cross sections as described 
 in the Section~\ref{results}.
 
\section{Results and discussion}
\label{results} 
Within the framework of JM06 model, we have succeeded 
in fitting the CLAS  $p\pi^+\pi^-$ electroproduction 
data at low W and Q$^{2}$ \cite{Fe07a,db07}.

 The band of cross sections selected in fitting procedure provided reasonable
 data description. It may be seen from intervals of 
$\chi^{2}$/d.p. values achieved in the CLAS data  fit and   
listed in Table~\ref{chi} for three $Q^{2}$ areas, where data
fits were carried out. The example of data description in single bin of
$W$ and $Q^2$ is shown in Fig.~\ref{fit_9sec}. For the first time 
we determined in the fit
the intervals of 1-fold differential $p\pi^+\pi^-$ cross sections that are
compatible to the data, accounting for  
the data uncertainties$^{1}$.\footnotetext[1]{To the best
of our knowledge,
in previous resonance analysis of electroproduction data the errors in the fit
parameters were estimated
assuming linear error propagation. This simplification may result in
unrealistically small uncertainties
in the resulting fit parameter.}

In the upper two rows of
Fig.~\ref{9sectok} we show some of the best fits (solid curves) to
the three invariant mass distributions (top row), angular distributions
of three outgoing particles (center row).
The contributions from the  $\pi^- \Delta^{++}$,
$\pi^+\Delta^0$ isobar channels and direct two-pion production term $T^{dir}$ 
are also
shown there. The minimal $\chi^{2}$/d.p.  achieved in the fits
of the mentioned above six 1-fold differential cross sections in each bin of
$Q^2$ covered by measurements are listed 
in the Table~\ref{pizdato}.
The larger
$\chi^{2}$/d.p. at smallest
photon
virtualities are mostly related to the smaller statistical uncertainties of the
data points and the fact that systematic uncertainties were not included in
the fit.

To check the robustness of JM06 mechanisms, we then use the determined
parameters to predict the distributions as functions
of the angles $\alpha_i$, defined in Section~\ref{xsect},  between 
various reaction planes. In the bottom row of Fig.~\ref{9sectok}, we see that the predicted distributions
(solid curves) are in good agreement with the data.
Clearly, a global fit to the differential
cross sections data enables us to establish all essential
$p\pi^+\pi^-$ electroproduction
mechanisms at phenomenological level in the area of 1.3$<$W$<$1.57 GeV and
0.25$<$$Q^{2}$$<$0.6 GeV$^{2}$ covered by the recent CLAS data \cite{Fe07a}. No additional mechanisms are needed to
describe the CLAS data in this kinematical domain.

\begin{figure*}
\begin{center}
\epsfig{file=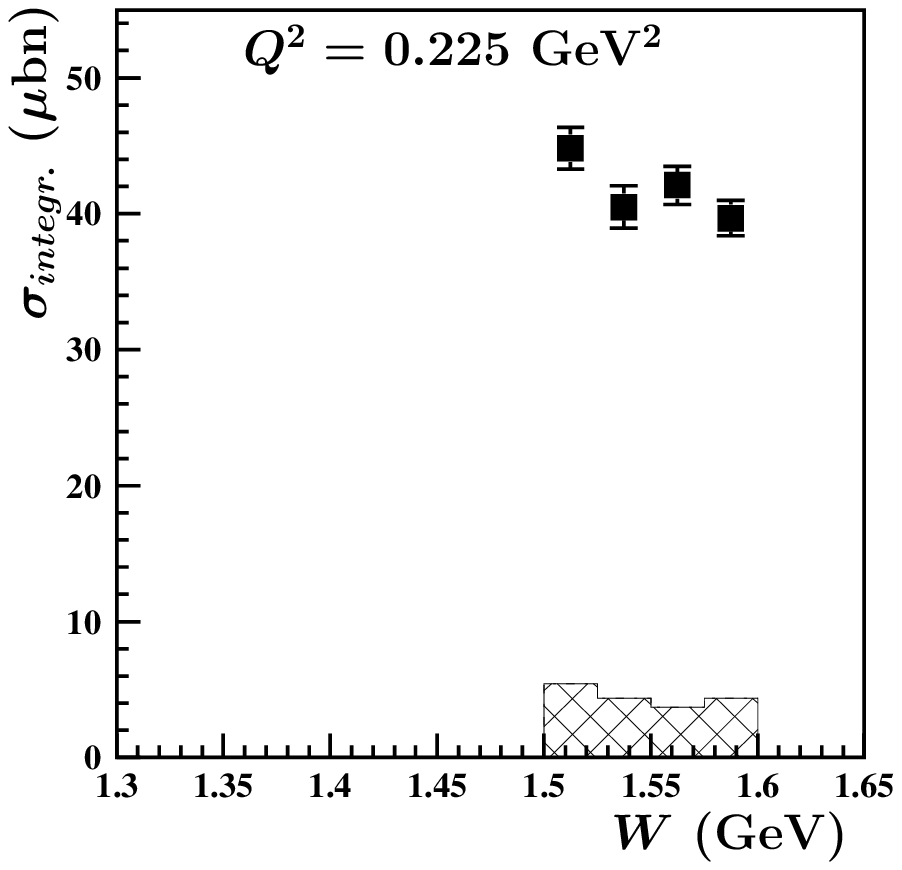,width=6cm}
\epsfig{file=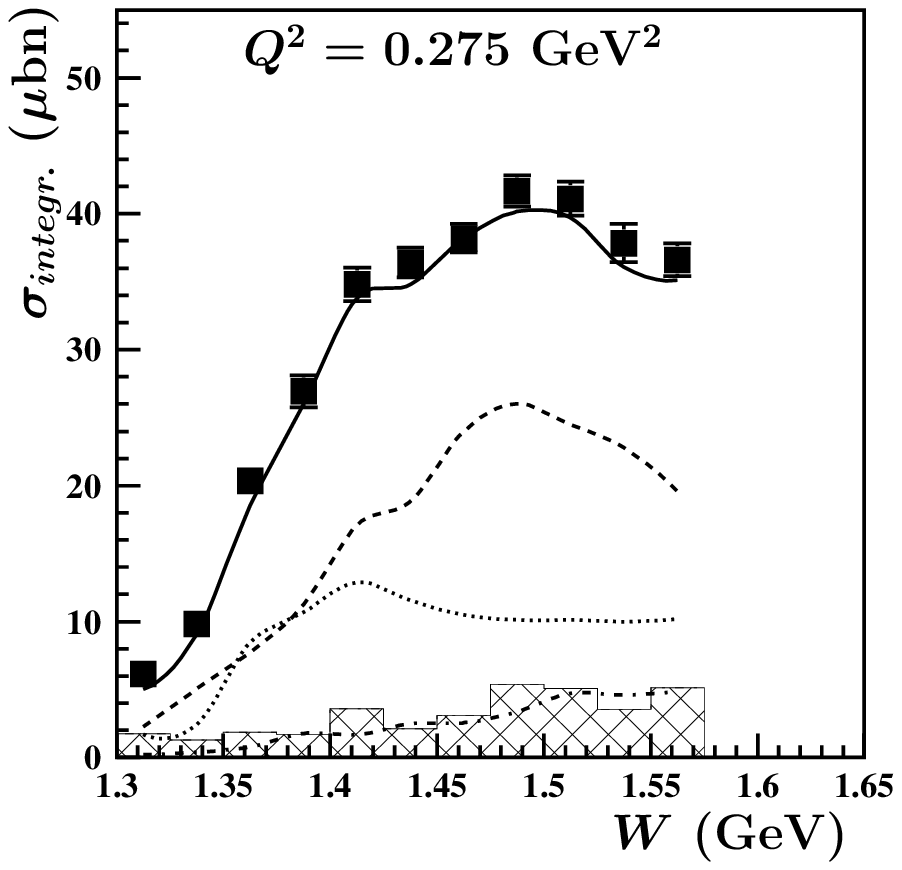,width=6cm}
\epsfig{file=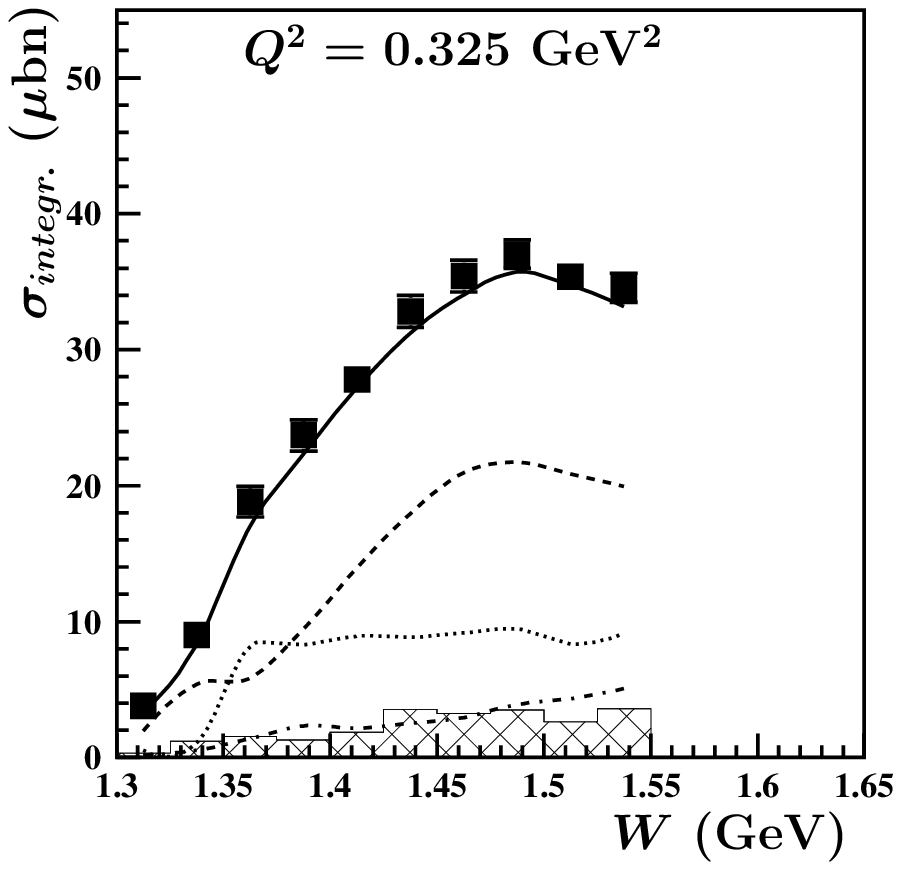,width=6cm}
\epsfig{file=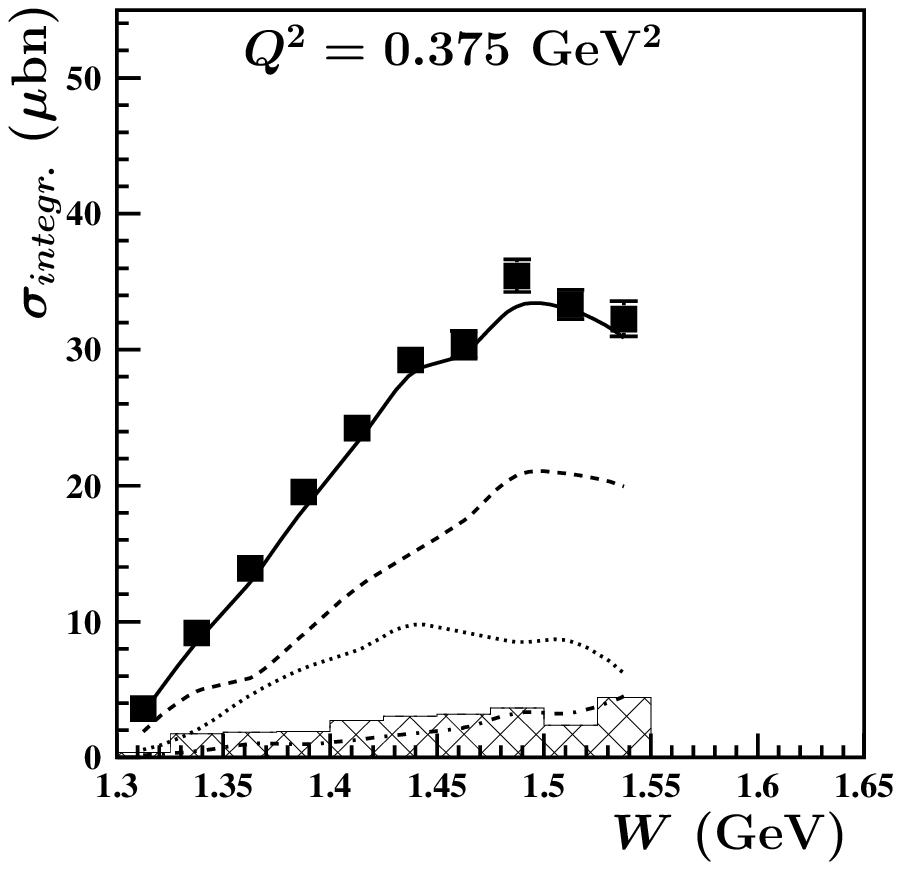,width=6cm}
\epsfig{file=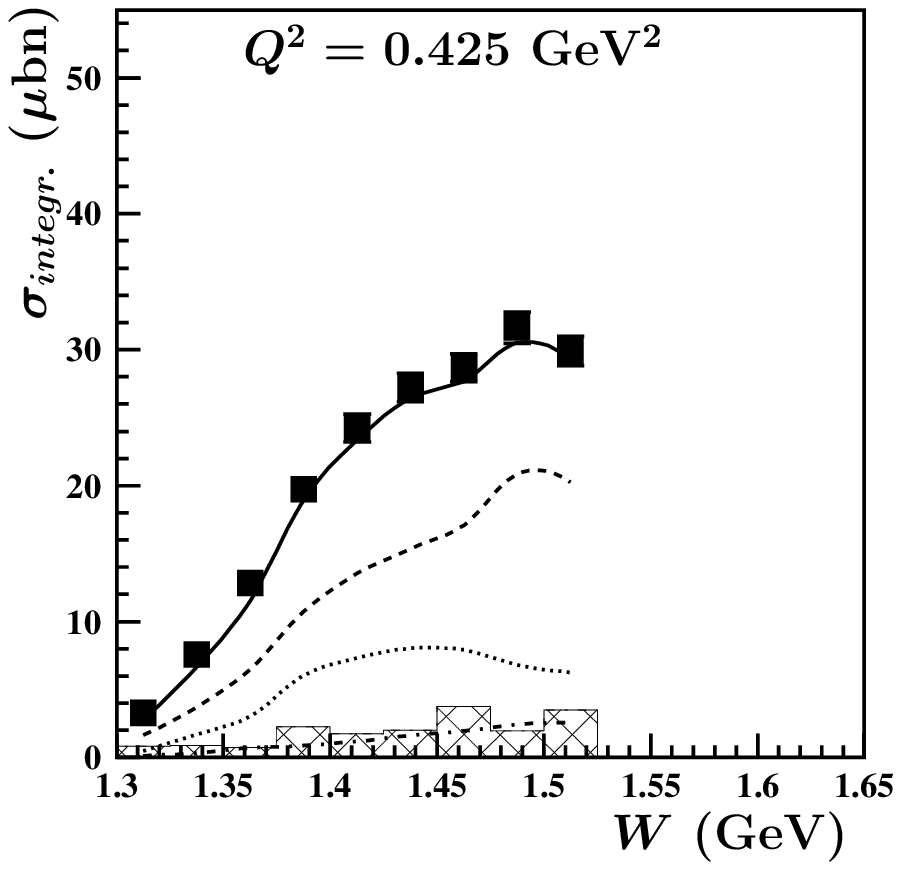,width=6cm}
\epsfig{file=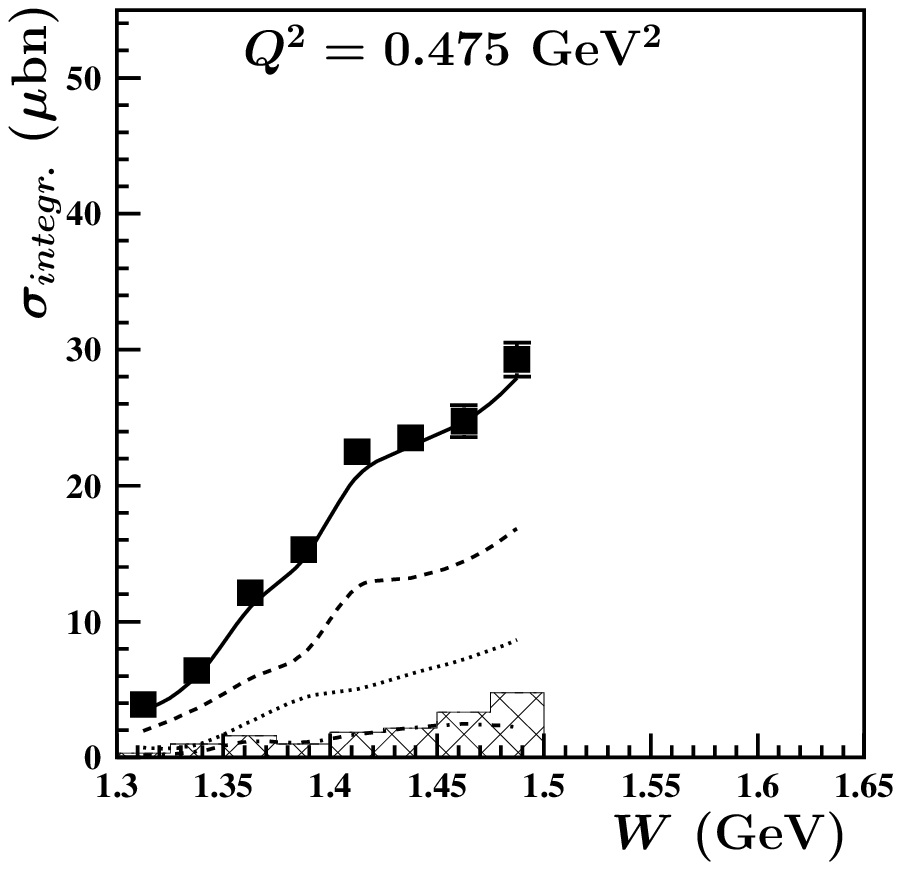,width=6cm}
\epsfig{file=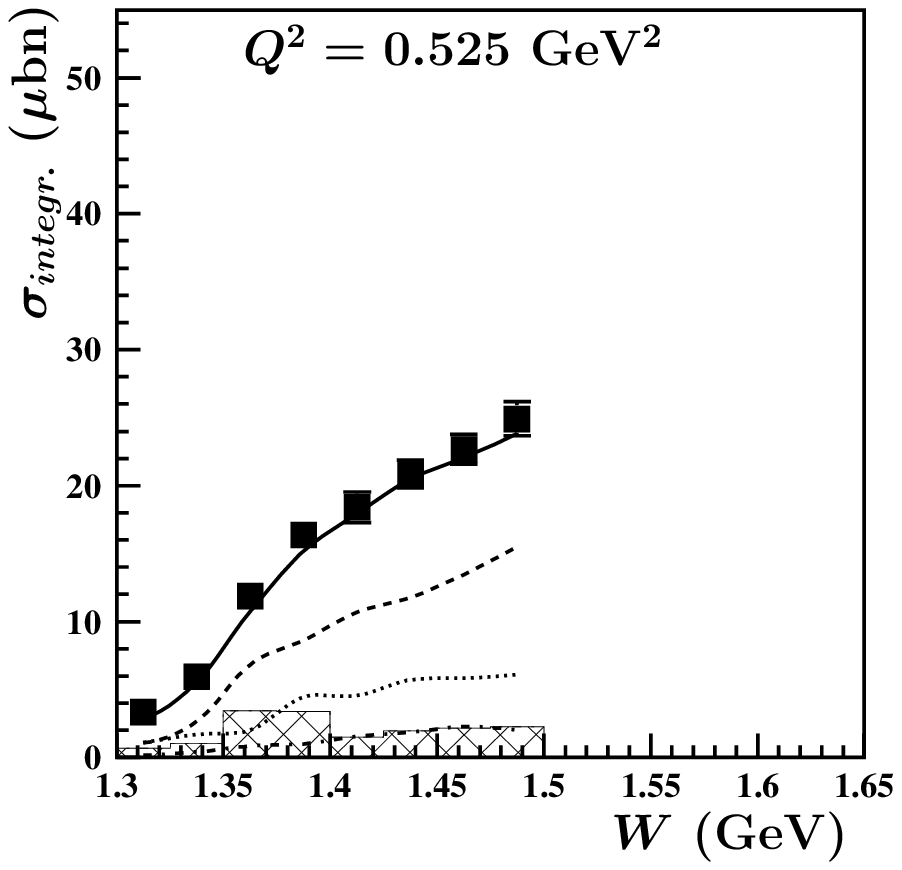,width=6cm}
\epsfig{file=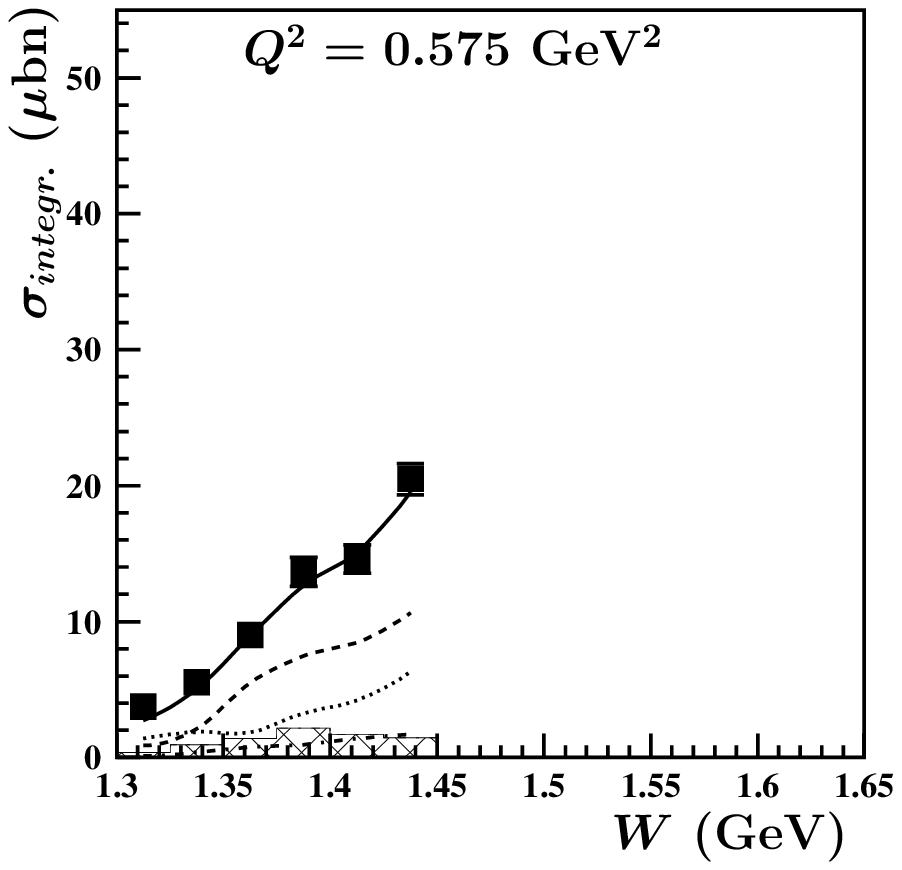,width=6cm}
\end{center}
\vspace{-0.6cm}
\caption{\small The contributions from 
various isobar channels to the fully integrated $p\pi^+\pi^-$ electroproduction 
cross sections. The recent CLAS data are
shown by full symbols. Shadowed
areas represent the systematical uncertainties. Full calculations within the
framework of JM06 are shown by solid lines. The
contributions from $\pi^{-} \Delta^{++}$, $\pi^{+} \Delta^{0}$ channels are
shown by dashed and dot-dashed lines, respectively. The contributions from direct
2$\pi$ production are shown by dotted lines.}
\label{systemerr}
\end{figure*}

The success of the   
JM06 model 
can also  be seen in Fig.~\ref{systemerr} comparing
the calculated fully
integrated cross sections with the data.

\begin{figure*}
\begin{center}
\epsfig{file=pictures/q0225.eps,width=6cm}
\epsfig{file=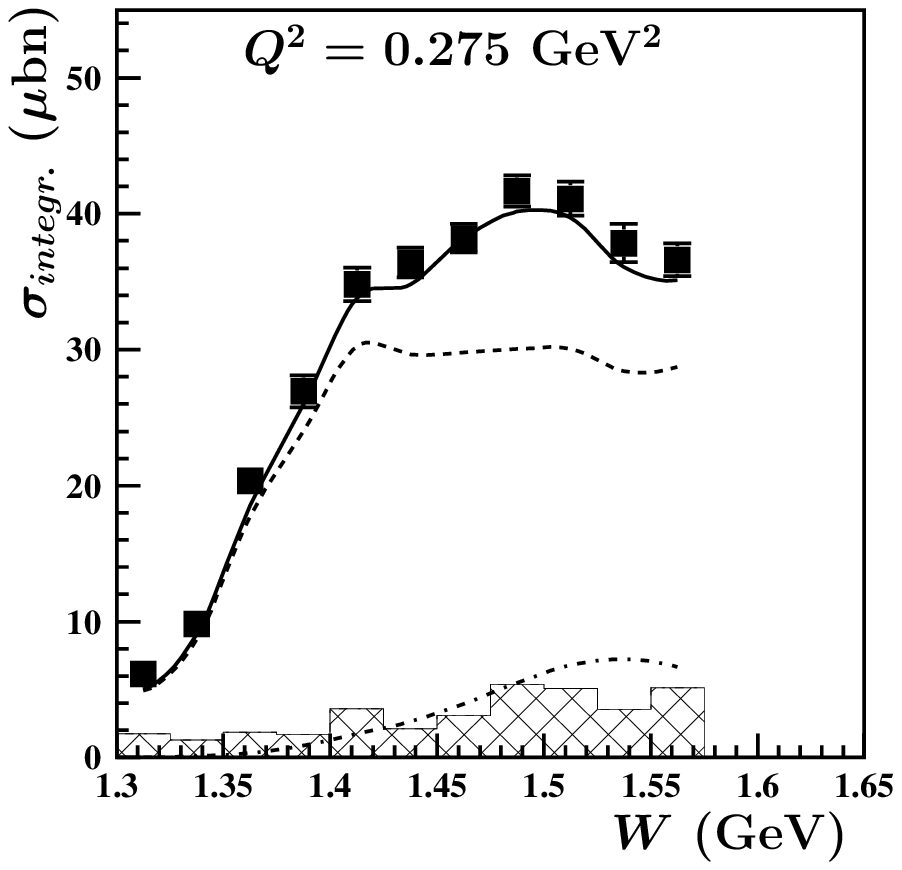,width=6cm}
\epsfig{file=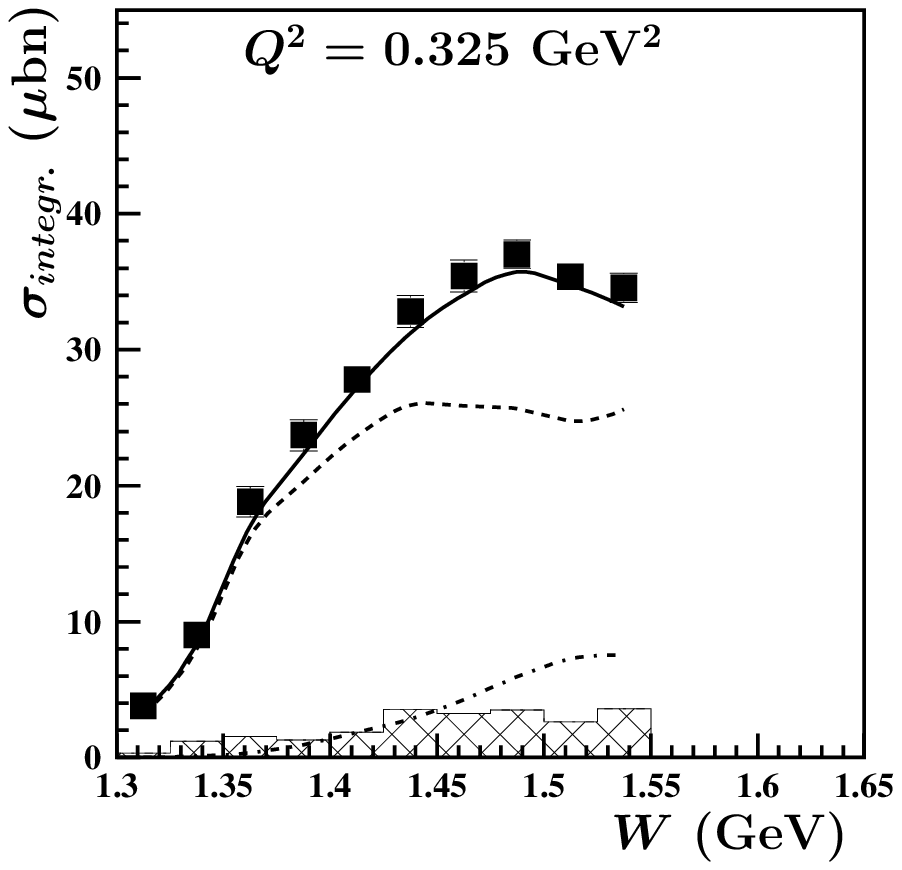,width=6cm}
\epsfig{file=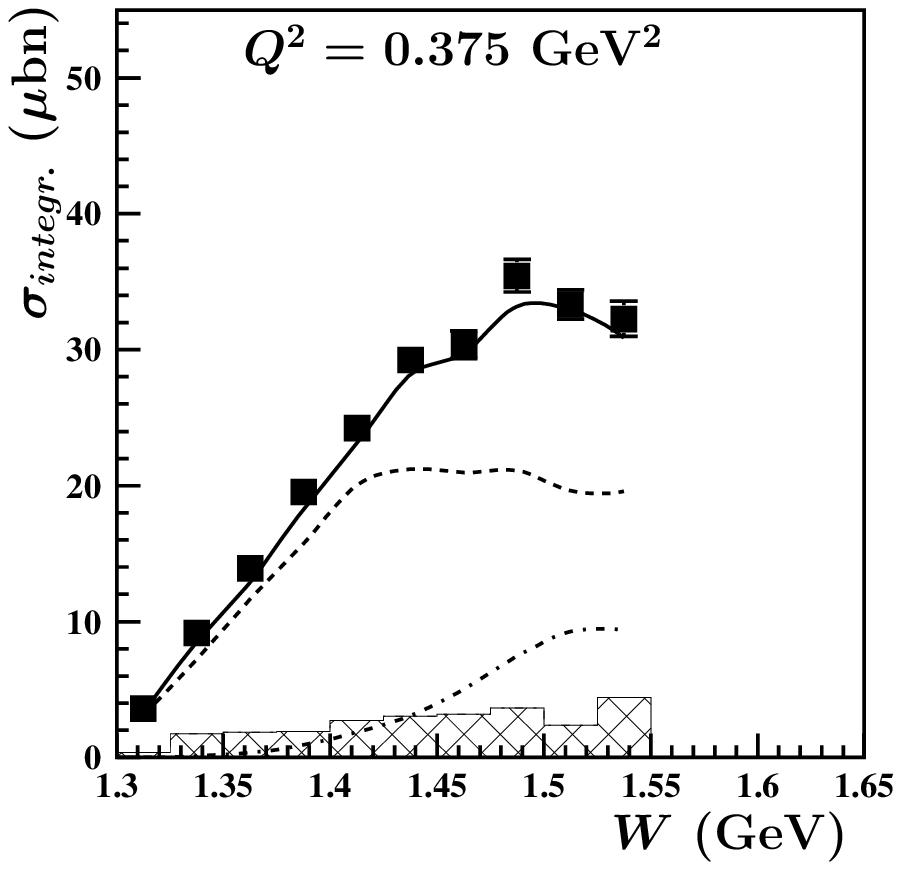,width=6cm}
\epsfig{file=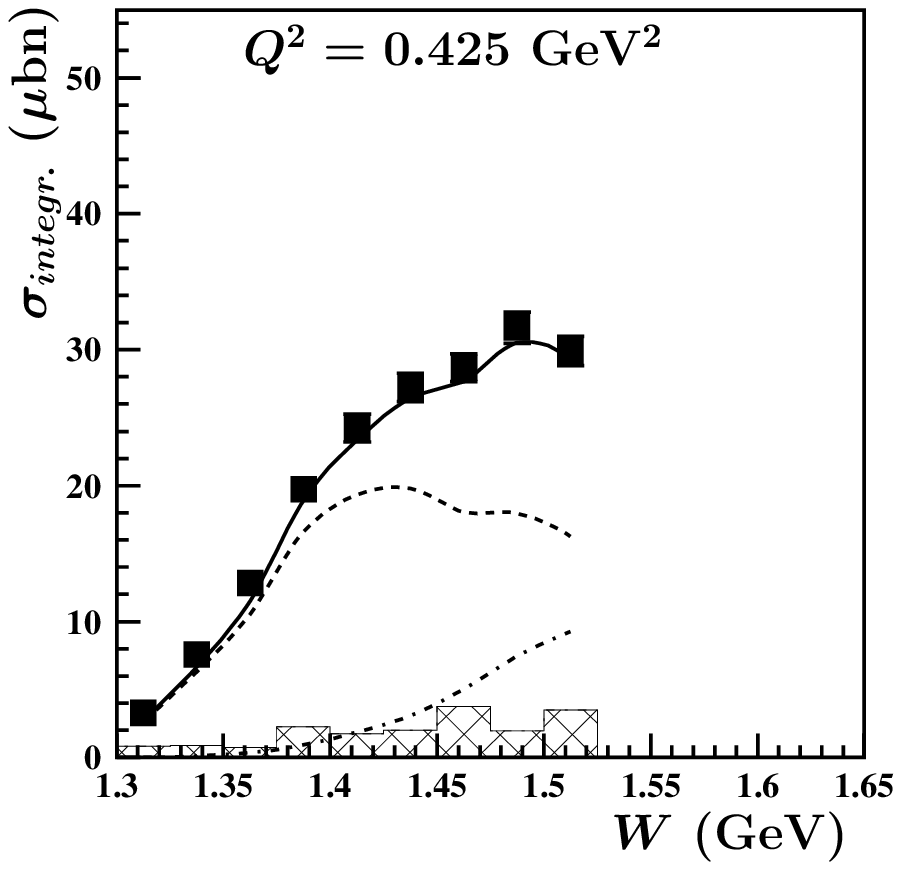,width=6cm}
\epsfig{file=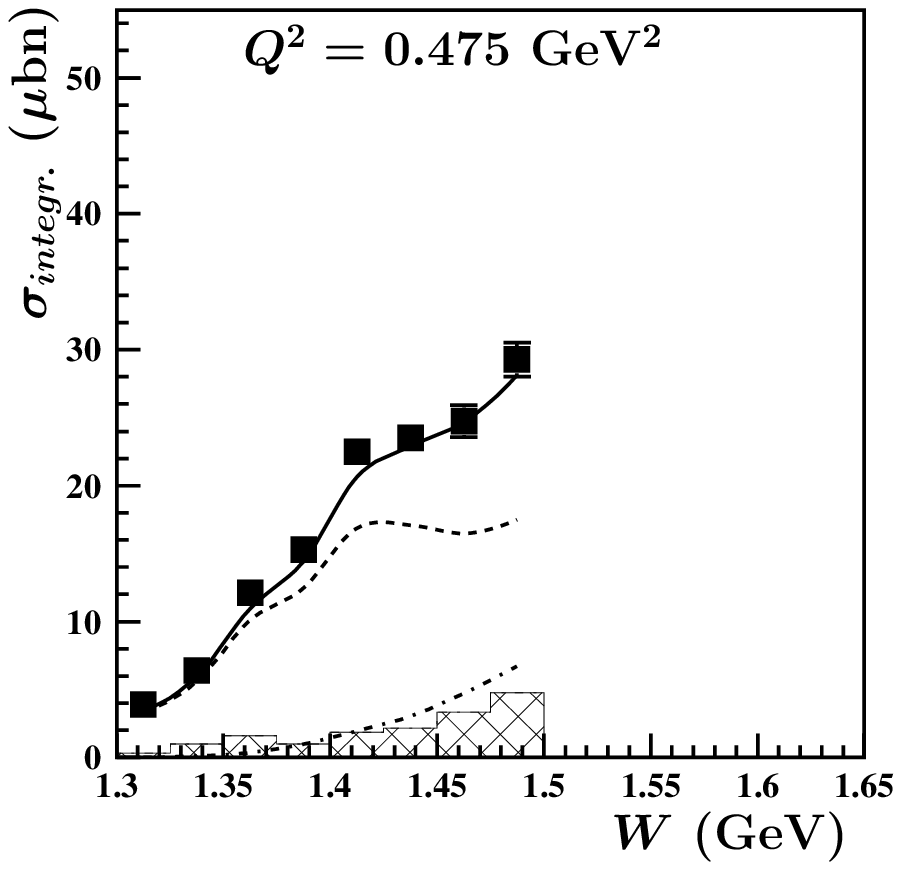,width=6cm}
\epsfig{file=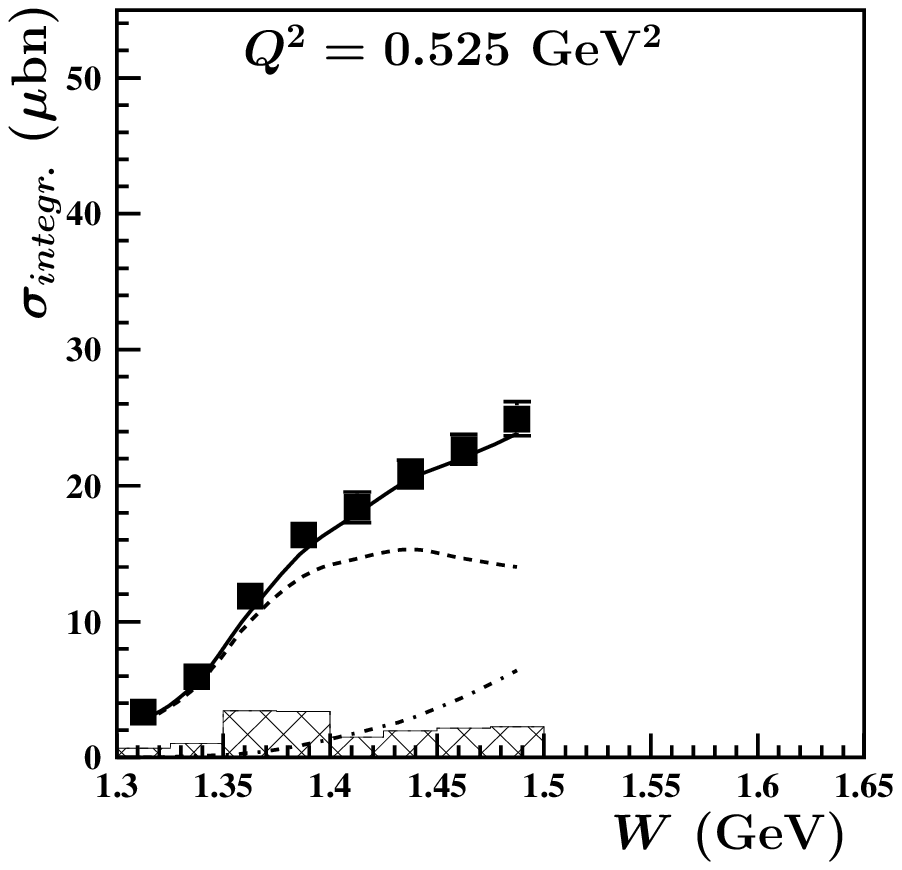,width=6cm}
\epsfig{file=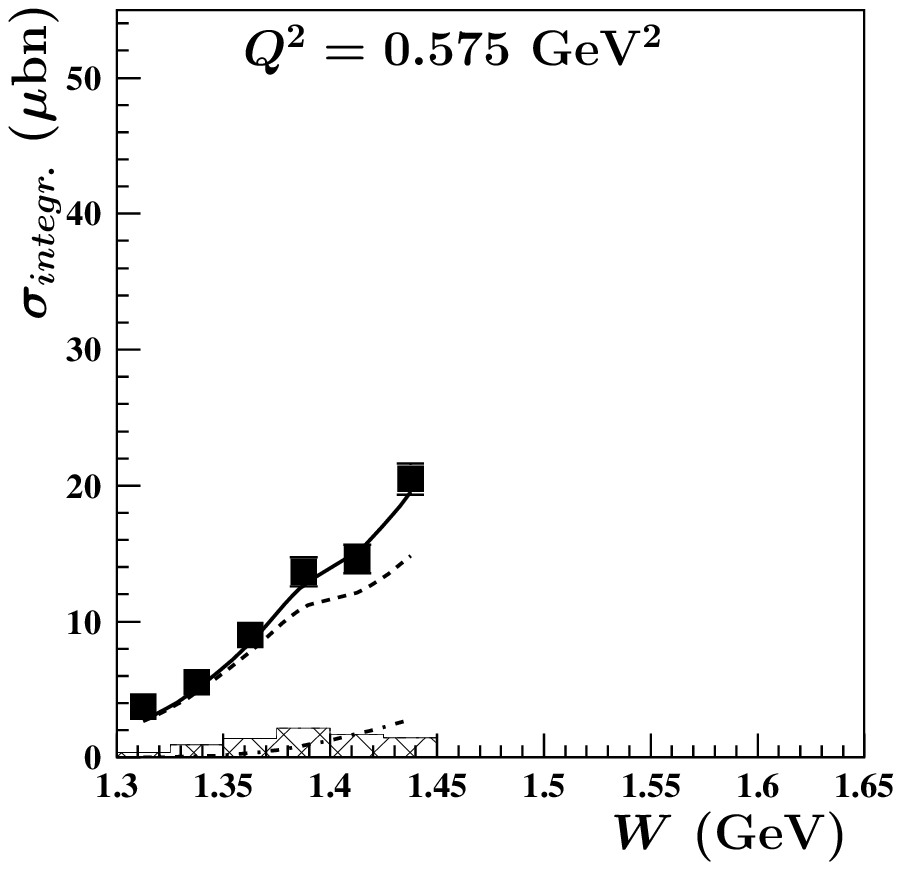,width=6cm}
\end{center}
\vspace{-0.6cm}
\caption{\small Resonant and non-resonant contributions, determined from the
recent CLAS data fit within the framework of JM06 model.
The data are shown by full symbols.
  Shadowed
areas represent the systematic uncertainties. Full JM06 calculations 
are shown by full solid lines. The
contributions from $N^{*}$'s and non-resonant mechanisms are shown by dot-dashed
and dashed lines, respectively.}
\label{systemerr_nst}
\end{figure*}

\begin{figure*}[htp]
\begin{center}
\includegraphics[width=15cm]{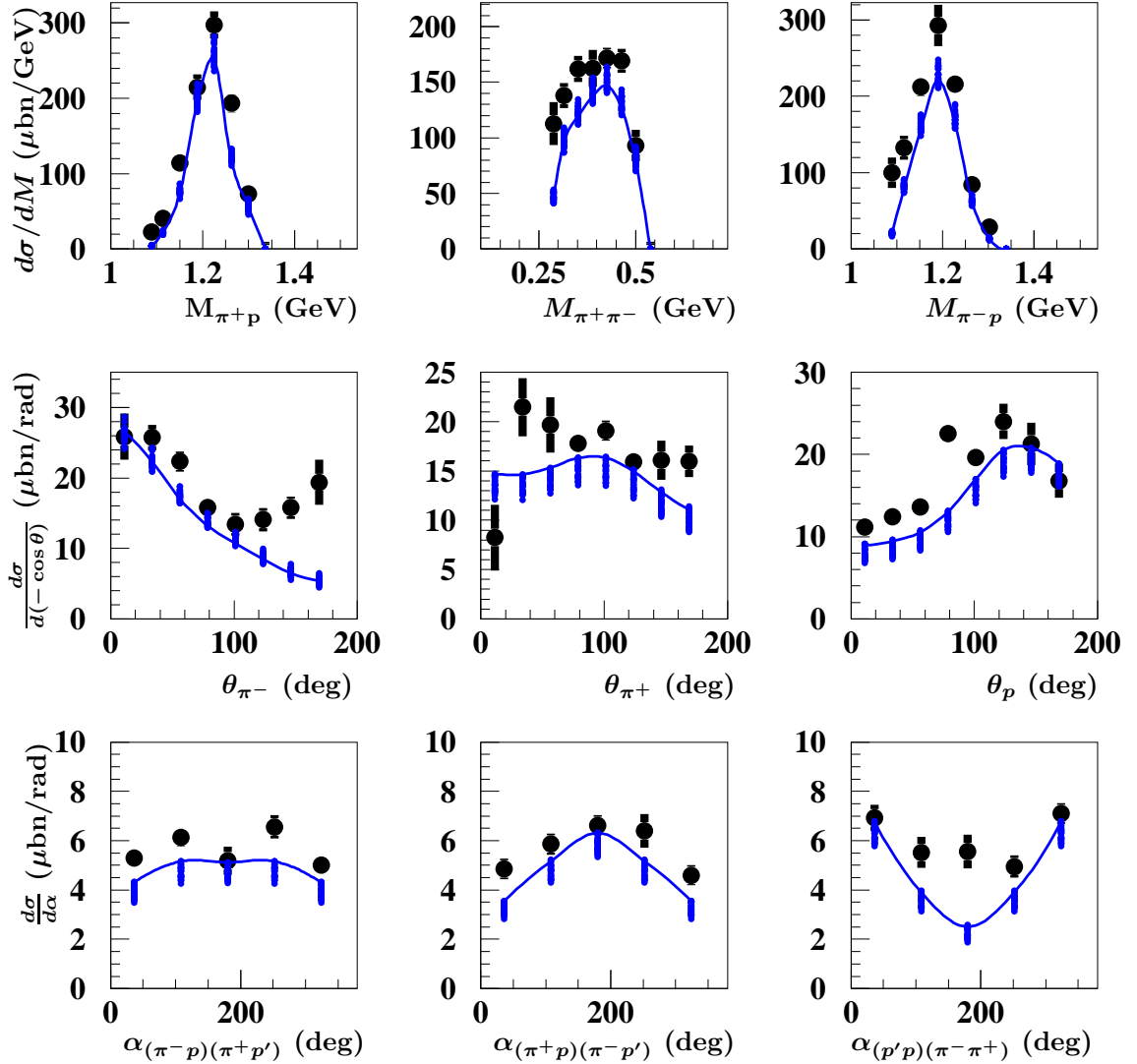}
\caption{\small(color online) Differential cross sections for the 
coherent sum of $\pi \Delta$ channel amplitudes
at W=1.46 GeV and $Q^{2}$=0.325 GeV$^{2}$ are shown by  bars connected by curves. 
 The bar sizes
represent uncertainties for these cross sections.
The cross section values corresponding to the best data fit 
are shown by solid lines. 
Full symbols are the measured cross sections.}
\label{isotot}
\end{center}
\end{figure*}

\begin{figure*}[htp]
\begin{center}
\includegraphics[width=15cm]{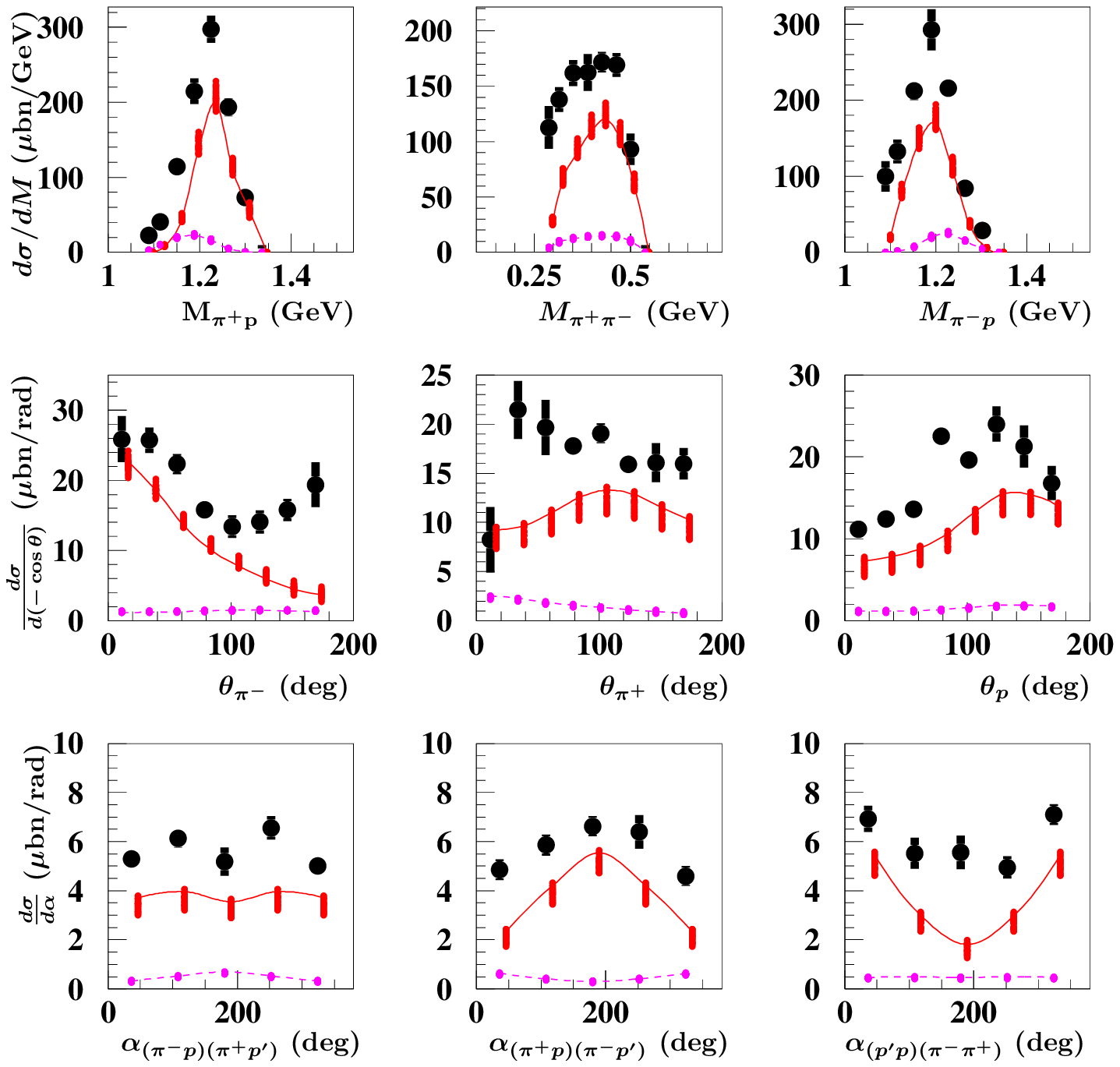}
\caption{\small(color online) Contributions from various isobar channels to 
charged double
pion electroproduction obtained from the CLAS data fit at W=1.46 GeV and 
$Q^{2}$=0.325 GeV$^{2}$. The cross sections for
$\pi^{-} \Delta^{++}$, $\pi^{+} \Delta^{0}$ isobar channels are shown by bars
connected by solid and dash lines, respectively. 
 Full symbols are the measured cross sections. The uncertainties of bars
 represent the spread of the cross sections calculated 
 with JM06 parameters
 determined from the fit of measured differential cross sections, as 
 described in the Section~\ref{fit}. The values connected by solid lines
 represent the $\pi^{-} \Delta^{++}$ and $\pi^+\Delta^0$
  channel cross sections, estimated with JM06 parameters, that correspond to the
  best data fit of minimal $\chi^2$/d.p.  }
\label{iso_9sec}
\end{center}
\end{figure*}

\subsection{$\pi \Delta$ contributions}\label{isobar}

With the parameters determined from the fits 
(see example in Fig.~\ref{fit_9sec})
, we now use the JM06 model to investigate
the $\pi \Delta$ contributions to the $p\pi^+\pi^-$ electroproduction
cross sections.
This is done by examining the calculations of nine differential 
cross sections
without including the direct two-pion
production term $T^{dir}$. For each selected in the data fit set of
differential cross
sections we evaluated the contribution from the coherent sum of the amplitudes from 
$\pi^{-} \Delta^{++}$ and $\pi^+\Delta^0$ channels.
In this way we determine the bands of
these partial cross sections imposed by the uncertainties of the experimental
data. The coherent sum of isobar channel contributions extracted in the fit
of  $p\pi^+\pi^-$
cross sections in single bin of $W$ and $Q^2$ is shown 
in Fig.~\ref{isotot}. 

We see in Fig.~\ref{isotot} that
the
$\pi \Delta$ contributions account for most of the measured cross sections.
 The differences
are due to the direct
2$\pi$ production mechanisms and their interference with $\pi \Delta$
production amplitudes. These contributions are most pronounced
for the $\pi^{-}$ production at backward angles where
the direct 2$\pi$ production
mechanisms give the main contributions. This distinctive signature
allows us to separate the cross sections generated by a 
coherent sum of isobar channel amplitudes and those 
by the  direct 2$\pi$ production.

In Fig.\ref{iso_9sec} we compare the contributions from each of the
two $\pi \Delta$ channels.
Clearly, the $\pi^{-} \Delta^{++}$ channel is 
a major contributor to the cross sections, as  
expected from isospin symmetry. The $\pi^{-} \Delta^{++}$ and 
$\pi^{+} \Delta ^{0}$ channels exhibit
rather different shapes in all nine differential cross sections, allowing us to
separate their respective contributions. 

Differential cross sections of individual isobar
channels extracted from the CLAS data \cite{Fe07a} 
may be found in \cite{Gleb_web}. 
This information will be useful
for developing  a more  microscopic understanding of
the isobar production mechanisms, such as those described in terms
of explicit meson
baryon diagrams \cite{Os1,Os2,Ki06,Ma06,Lee07,Lee07a}.

\subsection{Non-resonant partial-wave amplitudes in $\pi \Delta$ channels}
\label{pwpidel}

 Estimates for the amplitudes of mechanisms contributing to the
 charged double pion electroproduction are of particular
importance for N$^{*}$ studies in a global multi-channel analysis 
within the
framework of
advanced coupled channel formalism, such as that developed in 
\cite{Ma06,Lee07,Lee07a,Lee08,Lee09}. A first step foreseen in these efforts will be 
the combined analysis of major single and double pion electroproduction
channels. Analysis of the $p\pi^+\pi^-$ electroproduction data allowed 
us to obtain  
information on non-resonant amplitudes contributing to
$\pi \Delta$ isobar channels, decomposed  
into a set of partial waves, described  in the Appendix VI. They were 
estimated within the framework of JM06 model with parameters corresponding to
the minimal $\chi^{2}$/d.p.. Non-resonant parts of $\pi \Delta$ channel
amplitudes consist of Born terms and additional contact terms
described in Section~\ref{jlabmsugen}. The absorbtive coefficients applied to
the Born term in order to account for ISI$\&$FSI \cite{Ri00} were taken off. In
this way we get the amplitudes, that are non-distorted by the interactions 
with open inelastic channels. The
partial waves for non-resonant amplitudes in $\pi \Delta$ isobar channels,
derived in the CLAS data fit, are given  
in \cite{db_cal}.
In Tables~\ref{lg1lp12ld32},\ref{lg1lp12ld12},\ref{lg1lp12ldm12} and
\ref{lg1lp12ldm32} we give example of such partial waves at 
$Q^2$ = 0.275 GeV$^2$. The tables contain the partial waves for 4 of 12 
independent helicity amplitudes for the photon helicity 
$\lambda_{\gamma}$=1, at fixed values of $Q^{2}$ and running mass of 
$\pi^{+} p$ system as a function of $W$. It shall be noted that the partial waves 
of $J$=1/2 total angular momenta vanish for
the helicity amplitudes corresponding to the 3/2 total spin projections on 
the initial photon or the final pion momenta, as it is expected from angular
momentum conservation.

\begin{figure*}[htp]
\begin{center}
\includegraphics[width=15cm]{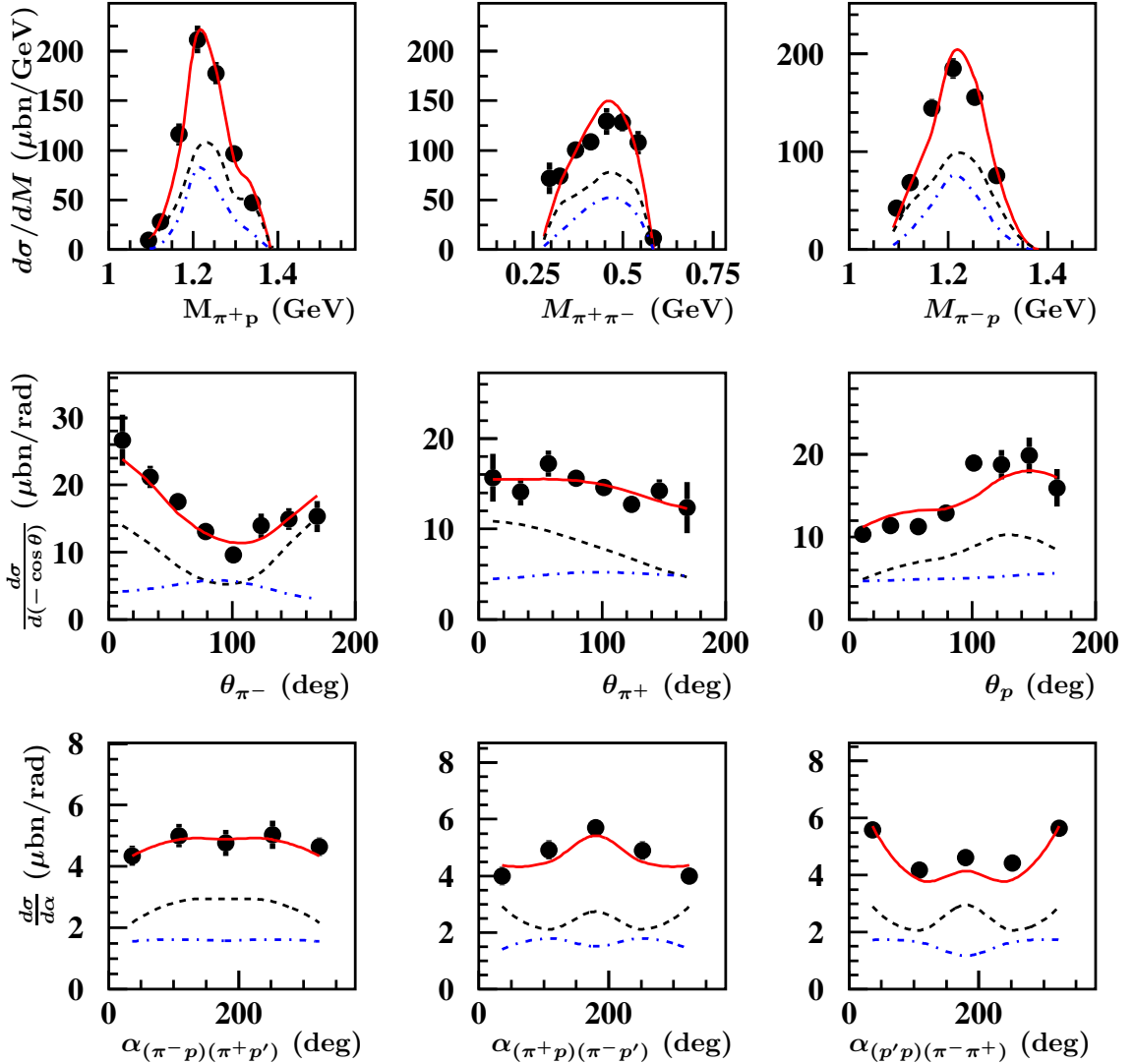}
\caption{\small(color on line) Resonant (dot-dash lines) and non-resonant 
(dot lines) mechanism
contributions to the 1-fold differential cross sections at W=1.51 GeV and
$Q^{2}$=0.425 GeV$^{2}$. The solid lines represent full JM06 calculation.} 
\label{9secnstbck}
\end{center}
\end{figure*}

\subsection{Resonant mechanisms}
 
With the parameters determined, we can examine 
 the resonant and
non-resonant contributions within JM06.
The results for the integrated cross sections are already shown in
Fig.~\ref{systemerr_nst}.
The results for the
nine  differential cross sections at W=1.51 GeV and
$Q^{2}$=0.425 GeV$^{2}$ are shown in Fig.~\ref{9secnstbck}.
 
In Fig.~\ref{systemerr_nst} we see that
the contribution from resonant cross sections becomes sizable at  W $>$ 1.40
GeV and increases with W. The 
resonant contributions also
increase with Q$^{2}$. At W $>$ 1.40 GeV 
the resonant contribution ranges
from 10 \% to 30 \%. In Fig.~\ref{9secnstbck} we see that the shapes 
of the resonant and non-resonant  differential cross sections  
are rather different, allowing us to isolate the resonant contribution in a
combined fit of nine 1-fold differential cross section, despite of relatively 
small resonant contributions to the fully integrated cross sections

From analyzing the results shown in Figs.\ref{systemerr_nst} 
and \ref{9secnstbck},
we find that the dominant part of resonant
cross sections  at  $W$ $<$ 1.6 GeV and Q$^{2}$
from 0.25 to 0.60 GeV$^{2}$ 
comes from the $P_{11}(1440)$ and 
$D_{13}(1520)$ resonant
states. Previously,  the $P_{11}(1440)$ and $D_{13}(1520)$ electrocouplings
 at a single $Q^{2}$ $=$ 0.4
GeV$^{2}$ were determined from the analysis of the 1$\pi$  CLAS 
data\cite{Az05-1}. 
 The analysis presented above
has extended the determination of these resonance parameters
to cover the region of photon
virtualities from 0.25 to 0.60 GeV$^{2}$.
Moreover, in $p\pi^+\pi^-$ production the 
contribution from $P_{11}(1440)$ 
does not interfere with that from the
tail of $P_{33}(1223)$, as it does in $N\pi$ electroproduction. This feature
makes $p\pi^+\pi^-$ exclusive channel particularly attractive 
for the studies of the $P_{11}(1440)$ resonance.

\section{\label{concl} Conclusions and outlook.}
   
The analysis of comprehensive CLAS data on charged double pion 
electroproduction at W $<$ 1.6 GeV and Q$^{2}$=0.25-0.60 GeV$^{2}$
within the framework of JM06 model allowed us to establish all essential 
contributing mechanisms at those
kinematics.
A good description of nine 
1-fold differential and fully integrated cross sections 
was achieved in the entire kinematics. The robustness of the 
established mechanisms
was demonstrated in the successful prediction of the three angular
distributions $\alpha_{i}$ (i=1,2,3), which were not included in the fit.

The CLAS data allowed us to evaluate the 
contributions from the $\pi \Delta$ channels, coherently or incoherently,
to nine 
differential cross sections.
 These $\pi \Delta$ differential
cross sections 
are determined in each bin of $W$ and $Q^{2}$ and may be found in \cite{Gleb_web}.
 This information as well the corresponding extracted partial wave
amplitudes  could be useful
for  a microscopic understanding of $\pi \Delta$
production mechanisms, such as that being
investigated at EBAC~\cite{Ma06,Lee07,Lee07a,Lee08}.

Successful description within
the framework of JM06 model of the large body of $p\pi^+\pi^-$
electroproduction data  allowed us to isolate the resonant parts of
cross sections, offering access to electrocouplings of 
$P_{11}(1440)$ and $D_{13}(1520)$
resonances at Q$^{2}$=0.25-0.60 GeV$^{2}$. Evaluation of these electrocouplings
is in progress \cite{Mo09,Fe07b}. In this
low $Q^{2}$ region,
the meson-baryon dressing of the $N$-$N^*$ transition amplitudes is expected 
to be large~\cite{Can98,Sa01,Lee08}.
Therefore, evolution of $P_{11}(1440)$ and $D_{13}(1520)$ electrocouplings 
at lower $Q^2$ $<$ 0.6 GeV$^2$ is of particular interest for studies of  
meson-baryon dressing and still 
unexplored in double pion electroproduction.

\section{\label{cha:ack}Acknowledgments}
This work was supported
in part by the U.S. Department of Energy and the National
Science Foundation, the Jefferson Lab, the Argonne National Lab, 
the Skobeltsyn Institute of Nuclear Physics and
Physics Department at Moscow State University and the Russian Federation Government
Grant 2009-1.1-125-055. 
Jefferson Science Associates, LLC, operates Jefferson Lab
under U.S. DOE contract DE-AC05-060R23177.

\section*{Appendix I: Born amplitudes.}
\begin{figure}[htp]
\begin{center}
\includegraphics[width=9cm]{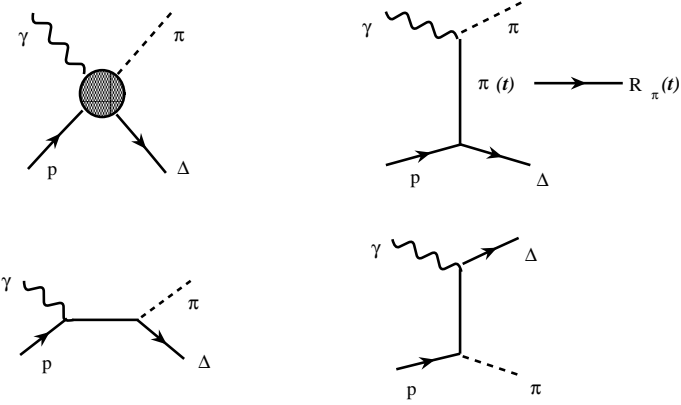}
\caption{\small Born amplitudes in JM06 model} 
\label{piborn}
\end{center}
\end{figure}

Here we present the list of amplitudes for the Reggeized
Born terms used in the JM06 model. These terms were derived from the 
diagrams shown in Fig.~\ref{piborn}. 

They consist of the contact
term, 
the  t-channel
pion-in-flight diagram,the  s-pole nucleon term, and the u-channel $\Delta$-in-flight
diagram \cite{So71,Bar72}.
In order to describe the off-shell pion or $\Delta$ interactions to the hadronic currents shown 
in the right side diagrams of the Fig.~\ref{piborn}, 
the respective hadronic vertex functions were used. 
The off-shell effects in virtual photon interactions to the hadronic currents depicted in the 
Fig.~\ref{piborn} were taken into account by implementing electromagnetic
vertex functions. For hadronic and electromagnetic  vertex functions we used a compilation of
experimental data \cite{Am89,Mac79,Beb78}.

The helicity amplitudes for
the contact term are given by
\begin{equation}
t_{\lambda_{\Delta}\lambda_{\gamma}\lambda_{p}}^{c} =
g_{c}(Q^{2},t)\overline{u}_{\mu}(p_{2},\lambda_{\Delta})
u(p_{1},\lambda_{p})\varepsilon_{\mu}(q,\lambda_{\gamma})
\end{equation}
where \(p_{1}\) and \(p_{2}\) are the target proton and
\(\Delta\) four momenta, \(q\) is the photon four momentum and
\(u_{\mu}\),
\(\varepsilon_{\mu}\) and \(u\) are
Rarita-Schwinger spinor-tensor for \(\Delta\) with
\(\lambda_{\Delta}\)
helicity,
the 4-vector of ingoing photon with \(\lambda_{\gamma}\) helicity, and the proton target spinor 
of \(\lambda_{p}\) helicity;
\(g_{c}(Q^{2},t)\) is an effective contact term vertex function.

The pion-in-flight contribution 
reads:
\begin{eqnarray}
t_{\lambda_{\Delta}\lambda_{\gamma}\lambda_{p}}^{pif}
= g_{\pi}(Q^{2},t)\frac{(2p_{\pi}^{\mu}-
q^{\mu})\varepsilon_{\mu}(q,\lambda_{\gamma})}
{t-m_{\pi}^{2}} \cdot\nonumber \\
\overline{u}_{\nu}(p_{2},\lambda_{\Delta})u(p_{1},\lambda_{p})
(q^{\nu}-p_{\pi}^{\nu})
\end{eqnarray}
where \(p_{\pi}\) is the pion momentum, \(m_{\pi}\) is the pion mass and \(g_{\pi}\)($Q^{2}$,$t$) is the product
of strong and electromagnetic vertex functions:
\begin{equation}
g_{\pi}(Q^{2},t) = G_{\pi ,em}(Q^{2})G_{\pi N\Delta}(t).
\end{equation}
The electromagnetic vertex function is described by the
pole fit of the pion form factor \cite{Am89}:
\begin{equation}
G_{\pi ,em}(Q^{2}) = \frac{1}{\left( 1+\frac{Q^{2}(GeV^{2})}{\Lambda_{\pi}^{2}}
\right) } \frac{1}{G_{\pi N \Delta}(t_{min})}
\end{equation}
 where $t_{min}$ corresponds to pion production in hadronic C.M. at
zero degree angle; the factor $\frac{1}{G_{\pi N \Delta}(t_{min})}$
reflects the way $G_{\pi, e m}(Q^{2})$ form factor was extracted from
single pion electroproduction data \cite{Beb78}.  
The analysis in \cite{Beb78} yielded \(\Lambda_{\pi}^{2} = 0.462\) \(GeV^{2}\)
which we used in our calculations. Concerning the t-dependence of $\pi N \Delta$ vertex, we introduce 
as vertex function the hadronic form factor successfully applied in \(NN \rightarrow N\Delta\)
relativistic transition potentials \cite{Mac79}:
\begin{equation}
G_{\pi N\Delta}(t) = g_{0}\frac{\Lambda^{2}-m_{\pi}^{2}}{\Lambda^{2}-t}
\end{equation}
The interaction constant \(g_{0}\) and cut-off parameter \(\Lambda\) are:
\(g_{0}=2.1/m_{\pi}\) and  $\Lambda = 0.75$ GeV.
Current conservation implies equality of the contact term coupling 
\(g_{c}(Q^{2},t)\) 
and the product of the hadronic and electromagnetic vertex functions  
\(g_{\pi}(Q^{2},t)\) in the pion-in-flight term.

The   s-channel nucleon contribution  is
given by:
\begin{eqnarray}
t_{\lambda_{\Delta}\lambda_{\gamma}\lambda_{p}}^{N} =
g_{N}(Q^{2})g_{0}\frac{(2p_{1}^{\mu}+q^{\mu})\varepsilon_{\mu}(q,\lambda_{\gamma
})}
{s-m_{N}^{2}} \cdot\nonumber \\
\overline{u}_{\nu}(p_{2},\lambda_{\Delta})u(p_{1},\lambda_{p})
p_{\pi}^{\nu}
\end{eqnarray}
where  s is Mandelstam invariant \(s=(q+p_{1})^{2}\), \(m_{N}\)
is the nucleon mass, \(g_{N}\) is
the electromagnetic vertex function,
described
by the dipole fit \cite{Am89}:
\begin{equation}
g_{N}(Q^{2}) = \frac{1}{\left( 1+\frac{Q^{2}(GeV^{2})}{0.71} \right)^{2} }
\end{equation}

The  last contribution to the Born terms that we
considered was the \(\Delta\)-in-flight, which is given by:
\begin{eqnarray}
t_{\lambda_{\Delta}\lambda_{\gamma}\lambda_{p}}^{\Delta}= 
2g_{\Delta}(Q^{2},t)\frac{(2p_{2}^{\mu}-
q^{\mu})\varepsilon_{\mu}(q,\lambda_{\gamma})}
{u-m_{\Delta}^{2}} \cdot\nonumber \\
\overline{u}_{\nu}(p_{2},\lambda_{\Delta})p_{\pi}^{\nu}
u(p_{1},\lambda_{p}),
\end{eqnarray}
where u is the Mandelstam variable corresponding to the crossed
invariant momentum transfer \(u=(p_{1}-p_{\pi})^2\),
 and \(g_{\Delta}(Q^{2},t)\)
is the product of
electromagnetic and strong vertex functions. The vertex function 
\(g_{\Delta}(Q^{2},t)\) is related to the \(g_{\pi}(Q^{2},t)\) and  \(g_{N}\)
vertex functions  by current conservation of
the total  Born amplitude:
\begin{equation}
g_{\Delta}(Q^{2},t) = \frac{g_{\pi}(Q^{2},t)+g_{N}(Q^{2})}{2} .
\end{equation}

Finally we substituted $\pi$ propagator in the pion in flight diagram (right top Fig.~\ref{piborn}) 
by the $\pi$-Regge trajectory, following to the approach proposed in \cite{Lag98}:

\begin{eqnarray}
\frac{1}{t-m^{2}_{\pi}} \rightarrow R_{\pi}(t) = \left( S \over S_{0}
\right)^{\alpha_{\pi}(t)}\nonumber \\
\frac{\pi \alpha^{'}_{\pi}}{\sin(\pi
\alpha_{\pi}(t))} \cdot \frac{1+e^{-i\pi \alpha_{\pi}(t)}}{2}
\cdot \frac{1}{\Gamma(1+\alpha_{\pi}(t))}
\label{regge}
\end{eqnarray}
where $S_{0}$=1.0 $GeV^{2}$, and $\alpha^{'}$=0.7 GeV$^{-2}$ is
the slope of the pion Regge trajectory, and $\Gamma$ is the gamma function.  
The pion Regge trajectory
$\alpha_{\pi}(t)$ is given by:
\begin{equation}
\alpha_{\pi}(t) = \alpha^{'}(t-m^{2}_{\pi}),
\end{equation} 
where  the dimension variables are in units of GeV$^{2}$. 
 We also implemented an additional W-independent multiplicative   
factor $\alpha_{R}$ to the 
Reggeized t-channel Born amplitudes in order to account for the differences in effective 
electromagnetic and 
hadronic couplings for Regge trajectory and for the pion. 
Parameter $\alpha_{R}$ was determined from the data fit and is equal to 1.35.

Reggeized Born terms were obtained by multiplying all Born amplitudes with a common factor:
\begin{equation}
\alpha_{R}(t-m^{2}_{\pi})R_{\pi}(t)
\label{transf}
\end{equation}
In this way we maintain the full Reggeized Born amplitude, which remains current
conserving, 
while the t-channel term acquires the propagator  $R_{\pi}(t)$ Eq.~(\ref{regge}), 
instead of the pion pole.

\section*{Appendix II: The amplitudes for non-resonant mechanisms in $\pi$ $\Delta$ 
channels beyond Reggeized Born terms.}
Here we present the expressions for non-resonant mechanisms 
in $\pi \Delta$ channels, complementary to the Reggeized Born terms, described
in Appendix I. These mechanisms
were parametrized by two additional contact terms:

\begin{eqnarray}
\label{extracontact} 
t^{c}=(A(W,Q^{2})\epsilon^{\gamma}_{\mu}\overline{u}^{\nu}_{\Delta}
\gamma^{\mu}up^{\pi}_{\nu}+\nonumber\\ 
B(W,Q^{2})
\epsilon^{\gamma}_{\nu}\overline{u}^{\nu}_{\Delta}\gamma^{\delta}u(2p_{\pi}-q_{\gamma})_{\delta})
\frac{1}{t-\Lambda^{2}} .
\end{eqnarray}
Parameters A($W$,$Q^{2}$), B($W$,$Q^{2}$) were fitted to the 
CLAS data 
in each bin of $W$ and $Q^{2}$ independently. The factor
$\frac{1}{t-\Lambda^{2}}$ in Eq. (\ref{extracontact}) allowed us 
to better describe the angular distributions of the final
hadrons. The cut-off parameter 
$\Lambda^{2}$ was determined to 1.64 GeV$^{2}$ from the data fit.

These terms could effectively account for the additional contributions to the $\pi \Delta$ final states from 
other open exclusive channels arising from hadronic final state interactions. These extra terms
may come from other mechanisms in  $\pi \Delta$ production in addition to  Reggeized Born terms.
Lorentz structure of the
transition $p \rightarrow \Delta$ current could be more complex than a simple tensor structure $g_{\mu\nu}$, which was used
for the Reggeized Born terms of the Appendix I.   
The two terms in Eq. (\ref{extracontact}) may originate from the additional tensor
structures, which may enter into this current:

\begin{eqnarray}
\label{newstr}
\gamma^{\mu}p^{\pi}_{\nu},\nonumber  \\
p_{c}^{\delta}\gamma_{\delta}g_{\mu\nu} ,
\end{eqnarray}
where $p_{c}=(2p_{\pi}-q_{\gamma})$ is the difference of the final pion momentum and 
momentum transfered $q_{\gamma}-p_{\pi}$. The tensor 
structures (\ref{newstr}) being contracted to the spin-tensors of the initial and the final particles
gives us the first and the second terms in Eq. (\ref{extracontact}), respectively. 

We found no evidence for contributions to the $p \rightarrow \Delta$ transition
current from other tensor structures.

The parameters A($W$,$Q^{2}$), B($W$,$Q^{2}$) in Eq. (\ref{extracontact}), obtained
 from the CLAS data fit, are
listed in the tables below for both $\pi^{-} \Delta^{++}$ and $\pi^{+}
\Delta^{0}$ isobar channels.In the fits we varied sum of A($W$,$Q^{2}$) and B($W$,$Q^{2}$), while their ratio was fixed in preliminary adjustment to the
CLAS data \cite{Fe07a}.

\begin{table}
\centering
\caption{The parameters A($W$,$Q^{2}$), B($W$,$Q^{2}$) 
in Eq. (\ref{extracontact}) derived from the CLAS
data fit at $Q^{2}$=0.275 GeV$^2$ \label{abpa1}}
\vspace{0.2cm}
\begin{tabular}{|c|c|c|c|c|}
\hline
W, GeV &  A(W,$Q^{2}$), & B(W,$Q^{2}$),  & A(W,$Q^{2}$),   &B(W,$Q^{2}$),  \\
 &GeV $\pi^{-} \Delta^{++}$  & GeV$ \pi^{-} \Delta^{++}$  &GeV $\pi^{+} \Delta^{0}$    &GeV  $\pi^{+} \Delta^{0}$\\
 & channel&                       channel&         channel &    channel\\ 
\hline
 1.31& 7.0&    63.0&    5.0& 45.0\\
 1.34& 4.5&    40.5&    5.0& 50.0\\
 1.36& 22.5&    22.5&    17.0& 17.0\\
 1.39& 22.5&    22.5&    30.0&30.0\\
 1.41& 15.0&    10.0&    36.0&24.0\\
 1.44& 16.0&    4.0&    62.4& 15.6\\
 1.46& 18.0&    2.0&    90.0& 10.0\\
 1.49& 14.4&    3.6&    88.0& 12.0\\
 1.51& 9.0&    1.0&    126.0&  14.0\\
 1.54& 7.2&    0.8&    126.0& 14.0\\
 1.56& 5.4&    0.6&    128.0& 14.5\\
\hline

\end{tabular}
\end{table}

\begin{table}
\centering
\caption{The parameters A($W$,$Q^{2}$), B($W$,$Q^{2}$) 
in Eq. (\ref{extracontact}) derived from the CLAS
data fit at $Q^{2}$=0.325 GeV$^2$ \label{abpa2}}
\vspace{0.2cm}
\begin{tabular}{|c|c|c|c|c|}
\hline
W, GeV &  A(W,$Q^{2}$), & B(W,$Q^{2}$),  & A(W,$Q^{2}$),   &B(W,$Q^{2}$),  \\
 &GeV $\pi^{-} \Delta^{++}$  & GeV$ \pi^{-} \Delta^{++}$  &GeV $\pi^{+} \Delta^{0}$    &GeV  $\pi^{+} \Delta^{0}$\\
 & channel&                       channel&         channel &    channel\\ 

\hline
 1.31& 8.0&    72.0&    3.7& 33.3\\
 1.34& 9.6&    86.4&    6.0& 54.0\\
 1.36& 25.0&    25.0&    77.5& 77.5\\
 1.39& 20.0&    20.0&    75.0&75.0\\
 1.41& 28.0&    12.0&    80.5&34.5\\
 1.44& 56.0&    14.0&    92.0& 23.0\\
 1.46& 56.0&    14.0&    104.0& 11.0\\
 1.49& 20.0&    5.0&    100.0& 25.0\\
 1.51& 12.0&    3.0&    135.0&  15.0\\
 1.54& 12.0&    3.0&    153.0& 17.0\\

\hline

\end{tabular}
\end{table}

\begin{table}
\centering
\caption{The parameters A($W$,$Q^{2}$), B($W$,$Q^{2}$) 
in Eq. (\ref{extracontact}) derived from the CLAS
data fit at $Q^{2}$=0.375 GeV$^2$ \label{abpa3}}
\vspace{0.2cm}
\begin{tabular}{|c|c|c|c|c|}
\hline
W, GeV &  A(W,$Q^{2}$), & B(W,$Q^{2}$),  & A(W,$Q^{2}$),   &B(W,$Q^{2}$),  \\
 &GeV $\pi^{-} \Delta^{++}$  & GeV$ \pi^{-} \Delta^{++}$  &GeV $\pi^{+} \Delta^{0}$    &GeV  $\pi^{+} \Delta^{0}$\\
 & channel&                       channel&         channel &    channel\\ 

\hline
 1.31& 7.0&    63.0&    3.7& 33.3\\
 1.34& 8.0&    72.0&    6.0& 54.0\\
 1.36& 29.0&    29.0&    35.0& 35.0\\
 1.39& 25.0&    25.0&    25.0&25.0\\
 1.41& 33.6&    14.4&    21.0&10.0\\
 1.44& 40.0&    10.0&    32.0& 8.0\\
 1.46& 22.5&    2.5&    45.5& 5.0\\
 1.49& 16.0&    4.0&    56.0& 14.0\\
 1.51& 16.0&    4.0&    81.0&  9.0\\
 1.54& 16.0&    4.0&    108.0& 12.0\\

\hline

\end{tabular}
\end{table}

\begin{table}
\centering
\caption{The parameters A($W$,$Q^{2}$), B($W$,$Q^{2}$) 
in Eq. (\ref{extracontact}) derived from the CLAS
data fit at $Q^{2}$=0.425 GeV$^2$ \label{abpa4}}
\vspace{0.2cm}
\begin{tabular}{|c|c|c|c|c|}
\hline
W, GeV &  A(W,$Q^{2}$), & B(W,$Q^{2}$),  & A(W,$Q^{2}$),   &B(W,$Q^{2}$),  \\
 &GeV $\pi^{-} \Delta^{++}$  & GeV$ \pi^{-} \Delta^{++}$  &GeV $\pi^{+} \Delta^{0}$    &GeV  $\pi^{+} \Delta^{0}$\\
 & channel&                       channel&         channel &    channel\\ 
\hline
 1.31& 9.8&    88.2&    3.7& 33.3\\
 1.34& 8.4&    75.6&    4.0& 36.0\\
 1.36& 55.0&    55.0&    30.0& 30.0\\
 1.39& 53.0&    53.0&    20.0&20.0\\
 1.41& 80.6&    34.5&    21.0&9.0\\
 1.44& 80.0&    20.0&    24.0& 6.0\\
 1.46& 54.0&    6.0&    45.0& 5.0\\
 1.49& 38.4&    9.6&    32.0& 8.0\\
 1.51& 18.0&    2.0&    26.0&  4.0\\

\hline

\end{tabular}
\end{table}

\begin{table}
\centering
\caption{The parameters A($W$,$Q^{2}$), B($W$,$Q^{2}$) 
in Eq. (\ref{extracontact}) derived from the CLAS
data fit at $Q^{2}$=0.475 GeV$^2$ \label{abpa5}}
\vspace{0.2cm}
\begin{tabular}{|c|c|c|c|c|}
\hline
W, GeV &  A(W,$Q^{2}$), & B(W,$Q^{2}$),  & A(W,$Q^{2}$),   &B(W,$Q^{2}$),  \\
 &GeV $\pi^{-} \Delta^{++}$  & GeV$ \pi^{-} \Delta^{++}$  &GeV $\pi^{+} \Delta^{0}$    &GeV  $\pi^{+} \Delta^{0}$\\
 & channel&                       channel&         channel &    channel\\     
\hline
 1.31& 14.5&    131.5&    7.7& 69.3\\
 1.34& 10.0&    90.0&    7.5& 67.5\\
 1.36& 55.0&    55.0&    70.0& 70.0\\
 1.39& 45.0&    45.0&    55.0&55.0\\
 1.41& 91.0&    39.0&    91.0&39.0\\
 1.44& 80.0&    20.0&    96.0& 24.0\\
 1.46& 45.0&    5.0&    90.0& 10.0\\
 1.49& 16.0&    4.0&    48.0& 12.0\\

\hline

\end{tabular}
\end{table}

\begin{table}
\centering
\caption{The parameters A($W$,$Q^{2}$), B($W$,$Q^{2}$) 
in Eq. (\ref{extracontact}) derived from the CLAS
data fit at $Q^{2}$=0.525 GeV$^2$ \label{abpa6}}
\vspace{0.2cm}
\begin{tabular}{|c|c|c|c|c|}
\hline
W, GeV &  A(W,$Q^{2}$), & B(W,$Q^{2}$),  & A(W,$Q^{2}$),   &B(W,$Q^{2}$),  \\
 &GeV $\pi^{-} \Delta^{++}$  & GeV$ \pi^{-} \Delta^{++}$  &GeV $\pi^{+} \Delta^{0}$    &GeV  $\pi^{+} \Delta^{0}$\\
 & channel&                       channel&         channel &    channel\\     
\hline
 1.31& 8.5&    76.5&    7.7& 69.3\\
 1.34& 8.0&    80.0&    7.5& 67.5\\
 1.36& 90.0&    90.0&    60.0& 60.0\\
 1.39& 75.0&    75.0&    50.0&50.0\\
 1.41& 87.5&    37.5&    91.0&39.0\\
 1.44& 88.0&    22.0&    104.0& 26.0\\
 1.46& 81.0&    9.0&    99.0& 9.0\\
 1.49& 30.4&    7.6&    38.4& 9.6\\

\hline

\end{tabular}
\end{table}

\begin{table}
\centering
\caption{The parameters A($W$,$Q^{2}$), B($W$,$Q^{2}$) 
in Eq. (\ref{extracontact}) derived from the CLAS
data fit at $Q^{2}$=0.575 GeV$^2$ \label{abpa7}}
\vspace{0.2cm}
\begin{tabular}{|c|c|c|c|c|}
\hline
W, GeV &  A(W,$Q^{2}$), & B(W,$Q^{2}$),  & A(W,$Q^{2}$),   &B(W,$Q^{2}$),  \\
 &GeV $\pi^{-} \Delta^{++}$  & GeV$ \pi^{-} \Delta^{++}$  &GeV $\pi^{+} \Delta^{0}$    &GeV  $\pi^{+} \Delta^{0}$\\
 & channel&                       channel&         channel &    channel\\     
\hline
 1.31& 7.5&    67.5&    4.7& 42.3\\
 1.34& 6.8&    61.2&    7.5& 67.5\\
 1.36& 80.0&    80.0&    60.0& 60.0\\
 1.39& 75.0&    75.0&    55.0&55.0\\
 1.41& 84.0&    36.0&    91.0&39.0\\
 1.44& 96.0&    24.0&    104.0& 26.0\\

\hline

\end{tabular}
\end{table}

\section*{Appendix III: Full three-body amplitudes for isobar channels.}
The three-body amplitudes for $\gamma p \rightarrow \pi \Delta \rightarrow p\pi^{+}\pi^{-} $ 
isobar channels $T^{\pi\Delta}_{\gamma^*N, \pi\pi N}$ 
(Eq. (\ref{eq:full-t}))  were calculated using a  Breit-Wigner
ansatz as a product of 
quasi-two-body $\pi \Delta$ production amplitudes $\langle \pi \lambda_{\Delta} \vert T \vert 
\lambda_{p} \lambda_{\gamma} \rangle$, shown in square brackets in Eq. 
(\ref{eq:pid-t}); $\Delta \rightarrow \pi N$ $\langle \pi \lambda_{p'} \vert T \vert 
\lambda_{\Delta} \rangle$
decay amplitudes ($\Gamma_{\Delta,\pi N}$ in Eq. (\ref{eq:pid-t})) and $\Delta$ 
propagator($G_{\Delta}$ in Eq. (\ref{eq:pid-t})) :
\begin{eqnarray}
\label{iso3body}
T^{\pi\Delta}_{\gamma^*N, \pi\pi N}=\sum_{\lambda_{\Delta}}\frac{\langle \pi \lambda_{p'} \vert T \vert 
\lambda_{\Delta} \rangle\langle \pi \lambda_{\Delta} \vert T \vert 
\lambda_{p} \lambda_{\gamma} \rangle
}{M^{2}_{\Delta}-M^{2}_{\pi p}-i\Gamma_{\Delta}(M_{\pi p})M_{\Delta}} , 
\end{eqnarray}
where $M_{\Delta}$ ,  $\Gamma_{\Delta}(M_{\pi p})$, are the $\Delta$ mass and 
total hadronic decay width, which is a function of running $M_{\pi p}$  masses.
The mass
dependence was obtained by calculating the total decay width $\Gamma_{\Delta}(M_{\pi p})$
 from the matrix elements 
(\ref{decdelta}). Therefore, expression Eq. (\ref{iso3body}) fulfills 
the unitarity conditions.
The sum is running over helicities $\lambda_{\Delta}$.

The $\langle \pi \lambda_{p'} \vert T \vert 
\lambda_{\Delta} \rangle$ decay amplitudes  were calculated as: 
\begin{eqnarray} 
\langle \pi \lambda_{p'} \vert T \vert 
\lambda_{\Delta}\rangle=g_{\Delta}F_{\Delta}(M_{\pi p})
\overline{u}_{p'}u^{\mu}_{\Delta}p^{\pi}_{\mu} , 
\label{decdelta}        
\end{eqnarray}
The coupling constant $g_{\Delta}$, 
was fitted to the total 
hadronic decay widths of the $\Delta$ at the central mass 1.23 GeV:
\begin{eqnarray}
g_{\Delta^{++} \rightarrow \pi^{+} p}=15,  \\
g_{\Delta^{+} \rightarrow \pi^{-} p}=15*(1/\sqrt{3}).
\end{eqnarray}
A numerical factor $1/\sqrt{3}$ represents isospin Clebsch-Gordan coefficient.

Mass dependence of the total hadronic decay widths $\Gamma_{\Delta}(M_{\pi p})$
was determined through the Lorentz structure of decay amplitudes (\ref{decdelta}), 
as well as the hadronic form factors  $F_{\Delta}(M_{\pi p})$ \cite{Lo}:
\begin{eqnarray}
F_{\Delta}(M_{\pi p})=\frac{F(P^{*}_{\pi R})}{F(P^{*}_{\pi})},\nonumber  \\ 
F(P^{*}_{\pi})=\sqrt{\frac{P^{*2}_{\pi}}{\Lambda^{2}_{\Delta}+P^{*2}_{\pi}}}\left\{\frac{P^{*}_{\pi
R}}{P^{*}_{\pi}}\right\}.          
\end{eqnarray}
where $P^{*}_{\pi}$, $P^{*}_{\pi R}$  are moduli of pion three-momenta in the rest frame of the 
$\Delta$ at current and the central $M_{\pi p}$ respectively.
$\Lambda_{\Delta}=0.235$ GeV.

\section*{Appendix IV: Cross-sections and amplitudes in JM06 model.}
All amplitudes were calculated for the S-matrix defined as:
\begin{equation}
S=I+(2\pi)^{4}\delta^{4}(P_{f}-P_{i})iT,
\end{equation} 
where $P_{f}$ and $P_{i}$ are total four momenta of the final and the initial
particles respectively. The Dirac spinors were normalized as:

\begin{equation}
\label{norm}
\overline{U_{p}}U_{p}=2M_{N},
\end{equation}
where $U_{p}$, ($\overline{U_{p}}$)  are Dirac (conjugated Dirac) spinors,  
$M_{N}$ is the nucleon mass. With this parameterization of the S-matrix and
Dirac spinor normalization,  the phase space element for the final particle $i$ 
with three-momentum vector $\vec{p_{i}}$ and energy $E_{i}$  is defined as:
\begin{equation}
d^{3}\vec{p_{i}}/(2E_{i}(2\pi)^{3}),
\end{equation}

All time-space tensors (currents, the particle
four-momenta) in JM model correspond to the  $g_{\mu \nu}$ tensor ($\mu$=0,1,2,3,4) with
the components: $g_{00}=1$, $g_{11}=g_{22}=g_{33}=-1$.

The cross-sections of $p\pi^{+}$$\pi^{-}$ exclusive reaction induced by virtual photons absorption off the
protons were determined 
in the single photon exchange approximation. These cross-sections are related to the measured exclusive electron scattering 
cross-sections according to Eq. (\ref{eq:sigma-e}) of Section~\ref{xsect}. This formalism is described in details
in \cite{Am89}. The kinematics of the $p\pi^{+}$$\pi^{-}$ final state 
are determined
unambiguously by the 5-fold  differential 
cross-section $d^{5}\sigma$ (see Section~\ref{xsect}). This 5-fold differential
cross-section for $p\pi^{+}$$\pi^{-}$ production by virtual photons off the protons was 
calculated as a contraction of leptonic and hadronic tensors divided by the
invariant virtual photon flux and multiplied by $d^{5}\Phi$ phase space 
differential for the 3-body final state. The leptonic tensor
$L_{\mu\nu}$ is
well known from QED~\cite{Am89}.  
The hadronic tensor represents a product of the hadronic currents 
$J_{\mu}^{*}$ and $J_{\nu}$ contracted to the spin-density matrices for the
initial and the final hadrons.

This paper deals with spin averaged differential
cross sections, that are independent from any polarization observable. 
The $d^{5}\sigma$ cross-section is computed for 
unpolarized electron beam, proton target and with the unity spin-density
matrices for the final state hadrons. In order to get rid of virtual photon
polarization degree of freedom, we integrate $d^{5}\sigma$ cross-section 
over the azimuthal $\phi$ angle of 
one of the final
hadrons, defined in the Section~\ref{xsect}.
This integration results in $\phi$-independent 4-fold differential 
cross-section.

The 4-fold differential $\gamma_{virt} p \rightarrow p\pi^{+}\pi^{-} $ 
cross-section after integration over the final $\pi^-$ azimuthal angle $\phi_{\pi^{-}}$ is given by: 
\begin{equation}
d^{4}\sigma=\frac{4\pi\alpha}{4K_{L}M_{N}}\left\{\frac{J_{x}^{*}J_{x}+J_{y}^{*}J_{y}}{2}+
\epsilon_{L} J_{z}^{*} J_{z} \right\} d^{4}\phi,
\label{difcrsect}
\end{equation}
where $\alpha$ is fine structure
constant, $\epsilon_{L}$ stands for degree of longitudinal polarization of virtual photons, as it was defined in \cite{Am89}. 
The factor $4K_{L}M_{N}$ is the invariant virtual photon flux, $M_{N}$ is nucleon mass, 
$K_{L}$ is the equivalent photon energy$:$
\begin{equation}
K_{L}=\frac{W^{2}-M_{N}^{2}}{2M_{N}},
\end{equation}
$d^{4}\phi$ stands for the 3-body phase space differential after integration over $\phi_{\pi^{-}}$. 
This
differential may be expressed in terms of the final state kinematic variables 
as:
\begin{eqnarray}
d^{4}\phi=2\pi\frac{1}{32W^{2}(2\pi)^{5}}ds_{\pi^{+}\pi^{-}}ds_{\pi^{+}p}
d\theta_{\pi^{-}}d\alpha_{[p'\pi^{+}][p\pi^{-}]},\nonumber  \\
ds_{\pi^{+}\pi^{-}}=dM_{\pi^{+}\pi^{-}}^{2},\nonumber  \\
ds_{\pi^{+}p}=dM_{\pi^{+}p}^{2},
\label{diff3b}
\end{eqnarray}
where $W$ is invariant mass of the final hadron system,
while the $M_{i j}$ are the invariant masses of the i,j pair of the final hadrons. All
angles for the final hadrons are defined in the hadronic C.M. frame
(see Section~\ref{xsect} for definition of kinematic variables).

The Eq.(\ref{difcrsect}) also gives the 1-fold differential 
cross-sections for isobar channels $\gamma p
\rightarrow \pi \Delta$ with the unstable final particle $\Delta$ of running 
mass $M_{\Delta}$. In order to get these cross sections, $d^{4}\phi$ differential 
in Eq. (\ref{difcrsect}) should be
replaced by 1-fold differential 
d$\Phi_{2b}$ describing phase space element for two body final state after 
integration over the final hadron 
$\phi$ angle: 
\begin{eqnarray}
d\Phi_{2b}=2\pi\frac{p_{f}}{4W}d\theta_{f},  \\
E_{f}=\frac{W^2+M_{f}^2-M_{f'}^2}{2W},  \\
p_{f}=\sqrt{E_{f}^2-M_{f}^2},
\label{diff2b}
\end{eqnarray}
where $E_{f}$, $P_{f}$ are the energy and momentum modules of one of the  final
hadron $f$ ($f$=$\pi$, $\Delta$), $M_{f}$ is its mass, 
while the index $f'$ stands for the other hadron. 
All frame dependent kinematic variables of the final hadrons are defined 
in the $\pi \Delta$ or $\gamma_{virt} p$ 
C.M. frame. 

The longitudinal
polarization parameter $\epsilon_{L}$ of the virtual photon is determined by 
the electron electromagnetic currents. The QED
calculations give \cite{Am89}:
\begin{equation}
\label{espl}
\epsilon_{L}=\sqrt{\frac{Q^{2}}{\nu^{2}}}\left\{1+2\frac{Q^{2}+\nu^{2}}{Q^{2}}tg^{2}
\frac{\theta_{e'}}{2}\right\}^{-1},
\end{equation} 
where $\nu$ is Lorentz invariant in electron scattering:
\begin{equation}
\nu=\frac{(qp)}{M_{N}},
\end{equation}
and q and p are the four-momenta of photon and target proton, respectively.
The $\nu$ value is equal to the 
energy transfered to the virtual photon in the lab 
frame. $\theta_{e'}$
is the electron scattering angle in the lab. frame. 

The
hadronic current $J_{\nu}$ and the virtual photon vectors $\epsilon(\lambda_{\gamma})$ ($\lambda_{\gamma}$=-1,0,+1)  are related to 
reaction  helicity amplitudes 
$\langle \lambda_{f} \vert T \vert \lambda_{p} \lambda_{\gamma} \rangle$ as:

\begin{eqnarray}
\label{curramp}
\epsilon_{\nu}(\lambda_{\gamma}=-1)J^{\nu}(\lambda_{p},\lambda_{f})=\langle \lambda_{f}
\vert T \vert \lambda_{p} \lambda_{\gamma}=-1 \rangle, \\
\epsilon_{\nu}(\lambda_{\gamma}=1)J^{\nu}(\lambda_{p},\lambda_{f})=\langle \lambda_{f}
\vert T \vert \lambda_{p} \lambda_{\gamma}=1 \rangle, \\
\epsilon_{\nu}(\lambda_{\gamma}=0)J^{\nu}(\lambda_{p},\lambda_{f})=\langle
\lambda_{f}
\vert T \vert \lambda_{p} \lambda_{\gamma}=0 \rangle,
\label{ampcurr1}
\end{eqnarray}
where$\lambda_{i}$ (i=$\gamma , p$) stand for the initial state photon and proton
helicities. The $\lambda_{f}$ is generic symbol for the helicities 
in the final state.  The vectors $\epsilon(\lambda_{\gamma})$ were estimated in the lab. frame in all JM calculations, 
resulting in the hadronic
currents and hadronic tensor determined in lab. frame. \footnotetext[1]{explicit expressions for the photon vectors $\epsilon_{\nu}(\lambda_{\gamma})$ 
may be found in \cite{Am89}.} Contracted components of the leptonic and the hadronic
tensors should be determined in the same frame. 
 Therefore, components of the leptonic tensor 
 should calculated
in the lab. frame. This is a reason why the  kinematical variables in Eq.
(\ref{espl}) were evaluated in the lab. frame.

The $J_{0}$ component of the hadronic current was obtained employing 
current conservation:
\begin{equation}
q_{0}J^{0}-q_{z}J^{z}=0.
\label{consevration}
\end{equation}

The hadronic currents $J_{\nu}$ were derived from Eqs. 
(\ref{curramp}-\ref{consevration}):
\begin{eqnarray}
J_{x}=-\frac{\langle \lambda_{f} \vert T \vert \lambda_{p} \lambda_{\gamma}=1 \rangle
-\langle \lambda_{f} \vert T \vert \lambda_{p} \lambda_{\gamma}=-1
\rangle}{\sqrt{2}},\nonumber  \\
J_{y}=i\frac{\langle \lambda_{f} \vert T \vert \lambda_{p} \lambda_{\gamma}=1 \rangle
+\langle \lambda_{f} \vert T^ \vert \lambda_{p} \lambda_{\gamma}=-1
\rangle}{\sqrt{2}},\nonumber  \\
J_{z}=\frac{\nu}{\sqrt{Q^2}}\langle \lambda_{f} \vert T \vert \lambda_{p}
\lambda_{\gamma}=0 \rangle.
\end{eqnarray}

Since the contraction of the leptonic and the hadronic tensors is Lorentz 
invariant, all expressions for cross sections listed above are valid in any frame. 
However, the expressions for Lorentz invariants in
cross sections incorporate the frame dependent observables defined 
in particular frames: electron scattering angle in the lab frame, 
the final hadron angles in C.M.frame. 

\section*{Appendix V: Parameters for the resonant mechanisms}

\begin{table}[h!]
\centering
\caption{The initial values of $P_{11}(1440)$ and $D_{13}(1520)$ electrocouplings
used in calculations of the contributions from various isobar channels. \label{abpar1}}
\vspace{0.2cm}
\label{elpar}
\begin{tabular}{|c|c|c|c|c|c|}
\hline
                   &      $P_{11}(1440)$&  $P_{11}(1440)$& $D_{13}(1520)$   & $D_{13}(1520)$& $D_{13}(1520)$\\
$Q^{2}$,  & $A_{1/2}$*10$^{3}$& $S_{1/2}$*10$^{3}$&$A_{1/2}$*10$^{3}$&$S_{1/2}$*10$^{3}$ & $A_{3/2}$*10$^{3}$ \\  
        GeV$^{2}$           & GeV$^{-1/2}$  & GeV$^{-1/2}$ &     GeV$^{-1/2}$ & GeV$^{-1/2}$ &GeV$^{-1/2}$  \\

\hline
 0.275& -33.0&    50.0&    -55.0& -47.0& 70.0 \\
 0.325& -25.0&   51.0 &    -58.0& -50.0& 70.0 \\
 0.375& -15.0&   36.0 &    -65.0& -62.0& 80.0 \\
 0.425&  0.0 &   43.0&     -80.0& -48.0& 70.0\\
 0.475&  17.0&   44.0 &    -65.0&  -48.0& 70.0\\
 0.525&  20.0&   45.0 &    -60.0&  -48.0&  70.0\\
 0.575&  27.0&    46.0 &   -55.0&  -48.0 &  63.0\\         
\hline

\end{tabular}
\end{table}

\begin{table}
\centering
\caption{The $P_{11}(1440)$ and $D_{13}(1520)$ hadronic
decay widths
used to extract contributions from different isobar channels.
  \label{hadpar}}
\vspace{0.2cm}
\begin{tabular}{|c|c|c|}
\hline
                   &      $P_{11}(1440)$&  $D_{13}(1520)$   \\ 
\hline
 $\Gamma_{tot}$, MeV &320.&    125.\\
 $\Gamma_{\pi \Delta}$, MeV &   77. &   29. \\
 $\Gamma_{\rho p}$, MeV  &    0.& 11. \\        
\hline

\end{tabular}
\end{table}

Here we present the initial values of the parameters of $P_{11}(1440)$ and $D_{13}(1520)$
resonances used in computing the various isobar channel contributions. They were obtained from interpolation 
of previously available CLAS/world $N^*$ electrocoupling
data \cite{Burk2003} to the kinematical area covered by the recent CLAS data \cite{Fe07a}.
The starting values of hadronic parameters were taken from previous analysis 
of the CLAS $N\pi\pi$ electroproduction data within the framework of JM05 model
\cite{Mo06}.  
Interpolated electrocoupling values were further adjusted 
to measured differential cross sections. 
Only listed in the Tables~\ref{elpar},\ref{hadpar} resonances have a
measurable impact on 
observables in the CLAS data. The contributions from other
resonances are well inside the data uncertainties. Listed in the tables below 
resonance parameters were varied as it was
described in the Section~\ref{fit}

These parameters should be used in combination 
with non-resonant JM06 parameters in order to describe superposition of resonant 
and non-resonant mechanisms in various isobar channels. They can not be 
interpreted as resonance parameter values derived
from the data fit, since no attempt was made to improve the determined $N^{*}$ parameters as well as to evaluate 
the uncertainties of $N^{*}$ parameters, related to the accuracies of 
measured cross sections.

\section*{Appendix VI: The set of partial waves, used for expansion of the
non-resonant helicity amplitudes in $\pi \Delta$
isobar channels}

Non-resonant helicity amplitudes in $\pi \Delta$ isobar channels were expanded
over the set of partial waves, corresponding to the  
quantum states with the total
angular momenta $J$ and it's projections $\mu$ and $\nu$ onto the directions along the initial and
final particle momenta in the C.M, frame. The defined partial waves $\langle \pi \lambda_{\Delta} | T^{J} |
\lambda_{\gamma} \lambda_{p} \rangle$ are related to the full helicity amplitude
$\langle \pi \lambda_{\Delta} | T | \lambda_{\gamma}
\lambda_{p} \rangle $ as:
\begin{eqnarray}
\label{jopa1} 
\langle \pi \lambda_{\Delta} | T^{J} |
\lambda_{\gamma} \lambda_{p} \rangle =  \int\limits^{\pi}_{0}
\frac{2J+1}{2} \langle \pi \lambda_{\Delta} | T | \lambda_{\gamma}
\lambda_{p} \rangle\cdot\nonumber \\ 
 d^{J}_{\mu\nu}
(\cos\theta_{\pi}) \sin\theta_{\pi} d\theta_{\pi}\nonumber  \\
\mu  =   \lambda_{\gamma} - \lambda_{p}\nonumber \\
\nu  =  -\lambda_{\Delta}.
\end{eqnarray}

The helicity amplitudes $\langle \pi \lambda_{\Delta} | T | \lambda_{\gamma}
\lambda_{p} \rangle $ may be written 
as infinite sums over the partial waves in Eq. (\ref{jopa1})

\begin{equation}
\label{jopa2} \langle \pi \lambda_{\Delta} | T | \lambda_{\gamma}
\lambda_{p} \rangle =  \sum\limits_{J}  \langle \pi
\lambda_{\Delta} | T^{J} | \lambda_{\gamma} \lambda_{p} \rangle
d^{J}_{\mu\nu} (\cos\theta_{\pi}),
\end{equation}
where $\theta_{\pi}$ is the pion polar emission angle in the CM frame. CLAS data
analysis showed that at $W$ $<$ 1.55 GeV the partial wave basis may be restricted to $J_{max}$=5/2.   
Contribution from partial waves with
$J$ $>$ 5/2 are inside the data uncertainties.

The helicity amplitudes depend on 
the $W$, $Q^{2}$
variables, describing the initial state and on the $\theta_{\pi}$ , 
$M_{\pi p}$ variables of the final state, where $M_{\pi p}$ stands 
for running mass of 
the $\Delta$ isobar. After integration (\ref{jopa1}) the partial waves 
$\langle \pi \lambda_{\Delta} | T^{J} |
\lambda_{\gamma} \lambda_{p} \rangle$ depend on $W$, $Q^{2}$ and $M_{\pi p}$ variables
only. All helicity amplitudes and partial waves, discussed 
here, contain the factor 
$e^{i(\lambda_{\gamma}-\lambda_{p})\phi_{\pi}}$, which describes  their
dependence on the pion azimuthal emission angle $\phi_{\pi}$.
This exponential factor is retained for any reaction dynamic, being a
consequence of the
rotational invariance of the production amplitudes.

Instead of helicity representation, $LS$ representation is frequently 
used for the description of 
the $\pi \Delta$ final
state. The partial waves $\langle \pi \lambda_{\Delta} | T^{J} | \lambda_{\gamma}
\lambda_{p} \rangle $ in helicity representation
 may be expressed through the partial waves
$\langle LS(\pi \Delta) | T^{J} |
\lambda_{\gamma} \lambda_{p} \rangle$, in  $LS$ representation as:

\begin{equation}
\label{jopa3} \langle \pi \lambda_{\Delta} | T | \lambda_{\gamma}
\lambda_{p} \rangle =  \sum\limits_{LS} \langle  \pi
\lambda_{\Delta} | LS \rangle \langle LS(\pi \Delta) | T^{J} |
\lambda_{\gamma} \lambda_{p} \rangle
\end{equation}

The transition coefficients $\langle  \pi
\lambda_{\Delta} | LS \rangle $ are given by the products of Clebsch-Gordan
coefficients and the factor accounting for different wave function
normalizations in the $LS$ and helicity representations

\begin{eqnarray}
\label{jopa4} \langle LS | \pi \lambda_{\Delta} \rangle =
\sqrt{\frac{2J+1}{2L+1}} \langle  L& 0& S=3/2 -\lambda_{\Delta} |
J -\lambda_{\Delta}\rangle\cdot\nonumber \\ 
\langle  0 & 0 & 3/2  -\lambda_{\Delta} |
S -\lambda_{\Delta} \rangle.
\end{eqnarray}

The Clebsch-Gordan coefficient $\langle  S_{\pi}=0, \lambda_{\pi}=0,  S_{\Delta }=3/2,  -\lambda_{\Delta} |S=3/2, -\lambda_{\Delta} \rangle$ describes 
the spinless $\pi$ and the $\Delta$ of 3/3 spin, which are coupled to the total spin $S$=3/2. 
Since the quantization axis is along  the $\pi$ momenta, the 
total spin projection is -$\lambda_{\Delta}$.
The Clebsch-Gordan coefficient $\langle  L 0 S -\lambda_{\Delta} |J  -\lambda_{\Delta}\rangle$ corresponds 
to the orbital angular momentum $L$ and the
total spin $S$=3/2 of the $\pi \Delta$ system  
coupled to the total angular momentum $J$. 
The projection of the orbital angular
momentum is equal to zero because of choice of the quantization axis mentioned above.  

\end{document}